 \documentclass[mnsc,nonblindrev]{informs3_hide}

\OneAndAHalfSpacedXI 

\usepackage{natbib, multirow, multicol, xspace, enumitem, amsmath, pifont, algorithm, algorithmic, subcaption}
\usepackage{accents}
\bibpunct[, ]{(}{)}{,}{a}{}{,}%
\def\bibfont{\small}%
\usepackage{bm}
\usepackage[normalem]{ulem}
\usepackage[dvipsnames]{xcolor}
\usepackage[all]{xy}

\newcommand{\mathbbm}[1]{\text{\usefont{U}{bbm}{m}{n}#1}} 

\newcommand{\bI}{\mathbbm{1}}
\newcommand{\bbI}{\bar{\mathbbm{1}}}
\newcommand{\bY}{\mathbb{Y}}
\newcommand{\bE}{\mathbb{E}}
\newcommand{\bR}{\mathbb{R}}
\newcommand{\bN}{\mathbb{N}}

\newcommand{\bK}{\mathbb{K}}
\newcommand{\bT}{\mathbb{T}}
\newcommand{\bQ}{\mathbb{Q}}
\newcommand{\tbT}{\tilde{\mathbb{T}}}
\newcommand{\ttt}{\tilde{t}}

\newcommand{\cY}{\mathcal{Y}}
\newcommand{\cQ}{\mathcal{Q}}

\newcommand{\bq}{\bar{q}}

\newcommand{\obs}{\mathsf{obs}}

\newcommand{\so}{\mathsf{o}}

\newcommand{\var}{\mathsf{Var}}

\TheoremsNumberedThrough     
\ECRepeatTheorems

\EquationsNumberedThrough    

\MANUSCRIPTNO{}

\begin{document}


\RUNAUTHOR{Bojinov, Simchi-Levi, Zhao}

\RUNTITLE{Switchback Experiments}

\TITLE{Design and Analysis of Switchback Experiments}

\ARTICLEAUTHORS{
\AUTHOR{Iavor Bojinov}
\AFF{Technology and Operations Management Unit, Harvard Business School, Boston, MA 02163, \EMAIL{ibojinov@hbs.edu}}
\AUTHOR{David Simchi-Levi}
\AFF{Institute for Data, Systems, and Society, Department of Civil and Environmental Engineering, and Operations Research Center, Massachusetts Institute of Technology, Cambridge, MA 02139, \EMAIL{dslevi@mit.edu}}
\AUTHOR{Jinglong Zhao}
\AFF{Boston University, Questrom School of Business, Boston, MA, 02215,
\EMAIL{jinglong@bu.edu}}
}

\ABSTRACT{
Switchback experiments, where a firm sequentially exposes an experimental unit to random treatments, are among the most prevalent designs used in the technology sector, with applications ranging from ride-hailing platforms to online marketplaces. Although practitioners have widely adopted this technique, the derivation of the optimal design has been elusive, hindering practitioners from drawing valid causal conclusions with enough statistical power. We address this limitation by deriving the optimal design of switchback experiments under a range of different assumptions on the order of the carryover effect --- the length of time a treatment persists in impacting the outcome. We cast the optimal experimental design problem as a minimax discrete optimization problem, identify the worst-case adversarial strategy, establish structural results, and solve the reduced problem via a continuous relaxation. For switchback experiments conducted under the optimal design, we provide two approaches for performing inference. The first provides exact randomization based $p$-values, and the second uses a new finite population central limit theorem to conduct conservative hypothesis tests and build confidence intervals. We further provide theoretical results when the order of the carryover effect is misspecified and provide a data-driven procedure to identify the order of the carryover effect. We conduct extensive simulations to study the numerical performance and empirical properties of our results, and conclude with practical suggestions.
}

\maketitle

\section{Introduction}
Academic scholars have appreciated the benefits that experimentation brings to firms for many decades \citep{march1991exploration, sitkin1992learning, sarasvathy2001causation, thomke2001enlightened, johari2015always, kohavi2017surprising, sun2018designing, xiong2019optimal}. However, widespread adoption of the practice has only taken off in the last decade, partly fueled by the rapid cost reductions achieved by firms in the technology sector \citep{kohavi2007practical, kohavi2009online, bakshy2014designing, azevedo2019b, kohavi2020trustworthy}. Most large firms now possess internal tools for experimentation, and a growing number of smaller and more conventional companies are purchasing the capabilities from third-party sellers that offer full-stack integration \citep{thomke2020experimentation}. These tools typically allow simple ``A/B'' tests that compare the standard offering ``A'' to a new or improved version ``B''. The comparisons are made across a range of different business outcomes, and the tests are usually conducted for at least a week \citep{kohavi2020trustworthy}. This simple practice has provided tremendous value to firms \citep{koning2019experimentation}.


However, some firms and authors have recognized the limitations of these simple A/B tests \citep{gupta2019top, bojinov2020avoid}; the two most prominent being handling interference (the scenario where the assignment of one subject impacts another's outcomes) and estimating heterogeneous (or personalized) effects. For example, many online platforms and retail marketplaces often observe varying levels of interference when conducting experiments (see \citet{chamandy2016experimentation, cui2017discrimination, kastelman2018switchback, farronato2018innovation, glynn2020adaptive, holtz2020reducing, li2021interference} for online platforms like Airbnb, DoorDash, Lyft, and Uber; and \citet{caro2012clearance, ferreira2016analytics, cui2019learning, ma2020dynamic} for retail markets like Amazon, AB InBev, Rue la la, Zara) and desire to estimate heterogeneous effects (see \citet{nie2018adaptively, deshpande2018accurate, mcfowland2018efficient, hadad2019confidence}). 

In this paper, we simultaneously tackle both of these two challenges by developing a theoretical framework for the optimal design and analysis of switchback experiments under the minimal amount of assumptions. In switchback experiments, we sequentially expose a unit to a random treatment, measure its response, and repeat the procedure for a fixed period of time \citep{robins1986new, bojinov2019time}. By administering alternate treatments to the same unit, we can directly estimate an individual level causal effect and alleviate the challenges posed by interference.

In addressing the two challenges, many works in the literature assume specific outcome models under interference.
\citet{wager2019experimenting, johari2020experimental, li2021interference} work on experimental design for two-sided online platforms, by assuming that the interference can be captured via game-theoretic modeling.
\citet{glynn2020adaptive} assumes an underlying Markov Chain model and formulates the experimental design problem as estimating the difference between two steady state reward distributions.
Some other literature directly models the interference through a network, \emph{e.g.} \citet{li2015value, athey2018exact, eckles2016design, sussman2017elements, basse2019randomization, puelz2019graph}.
In such models, a treatment assigned to one node of the network creates a ``spillover effect,'' which impacts the outcomes of the neighboring nodes.
All of the above methods make specific assumptions on the outcome models.
If these assumptions hold, the above methods correctly identify the causal effects (or the model parameters) with great precision;
if these assumptions do not hold, the estimates are likely biased.

Unlike the above works, we make no specific outcome model assumptions in this paper.
Instead, we make assumptions about the existence of the carryover effects, which refer to the persistence of past interventions in impacting the future outcomes.
More specifically, we make assumptions on the order of carryover effects, which refers to the duration of time periods of such persistence.
We then establish formal results on the optimal design of switchback experiments under different assumptions of the order of the carryover effects; we also propose a data-driven procedure to estimate the order of the carryover effects.

\noindent \textbf{Applications}. There are two classes of applications where switchback experiments are widely used in practice. The first arises when units interfere with each other either through a network or some more complicated unknown structure. For example, consider a ride-hailing platform that wants to test a new fare pricing algorithm's effectiveness in a large city \citep{farronato2018innovation}. Administering the test version to a subset of drivers can impact their behavior, which, in turn, could change the behavior of drivers that are receiving the old version. Directly comparing the revenue generated by the drivers across the two groups will likely provide a biased estimate of what would happen if everyone were assigned to the new version compared to the old. Instead, practitioners treat the city as a single aggregated unit and use a switchback experiment to estimate the intervention's effectiveness, thereby alleviating the problem caused by interference.
A similar issue often arises in revenue management when, for example, a retailer wants to test the effectiveness of a new promotion planning algorithm \citep{ferreira2016analytics}. Administering the new version to a subset of Stock Keeping Units (SKU's) cannibalizes the sales from the other SKU's. Again comparing the generated revenue across the two groups is unlikely to provide an accurate measure of the promotion's effectiveness. 
Again, practitioners treat all the SKU's as a single aggregated unit and use a switchback experiment to obtain accurate estimates of the promotion's effectiveness.

The second application arises when we have a limited number of experimental units, and we believe the effects are likely to be heterogeneous.
For example, \citet{bojinov2019time} used switchback experiments to make causal claims about the relative effectiveness of algorithms compared with humans at executing large financial trades across a range of financial markets. More generally, psychologists and biostatisticians regularly use switchback experiments whenever studying the effectiveness of an intervention on a single unit, \emph{e.g.}, \cite{lillie2011n} and \cite{boruvka2018assessing}.

\noindent \textbf{Main Contributions}. There are three significant challenges to using switchback experiments. The first is that causal estimators from switchback experiments have large variances as the precision is a function of the total number of assignments.
The second is that past interventions are likely to impact future outcomes; this is often referred to as a carryover effect. Typically, many authors assume that there are no carryover effects \citep{chamberlain1982multivariate, athey2018design, Imai_2019}, although some recent work has relaxed this assumption \citep{robins1986new, Sobel_2012, bojinov2020panel}.
The third is that standard super population inference --- where researchers either assume a model for the outcome, or that the units are sampled from an infinitely large population --- requires unrealistic assumptions that fail to capture the problem's personalized nature \citep{bojinov2019time}.

This paper's main contributions are to address these three challenges and present a framework that allows firms and researchers to run reliable switchback experiments.
First, we derive optimal designs for switchback experiments, ensuring that we select a design that leads to the lowest variance among the most popular class assignment mechanisms.
The designs are optimal in the sense that we search for both the optimal randomization points and the optimal randomization probabilities, which, together, capture the most general class of randomization mechanisms.
Second, we assume the presence of a carryover effect and show that our estimation and inference are valid both when the order of the carryover effect is correctly specified and misspecified, the latter leading to a minor increase in the variance. For practitioners, we also propose a method to identify the order of the carryover effect by running a series of carefully designed switchback experiments.  
Finally, we take a purely design-based perspective on uncertainty; that is, we treat the outcomes as unknown but fixed (or equivalently, we condition on the set of potential outcomes) and assume that the assignment mechanism is the only source of randomness \citep{fisher1937design, kempthorne1955, rubin1980randomization, abadie2020sampling}. The main benefit of a design-based perspective is that the inference, and in turn the causal conclusions, do not depend on our ability to correctly specify a model describing the phenomena we are studying, ensuring that our findings are wholly non-parametric and robust to model misspecification \citep[Chapter~5]{imbens_rubin_2015}.

\noindent \textbf{Roadmap}. The paper is structured as follows.
In Section~\ref{sec:Definitions} we define the notations, the assumptions, and the assignment mechanism that we focus on, which we will refer to as the regular switchback experiments.
In Section~\ref{sec:MinimaxOptimization}, we discuss how to design an effective regular switchback experiment under the minimax rule.
The design is optimal with respect to (i) the optimal treatment assignment probability, and (ii) the randomization frequency and randomization points.
We cast the design problem as a minimax discrete optimization problem, identify the worst-case adversarial strategy, establish structural results, and then explicitly find the optimal design.
In Section~\ref{sec:InferenceAndTesting}, we discuss how to perform inference and conduct statistical testing based on the results obtained from an optimally designed switchback experiment.
We propose an exact test for sharp null hypotheses, and an asymptotic test for testing the average treatment effect.
We also discuss how to make inference when the carryover effect is misspecified, and how to conduct hypothesis testing to identify the true order of the carryover effect.
In Section~\ref{sec:Numerical}, we run simulations to test the correctness and effectiveness of our proposed theoretical results under various simulation setups.
In Section~\ref{sec:PracticalImplications}, we give empirical illustrations on how to conduct a switchback experiment in practice and conclude with limitations which may lead to future research directions.
All technical proofs are in the Appendix.
 

\section{Notations, Assumptions, and Regular Switchback Experiments}
\label{sec:Definitions}

\subsection{Assignment Paths and Potential Outcomes}
\label{sec:AssignmentPathAndPotentialOutcomes}

We focus our discussion on a single experimental unit.
For example, this unit could be a ride-hailing platform testing the effectiveness of a new fare pricing algorithm in a city.
At each time point $t\in [T]=\{1,2,...,T\}$, we assign the unit to receive an intervention $W_t\in\{0,1\}$.
For example, one experimental period could be one to two hours for a ride-hailing platform and $T$ could be two weeks, i.e., $T=336$ when one period is one hour.
In some applications, the time horizon $T$ is pre-determined, \emph{e.g.}, a typical experimental duration for a ride-hailing platform is a few weeks; however, when $T$ is not pre-determined, Section~\ref{sec:PracticalImplications} provides details for how to choose an appropriate $T$.
Throughout most of this paper, with the exception being the derivation of our asymptotic results, we consider $T$ to be a known, fixed constant.

Following convention, we say that the unit is assigned to treatment if $W_t=1$ and control when $W_t=0$; in A/B testing terminology, ``A'' is control and ``B'' is treatment. For example, \cite{chamandy2016experimentation} studied how a new surge-pricing subsidy (the treatment) compared to the current setup without the subsidy (the control). 
The assignment path is then the collection of assignments and is denoted using a vector notation whose dimensions are specified in the subscript, $\bm{W}_{1:T} = (W_1, W_2, ..., W_T) \in \{0,1\}^T$.
We adopt the convention that $\bm{W}_{1:T}$ stands for a random assignment path, while $\bm{w}_{1:T}$ stands for one realization.

After administering the assigned intervention, we observe a corresponding outcome.
For example, this could be the average ride-matching rate (often defined as the proportion of requested rides that were successfully matched with a driver) during each two hour experimental period.
Following the extended potential outcomes framework, at time $t\in [T]$, we posit that for each possible assignment path $\bm{w}_{1:T}$ there exists a corresponding potential outcome denoted by $Y_t(\bm{w}_{1:T})$; the set of all potential outcomes are collected in $\bY = \{Y_t(\bm{w}_{1:T})\}_{t \in [T], \bm{w}_{1:T} \in \{0,1\}^T}$ with support $\bY \in \cY$.

\begin{figure}[!tb]
\centering
\includegraphics[width=0.7\textwidth,]{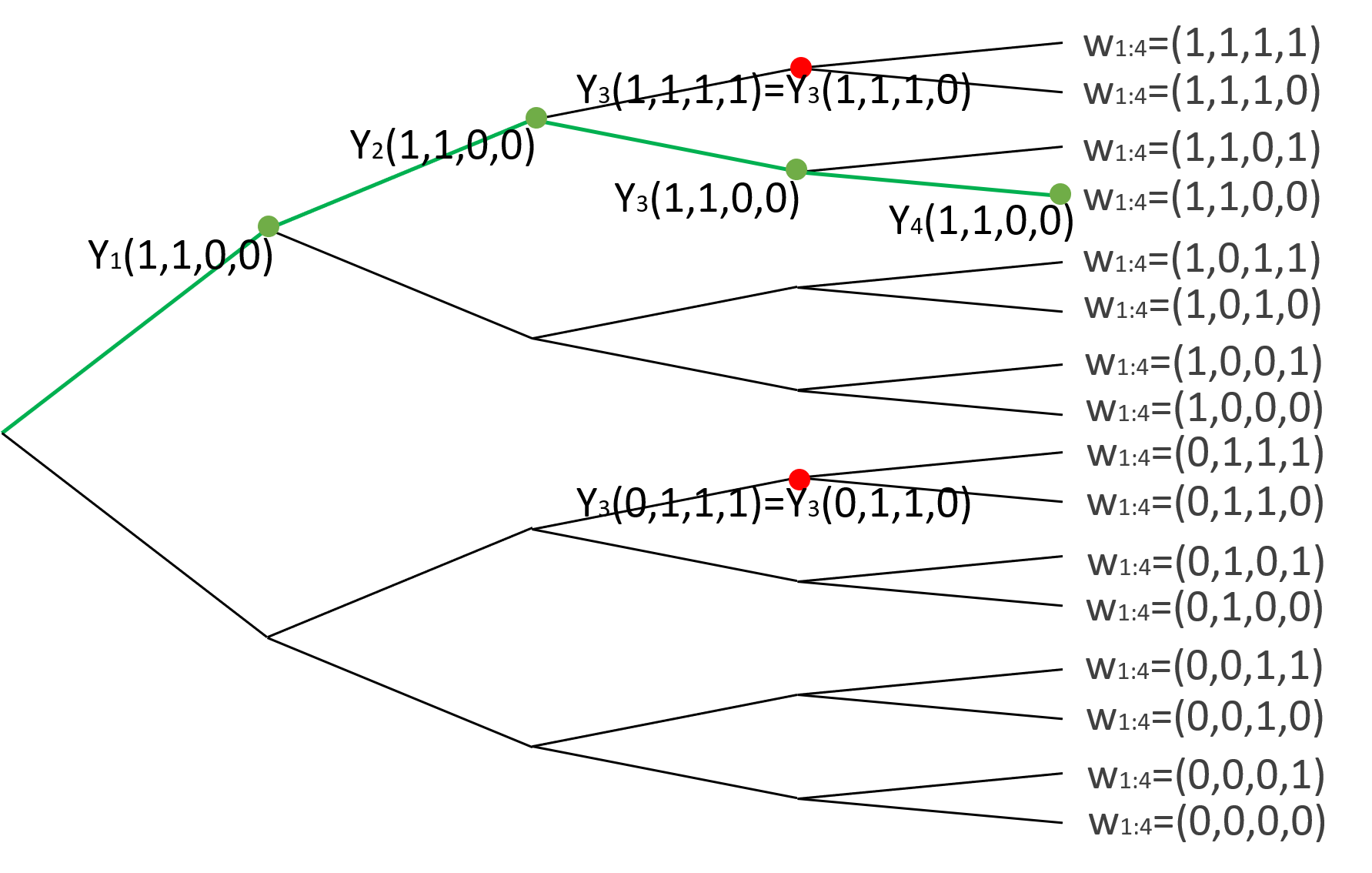}
\caption{Illustrator of assignment paths and potential outcomes when $T=4$. The green path stands for one assignment path $\bm{w}_{1:4} = (1,1,0,0)$. Following the green path there are four potential outcomes. The two red dots each stands for two potential outcomes that are equal under Assumption~\ref{asp:nonanticipating}. And the potential outcomes at the two red dots are equal if Assumption~\ref{asp:nomcarryover} is further assumed.}
\label{fig:AssignmentPath}
\end{figure}

\example
\label{exa:RE:basic}
When $T=4$, there are $16$ assignment paths as shown in Figure~\ref{fig:AssignmentPath}.
Associated with each assignment path $\bm{w}_{1:4}$ are four potential outcomes $Y_1(\bm{w}_{1:4}), Y_2(\bm{w}_{1:4}), Y_3(\bm{w}_{1:4}), Y_4(\bm{w}_{1:4})$.
\Halmos \endexample

Throughout this paper, we do not directly model the potential outcomes or impose a parametric relationship with the assignment path; instead, we treat them as unknown but fixed quantities, or, equivalently, we implicitly condition on $\bY$. Our setup does not preclude the possibility that the potential outcomes were generated through a dynamic process; however, it allows us to be completely agnostic to the data generating process, making our causal claims more objective. To make inference possible, we rely on the variation introduced by the random assignment path; this is commonly referred to as finite-sample or design-based perspective and has a long history going back to \cite{neyman1923application}, \cite{fisher1937design}, \cite{kempthorne1955}, and \cite{rubin1980randomization}. Unlike traditional sampling-based inference, the design-based approach does not require a hypothetical population from which to sample experimental units, see \cite{imbens_rubin_2015} and \citet{abadie2020sampling} for recent reviews. Instead, we make two assumptions that limit the dependence of the potential outcomes on assignment paths. Below let $\{t:t'\} = \{t,t+1,...,t'\}$, for any $t<t' \in [T]$.
\begin{assumption}
[Non-anticipating Potential Outcomes]
\label{asp:nonanticipating}
For any $t \in [T]$, $\bm{w}_{1:t} \in \{0,1\}^t$, and for any $\bm{w}'_{t+1:T}, \bm{w}''_{t+1:T} \in \{0,1\}^{T-t}$, $$Y_{t}(\bm{w}_{1:t},\bm{w}'_{t+1:T}) = Y_{t}(\bm{w}_{1:t},\bm{w}''_{t+1:T}).$$
\end{assumption}

Assumption~\ref{asp:nonanticipating} states that the potential outcomes at time $t$ do not depend on future treatments \citep{bojinov2019time, basse2019minimax, rambachan2019econometric}.
Since we control the assignment mechanism instead of letting the experimental units to administer future assignments (\emph{e.g.}, at a ride-hailing platform, a passenger does not know the price in the next hour), the design ensures that this assumption is satisfied.

\example
[Example~\ref{exa:RE:basic} Continued]
\label{exa:RE:nonanticipation}
Under Assumption~\ref{asp:nonanticipating}, $Y_3(1,1,1,1) = Y_3(1,1,1,0)$. In Figure~\ref{fig:AssignmentPath} the red dot at $Y_3(1,1,1)$ stands for both $Y_3(1,1,1,1)$ and $Y_3(1,1,1,0)$. 
\Halmos \endexample

\begin{assumption}
[$m$-Carryover Effects]
\label{asp:nomcarryover}
There exists a fixed and given $m$, such that for any $t \in \{m+1,m+2,...,T\}, \bm{w}_{t-m:T} \in \{0,1\}^{T-t+m+1}$, and for any $\bm{w}'_{1:t-m-1}, \bm{w}''_{1:t-m-1} \in \{0,1\}^{t-m-1}$, $$Y_{t}(\bm{w}'_{1:t-m-1}, \bm{w}_{t-m:T}) = Y_{t}(\bm{w}''_{1:t-m-1}, \bm{w}_{t-m:T}).$$
\end{assumption}

Assumption~\ref{asp:nomcarryover} restricts the order of the carryover effect \citep{laird1992analysis, senn1998robust, bojinov2019time, basse2019minimax}.
The validity of Assumption~\ref{asp:nomcarryover} depends on the setting and requires practitioners to use their domain knowledge to choose an appropriate $m$.
Examples arise in ride-hailing, in which the effect of surge pricing on a ride-hailing platform typically dissipates after one or two hours, depending on the city size \citep{garg2019driver}.
Moreover, in Section~\ref{sec:FigureOutm} we propose a data driven procedure for selecting an appropriate $m$.

Assumptions~\ref{asp:nonanticipating} and~\ref{asp:nomcarryover} allow us to simplify notation.
For any $t \in \{m+1,...,T\}$ and any two assignment paths $\bm{w}_{1:T}, \bm{w}'_{1:T} \in \{0,1\}^{m+1}$, whenever $\bm{w}_{t-m:t} = \bm{w}'_{t-m:t}$ this leads to $$Y_{t}(\bm{w}_{1:T}) = Y_{t}(\bm{w}'_{1:T}).$$
In the remainder of this paper, we will write $Y_{t}(\bm{w}_{t-m:t}) := Y_{t}(\bm{w}_{1:T})$ to emphasize the dependence on treatments $\bm{w}_{t-m:t}$.
For example, the potential outcomes at the two red dots in Figure~\ref{fig:AssignmentPath} are equal, i.e., $Y_3(1,1) := Y_3(1,1,1,1) = Y_3(1,1,1,0) = Y_3(0,1,1,1) = Y_3(0,1,1,0)$

\subsection{Causal Effects}
\label{sec:CausalEffects}
In the potential outcomes approach to causal inference, any comparison of potential outcomes has a causal interpretation.
In this paper, we focus on a special set of causal estimands that measure the relative effectiveness of persistently assigning a unit to treatment as opposed to control.
For any $p \in \{0, 1, ..., T-1\}$, let $\bm{1}_{p+1} = (1,1,...,1)$ be a vector of $(p+1)$ ones; let $\bm{0}_{p+1} = (0,0,...,0)$ be a vector of $(p+1)$ zeros.
Define the average lag-$p$ causal effect of consecutive treatments on the outcome, for any $p \in \{0, 1, ..., T-1\}$,
\begin{align}
\label{eqn:estimand}
\tau_p(\bY) = \frac{1}{T-p} \sum_{t=p+1}^{T} [Y_t(\bm{1}_{p+1}) - Y_t(\bm{0}_{p+1})].
\end{align}
This estimand captures the effects of permanently deploying a new policy
, and has been widely studied in the longitudinal experiments since the early work of \citet{robins1986new}.

\remark
Although we focus on an average causal effect, all of our results and analysis trivially extend to the total causal effect, which does not normalize, \emph{i.e.}, $(T-p)\tau_p(\bY)$.
The optimal design as we will show in Section~\ref{sec:MinimaxOptimization} will remain unchanged.
\endremark

In our setup, $p$ reflects the experimental designer's knowledge of the order of the carryover effect; see discussion below Assumption~\ref{asp:nomcarryover}.
Such a knowledge is either correct, which we refer to as the perfect knowledge case ($p=m$ ), or  incorrect, which we refer to as the ``misspecified'' $m$ case\footnote{Some authors specifically focus on $p<m$, particularly when $m$ is of the same order as $T$ \citep{bojinov2019time}.} ($p \ne m$).
In this section we focus on the $p=m$ case to derive the optimal design; Section~\ref{sec:Misspecifiedm} considers what happens when $m$ is misspecified by discussing the $p \ne m$ case.

The challenge of causal inference on switchback experiments is that we only observe one assignment path.
In other words, for each period $t$, we observe at most either $Y_t(\bm{1}_{p+1})$ or $Y_t(\bm{0}_{p+1})$ (and sometimes neither).
After conducting a switchback experiment, the observed data contains $\bm{w}^{\obs}_{1:T}$ the realized assignment path, and $Y_t^{\obs} = Y_t(\bm{w}^{\obs}_{1:T})$ the observed outcome at time $t$ under the realized assignment path $\bm{w}^{\obs}_{1:T}$.
To link the observed and potential outcomes, we assume there is only one version of the treatment\footnote{When combined with non-interference if there were multiple units, this is known as the stable unit treatment value assumption \citep{rubin1980randomization}.}, and that there is no non-compliance.

\subsection{Regular Switchback Experiments}
\label{sec:RegularSwitchbackExperiments}

The design of switchback experiment induces a probabilistic distribution over assignment paths $\bm{w}_{1:T} \in \{0,1\}^T$.
Formally, a design of switchback experiment is any $\eta: \{0,1\}^T \to [0,1]$ such that
\begin{align*}
\sum_{\bm{w}_{1:T} \in \{0,1\}^T} \eta(\bm{w}_{1:T}) = 1, & & \eta(\bm{w}_{1:T}) \geq 0, \ \forall \ \bm{w}_{1:T} \in \{0,1\}^T.
\end{align*}
Explicitly, $\eta(\cdot)$ is the underlying discrete distribution of the random assignment path $\bm{W}_{1:T}$.

In this paper, we narrow our scope to the family of regular switchback experiments.
This family of experiments are parameterized by $\bT$ and $\bQ$, defined as
$$\bT = \{ t_0 = 1 < t_1 < t_2 < ... < t_K \} \subseteq [T],$$
where $K < T$ is a positive integer, and $\bT$ contains a total of $K+1$ integers, which is a subset of all the time indices; and
$$\bQ = (q_0, q_1, ..., q_K) \in (0,1)^{K+1} := \cQ,$$
where $\bQ$ is a vector of $K+1$ real numbers between $(0,1)$.
For the ease of notations also denote $t_{K+1} = T+1$, though our time horizon is only $T$ periods.

\begin{definition}[Regular Switchback Experiments]
\label{defn:RegularSE}
For any $\bT = \{ t_0 = 1 < t_1 < ... < t_K \} \subseteq [T]$, and any $\bQ = (q_0, q_1, ..., q_K) \in (0,1)^{K+1}$, a regular switchback experiment $(\bT, \bQ)$ administers a probabilistic treatment at any time $t$, given by:
\begin{equation}
\label{eqn:CoinFlip}
\Pr(W_t = 1) =
\begin{aligned}
& q_k, & & \text{ \ if \ } t_k \leq t \leq t_{k+1} - 1
\end{aligned}
\end{equation}
\end{definition}

In words, the experimental designer jointly decides on a collection of randomization points, which consists of flipping biased coins at each period $t \in \{t_0,...,t_K\}$, as well as a collection of randomization probabilities behind the biased coins, $(q_0, ..., q_K)$.
If the resulting flip at period $t_k$ is \textsf{heads}, then the experimental designer assigns the unit to treatment during periods $(t_k, t_k+1, ..., t_{k+1}-1)$; otherwise, if \textsf{tails}, assigns the unit to control during periods $(t_k, t_k+1, ..., t_{k+1}-1)$.

\begin{figure}[bt]
\begin{subfigure}{.45\textwidth}
\centering
\includegraphics[width=\textwidth]{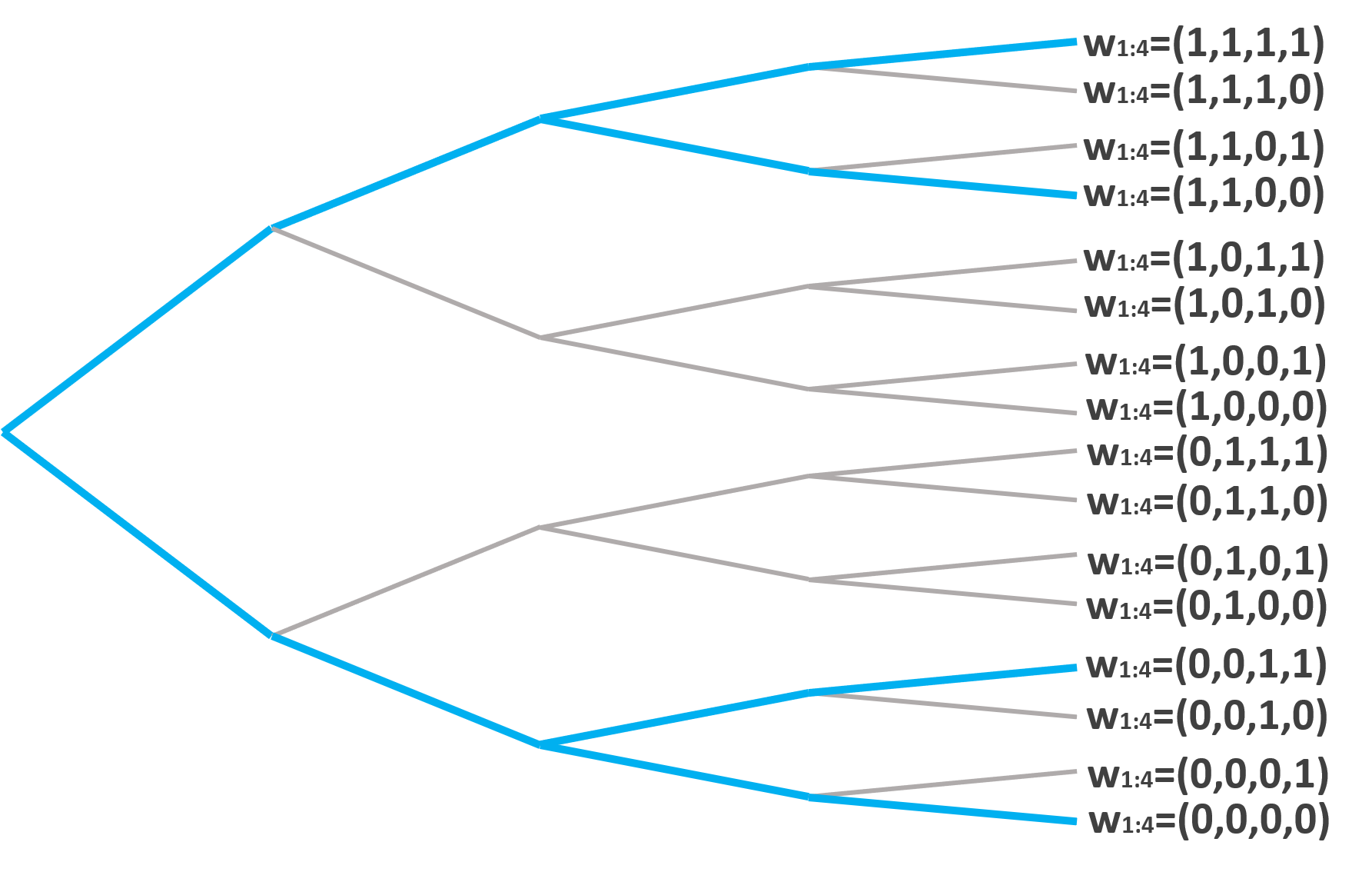}
\end{subfigure}\hfill
\begin{subfigure}{.45\textwidth}
\centering
\includegraphics[width=\textwidth]{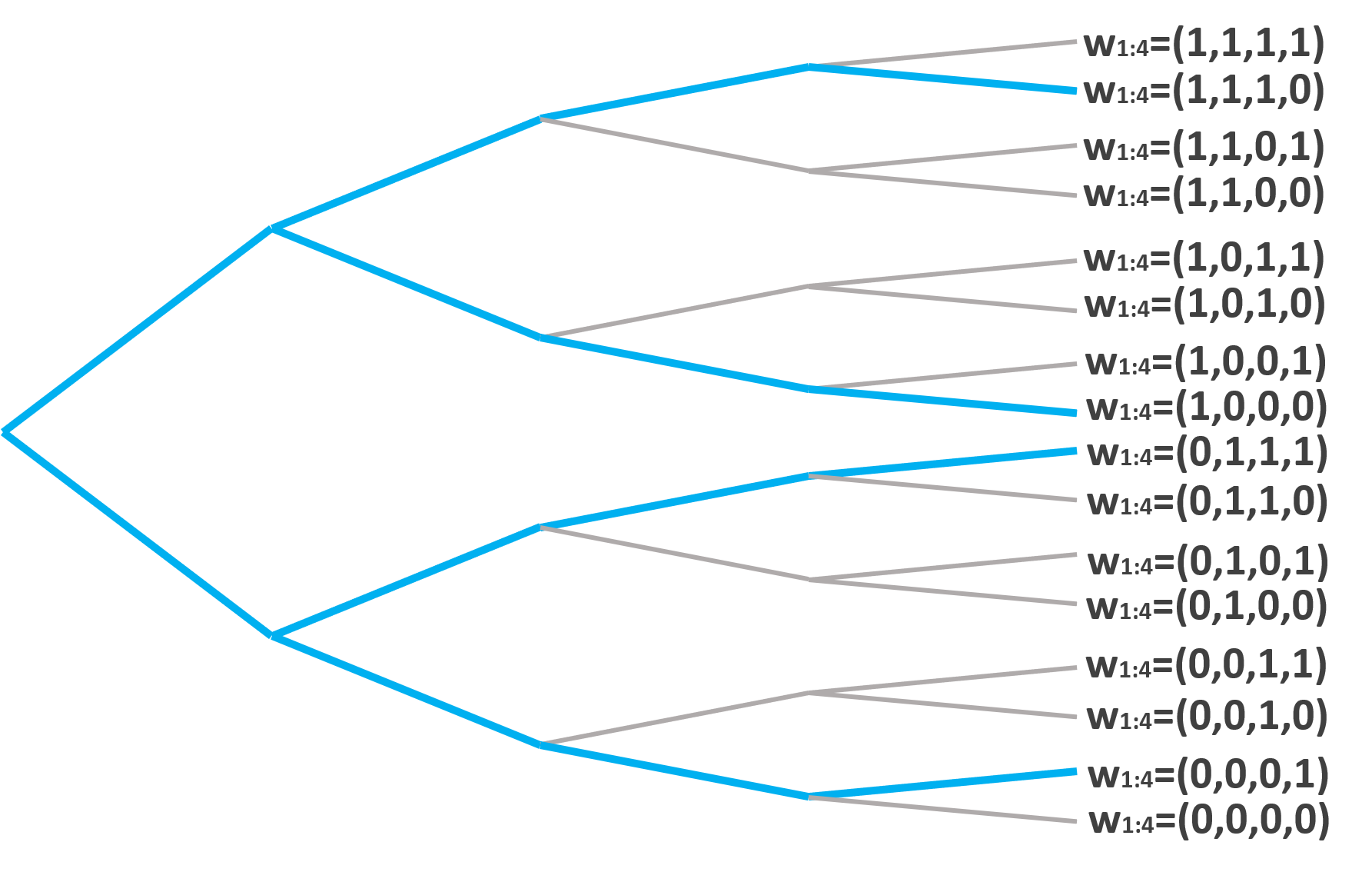}
\end{subfigure}
\caption{Two designs. The blue lines stand for the possible treatment assignments that a design could administer. Left: regular switchback experiment (Example~\ref{exa:RE:RegularSE}); Right: irregular switchback experiment (Example~\ref{exa:RE:IrregularSE}).}
\label{fig:RegularSwitchbackExperiments}
\end{figure}

\example
\label{exa:RE:RegularSE}
When $T=4$, $\bT=\{t_0=1, t_1=3\}, \bQ=(q_0,q_1)=(1/2, 1/2)$ corresponds to the following design:
with probability $1/4$, $\bm{W}_{1:4} = (1,1,1,1)$;
with probability $1/4$, $\bm{W}_{1:4} = (1,1,0,0)$;
with probability $1/4$, $\bm{W}_{1:4} = (0,0,1,1)$;
with probability $1/4$, $\bm{W}_{1:4} = (0,0,0,0)$.
See Figure~\ref{fig:RegularSwitchbackExperiments} (left figure) for the four assignment paths that are in the support of the discrete probability distribution.
\Halmos \endexample

\example
\label{exa:RE:IrregularSE}
Not all switchback experiments are regular.
For example, when $T=4$:
with probability $1/4$, $\bm{W}_{1:4} = (1,1,1,0)$;
with probability $1/4$, $\bm{W}_{1:4} = (1,0,0,0)$;
with probability $1/4$, $\bm{W}_{1:4} = (0,1,1,1)$;
with probability $1/4$, $\bm{W}_{1:4} = (0,0,0,1)$.
See Figure~\ref{fig:RegularSwitchbackExperiments} (right figure) for the four assignment paths that are in the support of the discrete probability distribution.
\Halmos \endexample

In Section~\ref{sec:MinimaxOptimization}, we show that fair coin flipping (i.e., $q_k = 1/2, \forall \ k \in \{0,1,...,K\}$) is indeed optimal, under a mild assumption.\footnote{Researchers have either shown that versions of completely randomized experiments (corresponding to ``fair coin flips'') are optimal, \emph{e.g.}, \citet{wu1981robustness, li1983minimaxity, basse2019minimax} where they make mild assumptions on permutation invariance; or have explicitly assumed that the coins flips be fair, \emph{e.g.}, \citet{bai2019optimality, harshaw2019balancing}.} The reason behind fair coin flips reflects our limited assumption on the outcome model and the inherent symmetry in the potential outcomes.

Note that we do not consider adaptive treatment assignments as most firms design the entire experiment before the experiment is launched; the treatment assignments are typically not updated based on the observed outcomes. We briefly outline adaptive experimental designs as future extensions in Section~\ref{sec:PracticalImplications}.

For any regular switchback experiment $(\bT, \bQ)$, we may use $\bT$ to refer to the same experiment when $\bQ$ is clear from the context.
We denote the underlying discrete probability distribution using $\eta_{\bT, \bQ}(\cdot)$.
For any $\bT$ and $\bQ$, the discrete probability distribution has a total of $2^{K+1}$ many supports.
The assignment path is random, and follows the discrete probability distribution $\eta_{\bT, \bQ}(\cdot)$:
\begin{equation}
\label{eqn:AssignmentProb}
\eta_{\bT, \bQ}(\bm{w}_{1:T}) =
\left\{
\begin{aligned}
& \prod_{k=0}^K q_{t_k}^{\bI\{ w_{t_k} = 1 \}} \cdot \bq_{t_k}^{\bI\{ w_{t_k} = 0 \}}, & & \text{ \ if \ } \forall \ k \in \{0,1,...,K\}, w_{t_k} = w_{t_k+1} = ... = w_{t_{k+1}-1}, \\
& 0, & & \text{ \ otherwise. \ }
\end{aligned}
\right.
\end{equation}
In the remainder of this paper, unless explicitly noted, all probabilities and expectations are taken with respect to this discrete probability distribution $\eta_{\bT, \bQ}(\cdot)$.

\subsection{Estimation}
\label{sec:HTEstimator}

Now that $\eta_{\bT, \bQ}(\cdot)$ is determined, following any realization of the assignment path $\bm{w}_{1:T}$, we use the Horvitz-Thompson estimator to estimate the causal effect:
\begin{align}
\label{eqn:estimator}
\widehat{\tau}_p (\eta_{\bT, \bQ}, \bm{w}_{1:T}, \bY) & = \frac{1}{T-p} \sum_{t=p+1}^T \left\{ Y_t^{\obs} \frac{\bI{\{\bm{w}_{t-p:t} = \bm{1}_{p+1}\}}}{\Pr(\bm{W}_{t-p:t} = \bm{1}_{p+1})} - Y_t^{\obs} \frac{\bI{\{\bm{w}_{t-p:t} = \bm{0}_{p+1}\}}}{{\Pr(\bm{W}_{t-p:t} = \bm{0}_{p+1})}} \right\}.
\end{align}
We emphasize that the estimator $\widehat{\tau}_p (\cdot, \cdot, \cdot)$ depends on (i) the probability distribution that the assignment path is sampled from, (ii) the realization of the assignment path, and (iii) the set of potential outcomes.

\example
\label{exa:RE:Estimator}
Suppose $T=4, p=m=1$.
Suppose the assignments are probabilistic and $\Pr(W_t=1) = \Pr(W_t=0) = 1/2, \forall t\in[4].$
With probability $1/16$ the green assignment path as in Figure~\ref{fig:AssignmentPath} is administered, $\bm{W}_{1:4} = (1,1,0,0)$.
The estimator is then $\widehat{\tau}_1 = \frac{1}{3}\left\{4 Y_2(1,1) + 0 - 4 Y_4(0,0)\right\}.$
\Halmos \endexample

Since the assignment path $\bm{W}_{1:T}$ is random, this Horvitz-Thompson estimator is also random. 
Moreover, when the assignment path satisfies a regular switchback, the probabilities in the denominator are known. As we will show in Theorem~\ref{thm:FairCoinOptimal}, under the optimal design, these probabilities will be multiplicatives of $1/2$, allowing us to avoid the known stability issues of the Horvitz-Thompson estimator when the probabilities are extreme (either close to 0 or close to 1).
It is well-known that the Horvitz-Thompson estimator is unbiased if the treatment and control probabilities are both non-zero.

\begin{proposition}
[Unbiasedness of the Horvitz-Thompson Estimator]
\label{thm:HTUnbiased}
In a regular switchback experiment, under Assumptions~\ref{asp:nonanticipating} and~\ref{asp:nomcarryover}, the Horvitz-Thompson estimator is unbiased for the average lag-$p$ causal effect of consecutive treatments on outcome, i.e., $$\bE[\widehat{\tau}_p(\eta_{\bT, \bQ}, \bm{W}_{1:T}, \bY)] = \tau_p(\bY).$$
\end{proposition}
The expectation $\bE[\cdot]$ is taken with respect to the random assignment $\bm{W}_{1:T} \sim \eta_{\bT, \bQ}(\cdot)$. when it is obvious we will compress the subscript in the expectation writing $\bE[\cdot]$ to mean $\bE_{\bm{W}_{1:T} \sim \eta_{\bT, \bQ}}[\cdot]$.
The proof to Proposition~\ref{thm:HTUnbiased} is standard, by checking the expectations. We defer its proof to Section~\ref{sec:proof:thm:HTUnbiased} in the Appendix.

\subsection{Evaluation of Experiments: the Decision-Theoretic Framework}
\label{sec:DecisionTheoreticFramework}

To evaluate the quality of a design of experiment, we adopt the decision-theoretic framework \citep{berger2013statistical, bickel2015mathematical}.
When the random design is $\eta_{\bT, \bQ}(\cdot)$, for any realization of the assignment path $\bm{w}_{1:T}$ and any set of potential outcomes $\bY$, we define the loss function
\begin{align*}
L(\eta_{\bT, \bQ}, \bm{w}_{1:T}, \bY) = \left( \widehat{\tau}_p(\eta_{\bT, \bQ}, \bm{w}_{1:T}, \bY) - \tau_p(\bY) \right)^2
\end{align*}
and the risk function
\begin{align}
\label{eqn:RiskFunction}
r(\eta_{\bT, \bQ}, \bY) = \sum_{\bm{w}_{1:T} \in \{0,1\}^T} \eta_{\bT, \bQ}(\bm{w}_{1:T}) \cdot \left( \widehat{\tau}_p(\eta_{\bT, \bQ}, \bm{w}_{1:T}, \bY) - \tau_p(\bY) \right)^2
\end{align}
Such a risk function quantifies the expected squared difference between our estimand and estimator.
Since the estimator is unbiased, the risk function also has a second interpretation: the variance of the estimator.
A design with a lower risk is also a design whose estimator has a lower variance.

\example
[Examples~\ref{exa:RE:RegularSE} and~\ref{exa:RE:Estimator} Revisited]
\label{exa:RE:LossNRisk}
Suppose $T=4$ and $p=m=1$.
As in Example~\ref{exa:RE:RegularSE}, $\bT=\{1,3\}$. With probability $1/4$, $\bm{W}_{1:4} = (1,1,0,0)$, $\widehat{\tau}_1(\bT) = \frac{1}{3}\{2Y_2(1,1)-2Y_4(0,0)\}$,
$L(\eta_{\bT, \bQ}, \bm{w}_{1:T}, \bY) = \frac{1}{9} \left( Y_2(1,1) + Y_2(0,0) - Y_3(1,1) + Y_3(0,0) - Y_4(1,1) - Y_4(0,0) \right)^2.$
As in Example~\ref{exa:RE:Estimator}, $\tbT=\{1,2,3,4\}$. With probability $1/16$, $\bm{W}_{1:4} = (1,1,0,0)$, $\widehat{\tau}_1(\tbT) = \frac{1}{3}\{4Y_2(1,1)-4Y_4(0,0)\}$,
$L(\eta_{\tbT, \bQ}, \bm{w}_{1:T}, \bY) = \frac{1}{9} \left( 3Y_2(1,1) + Y_2(0,0) - Y_3(1,1) + Y_3(0,0) - Y_4(1,1) - 3Y_4(0,0) \right)^2.$
\Halmos \endexample

Example~\ref{exa:RE:LossNRisk} suggests that, even if the two realizations of the assignment path are the same and the potential outcomes are the same, since the probability distributions $\eta_{\bT, \bQ}$ and $\eta_{\tbT, \bQ}$ are distinct, the corresponding estimators $\widehat{\tau}_1(\bT)$ and $\widehat{\tau}_1(\tbT)$ could be different, and the corresponding loss functions $L(\eta_{\bT, \bQ}, \bm{w}_{1:T}, \bY)$ and $L(\eta_{\tbT, \bQ}, \bm{w}_{1:T}, \bY)$ could also be different.
This observation suggests that there exists some design $\eta_{\bT^*}$ that has a small risk.
In the next section we find such a design when $m$ is correctly specified.

\section{Design of Regular Switchback Experiments under Minimax Rule}
\label{sec:MinimaxOptimization}

The goal of this section is to find the optimal design of regular switchback experiments, i.e., to select the optimal randomization points and the optimal randomization probabilities.
Throughout this section we assume $m$ is known and we set $p=m$.

We formalize our experimental design problem through the minimax framework.
The minimax decision rule \citep{berger2013statistical, wu1981robustness, li1983minimaxity} finds an optimal design of experiment such that the worst-case risk against an adversarial selection of potential outcomes is minimized,
\begin{align}
\label{eqn:TheProblem}
\min_{\bT \in [T], \bQ \in \cQ} \max_{\bY \in \cY} \ r(\eta_{\bT, \bQ}, \bY) = \min_{\bT \in [T], \bQ \in \cQ} \max_{\bY \in \cY} \ \sum_{\bm{w}_{1:T} \in \{0,1\}^T} \eta_{\bT, \bQ}(\bm{w}_{1:T}) \cdot \left( \widehat{\tau}_p(\bm{w}_{1:T}, \bY) - \tau_p(\bY) \right)^2.
\end{align}
One compelling reason to adopt the minimax framework, as commented in the seminal work of \citet{wu1981robustness}, is that ``the experimenter's information about the model is never perfect. When a model is proposed, there is always the possibility that the `true' model deviates from the assumed model.'' Instead of finding the best possible design by imposing a model, we try to derive the best possible design for the worse possible set of potential outcomes.

To overcome minimaxity and to lay out the foundation for inference, we impose an additional assumption on the support of the potential outcome. Since the potential outcomes are unknown but fixed, we assume that their absolute values are bounded from above, and that bound is attainable at every time period.
\begin{assumption}
[Bounded Potential Outcomes]
\label{asp:BoundedPO}
The potential outcomes are bounded by some constant, i.e., $\exists \ B>0, s.t. \ \forall \ t \in [T], \ \forall \ \bm{w} \in \{0,1\}^T$, $\left| Y_t(\bm{w}) \right| \leq B,$
or, equivalently, $\bY \in \cY = [-B, B]^{T}$.
\end{assumption}
Assumption~\ref{asp:BoundedPO} is often satisfied since it assumes that the potential outcomes are bounded by the same (possibly a large) constant, (\emph{e.g.}, the ride-matching rate from each experimental period is always a finite quantity) and that the extreme could possibly occur at any point in time (\emph{e.g.}, the maximum ride-matching rate could be observed at any time). In particular, knowledge about the magnitude of $B$ is not required, and, as we show below, the optimal design does not depend on $B$.

The reason to make Assumption~\ref{asp:BoundedPO} is two fold. First, for optimization purposes, Assumption~\ref{asp:BoundedPO} reflects the inherent symmetry in the potential outcomes under both treatment and control, which is in the same spirit as the permutation invariance assumption \citep{wu1981robustness, li1983minimaxity, basse2019minimax}. It is such symmetry that ensures the optimality of fair coin flipping. See Theorem~\ref{thm:FairCoinOptimal} below. Second, for inferential purposes, Assumption~\ref{asp:BoundedPO} ensures that the variance of the estimator is well-behaved, which is commonly assumed in the finite-sample inference literature \citep{aronow2017estimating, chin2018central, bojinov2019time, li2020rerandomization, han2021population}. It is the well-behaved variance that lays the foundation of our limiting distribution in Theorem~\ref{thm:AsymptoticNormality}.

To solve the minimax problem \eqref{eqn:TheProblem}, we start by focusing on the inner maximization part.
We characterize the worst-case potential outcomes by identifying two dominating strategies for the adversarial selection of potential outcomes.
Denote
$\bY^{+} = \left\{Y_t(\bm{1}_{m+1}) = Y_t(\bm{0}_{m+1}) = B \right\}_{t \in \{m+1:T\}}$
and
$\bY^{-} = \left\{Y_t(\bm{1}_{m+1}) = Y_t(\bm{0}_{m+1}) = -B \right\}_{t \in \{m+1:T\}}$

\begin{lemma}
\label{lem:AdversaryStrategy}
Under Assumptions~\ref{asp:nonanticipating}--\ref{asp:BoundedPO}, $\bY^{+}$ and $\bY^{-}$ are the only two dominating strategies for the adversarial selection of potential outcomes.
That is, for any $\bT \subseteq [T]$ and for any $\bY \in \cY$,
\begin{align*}
r(\eta_{\bT, \bQ}, \bY^+) \geq r(\eta_{\bT, \bQ}, \bY); & & r(\eta_{\bT, \bQ}, \bY^-) \geq r(\eta_{\bT, \bQ}, \bY).
\end{align*}
Moreover, for any $\bY \in \cY$ such that $\bY \ne \bY^{+}, \bY \ne \bY^{-}$, the above two inequalities are strict.
\end{lemma}
The proof of Lemma~\ref{lem:AdversaryStrategy} can be found in Section~\ref{sec:Proof:lem:AdversaryStrategy}.
Lemma~\ref{lem:AdversaryStrategy} simplifies the minimax problem in \eqref{eqn:TheProblem}, as it allows us to replace $\bY$ by $\bY^* = \bY^+$ or $\bY^* = \bY^-$, and reduce the minimax problem \eqref{eqn:TheProblem} into a minimization problem $$\min_{\bT\in[T], \bQ\in\cQ} r(\eta_{\bT,\bQ}, \bY^*).$$
Next we solve this minimization problem by first finding the optimal $\bQ$ values.

\begin{theorem}[Optimality of Fair Coin Flipping]
\label{thm:FairCoinOptimal}
Under Assumptions~\ref{asp:nonanticipating}--\ref{asp:BoundedPO}, any optimal design of experiment $(\bT, \bQ)$ must satisfy $q_0=q_1=...=q_K=1/2$.
\end{theorem}

The proof of Theorem~\ref{thm:FairCoinOptimal} can be found in Section~\ref{sec:proof:thm:FairCoinOptimal}. Theorem~\ref{thm:FairCoinOptimal} suggests that the optimal randomization probabilities should be $1/2$. So we can restrict our scope to only finding the experiments induced by fair coin flipping, and focus on the trade-off behind the number and timing of the randomization points.

The trade-off lies between having too many randomization points (corresponding to large $K$) and too few randomization points (corresponding to small $K$). Intuitively, too many decreases the probability of observing consecutive treatments $\bm{1}_{m+1}$ or controls $\bm{0}_{m+1}$, which, in turn, decreases the amount of useful data. On the other hand, too few decreases the number of independent observations and reduces our ability to produce reliable results. Both of these scenarios reduce our ability to draw valid causal claims. Theorem~\ref{thm:OptimalDesign} formalizes this trade-off.

\begin{theorem}
[Optimal Design]
\label{thm:OptimalDesign}
Under Assumptions~\ref{asp:nonanticipating}--\ref{asp:BoundedPO}, the optimal solution to the design of regular switchback experiment as we have introduced in \eqref{eqn:TheProblem} is equivalent to the optimal solution to the following subset selection problem.
\begin{align}
\min_{\bT \subset [T]} \left\{ 4 \sum_{k=0}^{K} (t_{k+1} - t_{k})^2 + 8 m (t_K - t_1) + 4 m^2 K - 4 m^2 + 4 \sum_{k=1}^{K-1} [(m-t_{k+1}+t_{k})^+]^2\right\} \label{eqn:SubsetSelection}
\end{align}
In particular, when $m=0$ then $\bT^* = \{1, 2, 3, ..., T\}$; when $m>0$, and if there exists $n \geq 4 \in \bN$, s.t. $T = n m$, then $\bT^* = \{1, 2m+1, 3m+1, ..., (n-2)m+1\}$.
\end{theorem}

The proof of Theorem~\ref{thm:OptimalDesign} is deferred to Section~\ref{sec:Proof:thm:OptimalDesign} in the appendix.
Theorem~\ref{thm:OptimalDesign} presents the optimal design in a class of perfect cases when the time horizon splits into several equal-length epochs\footnote{For other imperfect cases when $T$ is not divisible by $m$, we can also solve \eqref{eqn:SubsetSelection} and find the optimal design.
However, we do not present closed-form solutions to such subset selection problem due to integrality issues. Technical discussions about the optimal design in such imperfect cases are deferred to Section~\ref{sec:OptimalSolutionDiscussion} in the Appendix.}; see Table~\ref{tbl:exa:OptimalDesign} for an example.
In practice, we recommend selecting $T$ that satisfies the condition in Theorem~\ref{thm:OptimalDesign}; see Section~\ref{sec:PracticalImplications} for a discussion.

\begin{table}[!tb]
\TABLE{An example of the optimal design $\bT^*=\{1,5,7,9\}$ when $T=12$ and $p=m=2$. \label{tbl:exa:OptimalDesign}}
{
\begin{tabular}{| p{0.92cm} | p{0.92cm} | p{0.92cm} | p{0.92cm} | p{0.92cm} | p{0.92cm} | p{0.92cm} | p{0.92cm} | p{0.92cm} | p{0.92cm} | p{0.92cm} | p{0.92cm} | p{0.92cm} |}
\hline
& 1 & 2 & 3 & 4 & 5 & 6 & 7 & 8 & 9 & 10 & 11 & 12 \\
\hline
$\bT^*$ & $\checkmark$ & $-$ & $-$ & $-$ & $\checkmark$ & $-$ & $\checkmark$ & $-$ & $\checkmark$ & $-$ & $-$ & $-$ \\
\hline
\end{tabular}
}
{Each checkmark beneath a time period $t$ indicates that $t$ is a randomization point.}
\end{table}

There are two important implications of Theorem~\ref{thm:OptimalDesign}. First, the optimal randomization frequency depends on the physical duration of the carryover effect, regardless of the granularity of one single experimental period.
This observation suggests that practitioners should set each period to be almost as long as the order of the carryover effect, which sheds some light on the selection of granularity when practitioners design the experiment. 
See Example~\ref{exa:TimeGranularity}.
Second, a special case arises when there are no carryover effects $(m=0)$  or very little carryover effect $(m=1)$; in both cases the optimal designs are almost the same.
This observation suggests a layer of robustness when the granularity is set to be the same as the suspected order of the carryover effect; see Example~\ref{exa:LittleCarryover}.

\example
[Two Granularity Levels]
\label{exa:TimeGranularity}
In the ride-sharing application, suppose the firm has two options to treat one single time period either as 0.5 hour or 1 hour; and suppose the carryover effect lasts for 2 hours.
When one single experimental period corresponds to 0.5 hour, the carryover effect lasts for $m=4$ periods.
When one single experimental period corresponds to 1 hour, the carryover effect lasts for $m=2$ periods.
From Theorem~\ref{thm:OptimalDesign}, the optimal design exhibits an optimal structure that randomizes once every $m$ periods (except for the first and last epoch, which lasts for $2m$ time periods each).
In both cases, the optimal design would randomize once every two hours.
\Halmos \endexample


\example
[Little Carryover Effect]
\label{exa:LittleCarryover}
For example, Theorem~\ref{thm:OptimalDesign} suggests that the optimal design when $m=0$ is $\bT^* = \{1, 2, 3, ..., T\}$, and when $m=1$ is $\bT^* = \{1, 3, 4, ..., T-1\}$.
This suggests that the minimax optimal design in the absence of a carryover effect is robust to the existence of a short carryover effect.
\endexample

\section{Inference and Testing}
\label{sec:InferenceAndTesting}

\begin{figure}[!tb]
\centering
\includegraphics[width=\textwidth]{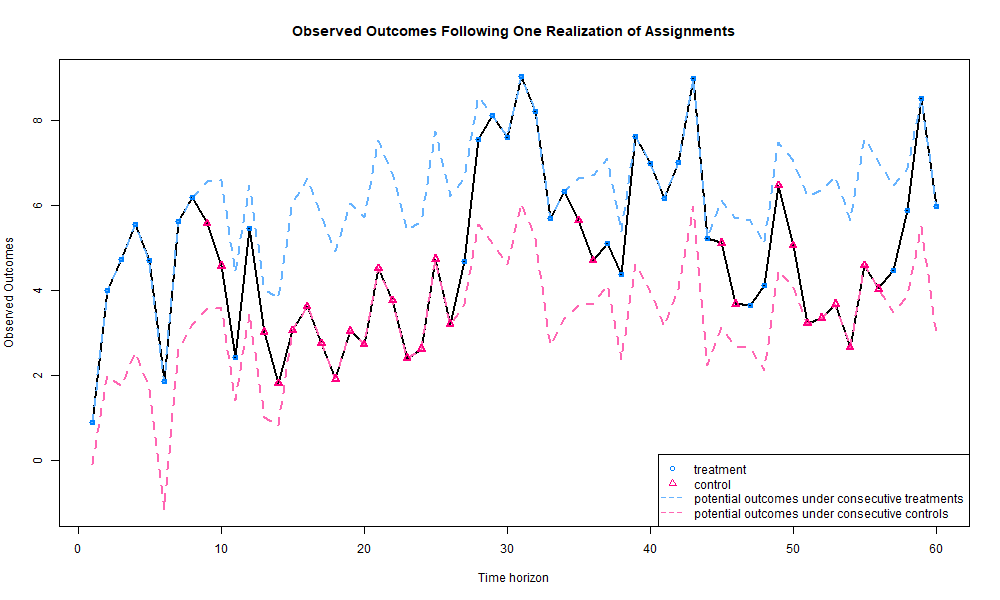}
\caption{Illustrator of the observed assignment path $\bm{w}_{1:T}^{\obs}$ (blue and red dots) and the observed outcomes $\bm{Y}_{p+1:T}^{\obs}$ (black curve). The dashed lines are the potential outcomes under consecutive treatments / controls.}
\label{fig:TwoSequences}
\end{figure}

After designing and running the experiment, we obtain two time series.
The first is the observed assignment path $\bm{w}_{1:T}^{\obs}$, and the second is the corresponding observed outcomes $\bm{Y}_{1:T}^{\obs}$.
See Figure~\ref{fig:TwoSequences}.
To draw inference from this data we propose two methods, an exact randomization based test and a finite population conservative test that establishes asymptotic result.

In Sections~\ref{sec:ExactInference} and~\ref{sec:AsymptoticInference}, we assume perfect knowledge of $m$, i.e., $p=m$, and we will write $\tau_m$ and $\widehat{\tau}_m$ to stand for $\tau_p$ and $\widehat{\tau}_p$, respectively.
We discuss in Section~\ref{sec:Misspecifiedm} the case when $p \ne m$ and show that our inference methods are still valid.
To conclude this section, we provide in Section~\ref{sec:FigureOutm} a data-driven procedure to identify a possible value for the carryover effect by running multiple experiments. Such a procedure relaxes Assumption~\ref{asp:nomcarryover} and is of great practical relevance.

\subsection{Exact Inference} \label{sec:ExactInference}
We propose an exact non-parametric test for the sharp null of no effect at every time point \citep{fisher1937design,rubin1980randomization, bojinov2019time}:
\begin{equation}
H_{0}:Y_{t}(\bm{w}_{t-m:t}) - Y_{t}(\bm{w}'_{t-m:t}) = 0 \quad \text{for all } \bm{w}_{t-m:t},\bm{w}_{t-m:t}^{\prime },\quad 
t \in \{m+1:T\}.  \label{Fisher null}
\end{equation}
The sharp null hypothesis implies that $Y_{t}(\bm{w}_{t-m:t}^{\obs})=Y_{t}(\bm{w}_{t-m:t})$ for all $\bm{w}_{t-m:t} \in \{0,1\}^t$.
That is, regardless of the assignment path $\bm{w}_{t-m:t}$ we would have observed the same outcomes.

We can conduct exact tests by using the known assignment mechanism to simulate new assignment paths; see Algorithm~\ref{alg:FisherExact} for details.
The test depends on the observation that, under the sharp null hypothesis of no treatment effect \eqref{Fisher null}, any assignment path $\bm{w}^{[i]}_{1:T}$ leads to the same observed outcomes.
In particular, in Step 3, we assume the observed outcomes remain unchanged.
Thus all treatment paths lead to the same observed outcomes $Y^\obs_{m+1:T}$.
To obtain a confidence interval, we propose inverting a sequence of exact hypothesis tests to identify the region outside of which (\ref{Fisher null}) is violated at the prespecified nominal level \citep[Chapter~5]{imbens_rubin_2015}. In practice, obtaining confidence intervals through this approach is somewhat challenging; instead, we refer the reader to the subsequent section that provides a less computationally intensive approach. 

\begin{algorithm}[t]
\begin{algorithmic}[1]
\caption{Algorithm for performing a sharp-null hypothesis test}
\label{alg:FisherExact}
\REQUIRE Fix $I$, total number of samples drawn.
\FOR{i in $1:I$}
\STATE Sample a new assignment path $\bm{w}_{1:T}^{[i]}$ according to the assignment mechanism.
\STATE Hold $Y^\obs_{p+1:T}$ unchanged. Compute $\widehat \tau^{[i]} $ according to (\ref{eqn:estimator}),
\begin{align*}
    \widehat{\tau}^{[i]} & = \frac{1}{T-m} \sum_{t=m+1}^T \left\{ Y_t^{\obs} \frac{\bI{\{\bm{w}^{[i]}_{t-m:t} = \bm{1}_{m+1}\}}}{\Pr(\bm{W}_{t-m:t} = \bm{1}_{m+1})} - Y_t^{\obs} \frac{\bI{\{\bm{w}^{[i]}_{t-m:t} = \bm{0}_{m+1}\}}}{{\Pr(\bm{W}_{t-m:t} = \bm{0}_{m+1})}} \right\}.
\end{align*}
\ENDFOR
\STATE Compute $\widehat{p}_{\mathsf{F}} = I^{-1} \sum_{i=1}^I \bI\left\{\left|\widehat{\tau}^{[i]}\right| > \left| \widehat{\tau}\right|\right\}$
\RETURN $\widehat{p}_{\mathsf{F}}$, the estimated $p$-value. For large $I$, this is exact.
\end{algorithmic}
\end{algorithm}

\subsection{Asymptotic Inference} \label{sec:AsymptoticInference}
We now introduce a conservative test for the null of no average treatment effect:
\begin{equation}
H_{0}: \tau_m = \frac{1}{T-m} \sum_{t=m+1}^{T} [Y_t(\bm{1}_{m+1}) - Y_t(\bm{0}_{m+1})] = 0. \label{Neyman null}
\end{equation}
To test such a null, we derive a finite population central limit theorem to approximate the distribution of the Horvitz-Thompson estimator.

Assume $n = T/m \geq 4$ is an integer, then under the optimal design as shown in Theorems~\ref{thm:FairCoinOptimal} and~\ref{thm:OptimalDesign}, the assignment path is determined by the realizations at $W_1, W_{2m+1}, ..., W_{(n-2)m+1}$.
To make the dependence on randomization clear, we introduce the following notations.
For any $k \in \{0,1,...,n-2\}$, let $\bar{Y}_k(\bm{1}_{m+1}) = \sum_{t=(k+1)m+1}^{(k+2)m} Y_t(\bm{1}_{m+1})$ and $\bar{Y}_k(\bm{0}_{m+1}) = \sum_{t=(k+1)m+1}^{(k+2)m} Y_t(\bm{0}_{m+1})$.
Moreover, for any $k \in \{0,1,...,n-2\}$, let $\bar{Y}_k^\obs = \sum_{t=(k+1)m+1}^{(k+2)m} Y_t^\obs$ be the sum of the observed outcomes.

\begin{lemma}
[Variance of the Horvitz-Thompson Estimator Under the Optimal Design]
\label{lem:HTvariance}
Under Assumptions~\ref{asp:nonanticipating}--\ref{asp:BoundedPO} and under the optimal design as shown in Theorems~\ref{thm:FairCoinOptimal} and~\ref{thm:OptimalDesign}, if $n = T/m \geq 4$ is an integer, then the variance of the Horvitz-Thompson estimator, $\var(\widehat{\tau}_m)$, is
\begin{align}
\var(\widehat{\tau}_m) = \frac{1}{(T-m)^2} & \left\{ \bar{Y}_0(\bm{1}_{m+1})^2 + \bar{Y}_0(\bm{0}_{m+1})^2 + 2 \bar{Y}_0(\bm{1}_{m+1}) \bar{Y}_0(\bm{0}_{m+1}) \vphantom{\sum_{k=0}^{n-3} 2 \left[ \bar{Y}_k(\bm{1}_{m+1}) \right] } \right. \nonumber \\
& + \sum_{k=1}^{n-3} \left[ 3 \bar{Y}_k(\bm{1}_{m+1})^2 + 3 \bar{Y}_k(\bm{0}_{m+1})^2 + 2 \bar{Y}_k(\bm{1}_{m+1}) \bar{Y}_k(\bm{0}_{m+1}) \right] \nonumber \\
& + \bar{Y}_{n-2}(\bm{1}_{m+1})^2 + \bar{Y}_{n-2}(\bm{0}_{m+1})^2 + 2 \bar{Y}_{n-2}(\bm{1}_{m+1}) \bar{Y}_{n-2}(\bm{0}_{m+1}) \nonumber \\
& + \left.\sum_{k=0}^{n-3} 2 \left[ \bar{Y}_k(\bm{1}_{m+1}) + \bar{Y}_k(\bm{0}_{m+1}) \right] \cdot \left[ \bar{Y}_{k+1}(\bm{1}_{m+1}) + \bar{Y}_{k+1}(\bm{0}_{m+1})\right] \right\} \label{eqn:HTVariance}
\end{align}
\end{lemma}

Lemma~\ref{lem:HTvariance} provides the variance of the Horvitz-Thompson estimator under the optimal design.
Since we never observe all the potential outcomes, most of the cross-product terms in \eqref{eqn:HTVariance} can not be directly estimated.
Instead, we provide the following upper bound to \eqref{eqn:HTVariance} and propose an unbiased estimator.

\begin{corollary}
\label{coro:VarianceUpperBound}
Under the conditions in Lemma~\ref{lem:HTvariance}, there exists an upper bound for the variance of the Horvitz-Thompson estimator, $\var(\widehat{\tau}_m) \leq \var^\mathsf{U}(\widehat{\tau}_m)$, which can be estimated by $\widehat{\sigma}^2_\mathsf{U}$, defined as:
\begin{align*}
\widehat{\sigma}^2_\mathsf{U} = \frac{1}{(T-m)^2} \left\{ 8 (\bar{Y}_0^\obs)^2 + \sum_{k=1}^{n-3} 32 (\bar{Y}_k^\obs)^2 \bI\{W_{km+1} = W_{(k+1)m+1}\} + 8 (\bar{Y}_{n-2}^\obs)^2 \right\}.
\end{align*}
Moreover, $\widehat{\sigma}^2_\mathsf{U}$ is unbiased, i.e., $\bE[\widehat{\sigma}^2_\mathsf{U}] = \var^\mathsf{U}(\widehat{\tau}_m)$.
\end{corollary}

Corollary~\ref{coro:VarianceUpperBound} provides the foundation to make conservative inference.
We make the following technical assumption for the asymptotic normal distribution to hold.

\begin{assumption}
[Non-negligible Variance]
\label{asp:NonNegligibleVariance}
Assume that the randomization distribution has a non-negligible variance, i.e.,
\begin{align}
\var(\widehat{\tau}_m) \geq \Omega(n^{-1}). \label{eqn:NonNegligibleVariance}
\end{align}
In particular, one sufficient condition for \eqref{eqn:NonNegligibleVariance} is to assume that all the potential outcomes are positive, i.e., there exists some constant $b>0$, such that $\forall t \in [T], \forall \bm{w}_{1:T} \in \{0,1\}^T$, $Y_t(\bm{w}_{1:T}) \geq b$.
\end{assumption}

Intuitively, the key to most central limit theorems is that all the variables roughly have variances of the same order. In other words, there cannot be a small number of variables that compromise the majority of the variance.
Since under Assumption~\ref{asp:BoundedPO} the potential outcomes are bounded, each variable contributes to the total variance of order $O(n^{-2})$.
Assumption~\ref{asp:NonNegligibleVariance} suggests that the total variance is large enough, such that it cannot come from only a few of the time periods.

\begin{theorem}
[Asymptotic Normality]
\label{thm:AsymptoticNormality}
Let $m$ be fixed. For any $n \geq 4 \in\bN$, define an $n$-replica experiment such that there are $T = n m$ time periods.
We take the optimal design as in Theorem~\ref{thm:OptimalDesign} whose randomization points are at $\bT^* = \{1, 2m+1, 3m+1, ..., (n-2)m+1\}$.
Under Assumptions~\ref{asp:nonanticipating}--\ref{asp:nomcarryover}, and under Assumption~\ref{asp:NonNegligibleVariance}, the limiting distribution of the Horvitz-Thompson estimator in the $n$-replica experiment has an asymptotic normal distribution.
That is, let $\var(\widehat{\tau}_m)$ be defined in Lemma~\ref{lem:HTvariance}.
As $n \to +\infty$,
\begin{align*}
\frac{\widehat{\tau}_m - \tau_m}{\sqrt{\var(\widehat{\tau}_m)}} \xrightarrow[]{D} \mathcal{N}(0,1).
\end{align*}
\end{theorem}

Theorem~\ref{thm:AsymptoticNormality} is in the spirit of the finite population central limit theorems as in \citet{li2017general,aronow2017estimating, chin2018central,bojinov2020panel, han2021population}.
Note that, Theorem~\ref{thm:AsymptoticNormality} does not require $\var(\widehat{\tau}_m)$ to converge as $n \to +\infty$.

To conduct inference, we replace $\var(\widehat{\tau}_m)$ by $\widehat{\sigma}_{\mathsf{U}}^2$ as provided in Corollary~\ref{coro:VarianceUpperBound}.
Define the test statistic to be $z = \left| \widehat{\tau}_m \right| / \sqrt{\widehat{\sigma}_{\mathsf{U}}^2}$.
When the alternative hypothesis is two-sided, the estimated $p$-value is given by $\widehat{p}_{\mathsf{N}} =  2 - 2 \Phi(z)$, where $\Phi$ is the CDF of a standard normal distribution.

The proofs of Lemma~\ref{lem:HTvariance}, Corollary~\ref{coro:VarianceUpperBound}, and Theorem~\ref{thm:AsymptoticNormality} are deferred to Sections~\ref{sec:proof:thm:Variance},~\ref{sec:proof:coro:VarianceUpperBound}, and~\ref{sec:proof:thm:AsymptoticNormality} in the Appendix, respectively.

\subsection{Inference under Misspecified $m$}
\label{sec:Misspecifiedm}
Up to now, we assumed that we knew the order of the carryover effect $m$, and set $p=m$. In practice, we may not know the exact value of the carryover effect, and we have to select $p$ either based on domain knowledge or the procedure we recommend in Section~\ref{sec:FigureOutm}. In this section, we consider what happens when $p \ne m$ and show that the estimation and inference are still valid and meaningful, although the design from Theorem~\ref{thm:OptimalDesign} is no longer optimal. Below we distinguish two cases: $p>m$ and $p<m$.

When $p>m$, due to Assumption~\ref{asp:nomcarryover}, $Y_t(\bm{1}_{p+1}) = Y_t(\bm{1}_{m+1}), \forall t \in \{p+1:T\}$, and the lag-$p$ causal effect is essentially the lag-$m$ causal effect. So all the estimation and inference results still hold.

However, when $p<m$, the Horvitz-Thompson estimator \eqref{eqn:estimator} will be biased for the causal estimand. See Section~\ref{sec:FurtherMisspecified} for more discussions.
When $p<m$, the exact inference procedure as in Section~\ref{sec:ExactInference} remains valid.
For the asymptotic inference procedure, a similar result to Theorem~\ref{thm:AsymptoticNormality} still holds when $m$ is misspecified, as we state in Corollary~\ref{coro:MisspecifiedmCLT}.
The only difference is that when $p<m$, the asymptotic normal distribution will not be centered around the causal estimand as we defined in \eqref{eqn:estimand}, but some quantity that we will discuss in Section~\ref{sec:FurtherMisspecified}.
The proof is deferred to Section~\ref{sec:proof:coro:AsymptoticNormality} in the Appendix.

\begin{corollary}
[Asymptotic Normality when $m$ is Misspecified]
\label{coro:MisspecifiedmCLT}
For any $n \geq 4 \in\bN$, define an $n$-replica experiment such that there are $T = n p$ time periods.
Take the optimal design as in Theorem~\ref{thm:OptimalDesign} whose randomization points are at $\bT^* = \{1, 2p+1, 3p+1, ..., (n-2)p+1\}$.
We have the following two observations.
\begin{enumerate}[label=\roman*]
\item When $p>m$, under Assumptions~\ref{asp:nonanticipating}--\ref{asp:nomcarryover}, the variance of the Horvitz-Thompson estimator, $\var(\widehat{\tau}_p)$, is explicitly given by \eqref{eqn:HTVariance}.
\item Furthermore, no matter if $p>m$ or $p<m$, under Assumptions~\ref{asp:nonanticipating}--\ref{asp:BoundedPO} and assume $\var(\widehat{\tau}_p) \geq \Omega(n^{-1})$, the limiting distribution of the Horvitz-Thompson estimator in the $n$-replica experiment has an asymptotic normal distribution.
That is, as $n \to +\infty$,
\begin{align*}
\frac{\widehat{\tau}_p - \bE[\widehat{\tau}_p]}{\sqrt{\var(\widehat{\tau}_p)}} \xrightarrow[]{D} \mathcal{N}(0,1).
\end{align*}
\end{enumerate}
\end{corollary}

Corollary~\ref{coro:MisspecifiedmCLT}, together with Theorem~\ref{thm:AsymptoticNormality}, is the key to identification of $m$, the order of the carryover effect.
In Section~\ref{sec:FigureOutm} we provide a procedure to identify $m$.

\subsection{Identifying the Order of the Carryover Effect}
\label{sec:FigureOutm}

Using Theorem~\ref{thm:AsymptoticNormality} and Corollary~\ref{coro:MisspecifiedmCLT} we can define a hypothesis testing procedure, which, combined with a searching method, yields an estimate of the order of the carryover effect.

To build intuition, suppose we have access to two comparable experimental units. The two experimental units could be two separate units or two non-overlapping time epochs on one experimental unit such that the two epochs are far enough such that the carryover effect from one does not affect the outcomes of the other. Suppose, on the first experimental unit, we design an optimal experiment under $p=p_1$ and on the second unit, we use $p=p_2$; without loss of generality let $p_1 < p_2$.

After running the experiment and collecting the results, consider the following two statistics. For the first unit, we calculate $\widehat{\tau}_{p_1}$, the sampling average, and $\widehat{\sigma}^2_{p_1}$, the conservative sampling variance as suggested by Corollary~\ref{coro:VarianceUpperBound}.
For the second unit, we calculate $\widehat{\tau}_{p_2}$ and $\widehat{\sigma}^2_{p_2}$.

Define a procedure that tests the following null hypothesis:
\begin{align}
H_0: \ m \leq p_1 \label{SubRoutineNull}
\end{align}
Under the null hypothesis \eqref{SubRoutineNull}, $\tau_{p_1} = \tau_{p_2} = \tau_m$, and so both $\widehat{\tau}_{p_1}$ and $\widehat{\tau}_{p_2}$ are unbiased estimators of $\tau_m$.
Furthermore, given that the two estimators both conform asymptotic normal distributions, and that the two experimental units are independent, the difference between the two estimators should be an asymptotic normal distribution centered around zero, i.e., $(\widehat{\tau}_{p_1} - \widehat{\tau}_{p_2}) / \sqrt{\var(\tau_{p_1}) + \var(\tau_{p_2})} \xrightarrow[]{D} \mathcal{N}(0, 1)$.
To test the null hypothesis \eqref{SubRoutineNull}, define the test statistic to be $z = \left| \widehat{\tau}_{p_1} - \widehat{\tau}_{p_2} \right| / \sqrt{\widehat{\sigma}^2_{p_1} + \widehat{\sigma}^2_{p_2}}$.
The estimated $p$-value is given by $\widehat{p} =  2 - 2 \Phi(z)$, where $\Phi$ is the CDF of a standard normal distribution.

The above procedure enables us to test the null hypothesis \eqref{SubRoutineNull}.
We can combine such a procedure with any searching method to identify $m$.

\section{Simulation Study}
\label{sec:Numerical}

There are five goals for this simulation study.
First, to show that the optimal design in Theorem~\ref{thm:OptimalDesign} has the smallest risk compared against two benchmarks.
There are two dimensions for our comparison: the worst-case risk and the risk under a specific outcome model.
Second, to verify the asymptotic normal distribution under a non-asymptotic setup, and to study the quality of the upper bound proposed in Corollary~\ref{coro:VarianceUpperBound}.
Third, to understand the rejection rate and its dependence on the length of time horizon.
Fourth, to study the performance of the optimal design under a misspecified $m$, and to compare the difference of the two inference methods proposed in Section~\ref{sec:InferenceAndTesting}.
Fifth, to study the performance of the hypothesis testing procedure as proposed in Section~\ref{sec:FigureOutm}, which identifies $m$ the length of the carryover effect.

We start with a simple linear additive carryover effect model which originates from \citet{oman1988switch, hedayat1978repeated, jones2014design}.
\begin{align}
Y_t(\bm{w}_{1:t}) = \mu + \alpha_t + \delta^{(1)} w_t + \delta^{(2)} w_{t-1} + ... + \delta^{(t)} w_1 + \epsilon_t \label{eqn:simu:MoreGeneral}
\end{align}
where $\mu$ is a fixed effect; $\alpha_t$ is a fixed effect associated to period $t$; $\delta^{(1)}, \delta^{(2)}, ..., \delta^{(t)}$ are non-stochastic coefficients; $w_t, w_{t-1}, ..., w_1$ are the treatment indicators; $\epsilon_t$ is the random noise in period $t$.
We will run many simulations based on this model.
For a more detailed discussion of the flexibility of the potential outcome framework, see Section~\ref{sec:simu:Appendix:OutcomeModels} in the Appendix.

\subsection{Comparison of the Risk Functions for Different Designs}
\label{sec:simu:RiskFunctions}

\subsubsection{Simulation setup.}
We consider two setups.
The first setup is for the worst-case risk.
We consider $T=120$, $p=m=2$ where $m$ is correctly identified, and $Y_t(\bm{1}_3) = Y_t(\bm{0}_3) = 10$.
We compare three different designs of switchback experiments.
The first one is our proposed optimal design as in Theorem~\ref{thm:OptimalDesign}, such that $\bT^*=\{1,5,7,...,117\}$.
The second one is the most common and naive switchback experiment, which independently assign treatment/control in every period with half-half probability.
It is parameterized by $\bT^\mathsf{H1}=\{1,2,3,...,120\}$.
The third one is the ``intuitive'' experiment discussed in Table~\ref{tbl:exa:OptimalDesign}, which divides the time horizon into several epochs each with length $m+1=3$.
It is parameterized by $\bT^\mathsf{H2}=\{1,4,7,...,118\}$.

Second, we run simulations based on the outcome model as in \eqref{eqn:simu:MoreGeneral}.
Similar to the first setup, we consider again $T=120, p=m=2$ where $m$ is correctly identified.
For the outcome model, we consider $\mu = 0$, $\alpha_t = \log{(t)}$, and $\epsilon_t \sim N(0,1)$ are i.i.d. standard normal distributions.
For any $t >3$, let $\delta^{(t)} = 0$.
We will vary the values of $\delta^{(1)}, \delta^{(2)}, \delta^{(3)} \in \{1, 2\}$ and conduct experiments under $2^3=8$ different scenarios.
Again we compare the same three different designs of switchback experiments.
$\bT^*=\{1,5,7,...,117\}, \bT^\mathsf{H1}=\{1,2,3,...,120\}$, and $\bT^\mathsf{H2}=\{1,4,7,...,118\}$.

We simulate one assignment path at a time, and conduct an experiment following this assignment path.
Since the outcome model is prescribed, we can calculate both the causal estimand and and the observed outcomes (along the simulated assignment path).
Then, we calculate the Horvitz-Thompson estimator based on the simulated assignment path and the simulated observed outcomes.
With both the estimand and estimator, we can calculate the loss function.
We repeat the above procedure enough ($100000$) times to obtain an accurate approximation of the risk function.

\subsubsection{Simulation results.}
First, we calculate the worst-case risk functions via simulations.
Notice that, when $p=m=2$, we could explicitly calculate the worst-case risk functions under the three different designs of switchback experiments $\bT^*, \bT^\mathsf{H1}$, and $\bT^\mathsf{H2}$.
Even though we can explicitly calculate them via the following expression (See Lemma~\ref{lem:RiskFunctionExplicit} in the Appendix for details),
\begin{align}
\frac{B^2}{(T-m)^2} \left\{ 4 \sum_{k=1}^{K+1} (t_{k} - t_{k-1})^2 + 8 m (t_K - t_1) + 4 m^2 K - 4 m^2 + 4 \sum_{k=2}^{K} [(m-t_k+t_{k-1})^+]^2\right\}, \label{eqn:simu:RiskExpression}
\end{align}
we still use the simulation to confirm this result.
See Table~\ref{tbl:simu:WorstCaseRisk} for our simulation results.

The causal effect is $\tau_2 = 0$ because $Y_t(\bm{1}_3) = Y_t(\bm{0}_3) = 10$.
The simulated estimator is $\bE[\widehat{\tau}^*_2] = -0.0291$ for our proposed optimal design, and $\bE[\widehat{\tau}^\mathsf{H1}_2] = 0.0104$ and $\bE[\widehat{\tau}^\mathsf{H2}_2] = -0.0478$ for the two benchmarks, respectively.
The risk function is $r(\eta_{\bT^*}) = 26.78$ for our proposed optimal design, and $r(\eta_{\bT^\mathsf{H1}}) = 33.67$ and $r(\eta_{\bT^\mathsf{H1}}) = 27.85$ for the two benchmarks, respectively.
Such simulation results suggest that our proposed optimal design have the smallest risk, under the worst case outcome model.
In the last three columns are the risk functions of the three designs, all suggested by expression \eqref{eqn:simu:RiskExpression}.
The risk functions calculated from theory take values that are very close to the risk functions calculated from expression \eqref{eqn:simu:RiskExpression}, which verifies our theory.
\begin{table}[!htb]
\TABLE{Simulation results for the worst-case risk function. \label{tbl:simu:WorstCaseRisk}}
{
\begin{tabular}{| p{1.25cm} | p{1.25cm} | p{1.25cm} | p{1.25cm} | p{1.25cm} | p{1.25cm} | p{1.25cm} | p{1.25cm} | p{1.25cm} | p{1.25cm} |}
\hline
$\tau_2$ & $\bE[\widehat{\tau}^*_2]$ & $\bE[\widehat{\tau}^\mathsf{H1}_2]$ & $\bE[\widehat{\tau}^\mathsf{H2}_2]$ & $r(\eta_{\bT^*})$ & $r(\eta_{\bT^\mathsf{H1}})$ & $r(\eta_{\bT^\mathsf{H2}})$ & $\tilde{r}(\eta_{\bT^*})$ & $\tilde{r}(\eta_{\bT^\mathsf{H1}})$ & $\tilde{r}(\eta_{\bT^\mathsf{H2}})$ \\
\hline
$0$ & $0.0250$ & $0.0200$ & $0.0059$ & $26.78$ & $33.67$ & $27.85$ & 26.67 & 33.96 & 27.81 \\
\hline
\end{tabular}
}
{The optimal design $\bT^*$ as suggested by Theorem~\ref{thm:OptimalDesign} yields the smallest risk, both in theory and confirmed by simulations.}
\end{table}

Second, we calculate the risk functions based on the outcome model in \eqref{eqn:simu:MoreGeneral}.
See Table~\ref{tbl:simu:OneModelRisk}.
As we vary the values of $\delta^{(1)}$, $\delta^{(2)}$ and $\delta^{(3)}$, the average lag-$2$ causal effect is being changed.
All three estimators are able to reflect the change as the estimand changes.
The risk function can be simulated and we see that the risk function associated with the first benchmark $\bT^\mathsf{H1}$ is $28\% \sim 32\%$ larger than the optimal design; and the second benchmark $\bT^\mathsf{H2}$ is $1\% \sim 2\%$ larger.
Such simulation results suggest again that our proposed optimal design have the smallest risk.
\begin{table}[!htb]
\TABLE{Simulation results for the risk function based on the outcome model in \eqref{eqn:simu:MoreGeneral}. \label{tbl:simu:OneModelRisk}}
{
\begin{tabular}{| p{1cm} | p{1cm} | p{1cm} | p{1cm} | p{1.5cm} | p{1.5cm} | p{1.5cm} | p{1.5cm} | p{1.5cm} | p{1.5cm} |}
\hline
$\delta^{(1)}$ & $\delta^{(2)}$ & $\delta^{(3)}$ & $\tau_2$ & $\bE[\widehat{\tau}^*_2]$ & $\bE[\widehat{\tau}^\mathsf{H1}_2]$ & $\bE[\widehat{\tau}^\mathsf{H2}_2]$ & $r(\eta_{\bT^*})$ & $r(\eta_{\bT^\mathsf{H1}})$ & $r(\eta_{\bT^\mathsf{H2}})$ \\ \hline
1 & 1 & 1 & 3 & 3.016 & 3.012 & 3.002 & 7.96  & 10.22 & 8.11 \\ \hline
1 & 1 & 2 & 4 & 4.018 & 4.013 & 4.002 & 9.57  & 12.39 & 9.74 \\ \hline
1 & 2 & 1 & 4 & 4.018 & 4.013 & 4.002 & 9.57  & 12.39 & 9.74 \\ \hline
2 & 1 & 1 & 4 & 4.018 & 4.013 & 4.002 & 9.57  & 12.39 & 9.74  \\ \hline
1 & 2 & 2 & 5 & 5.020 & 5.015 & 5.003 & 11.34 & 14.81 & 11.52 \\ \hline
2 & 1 & 2 & 5 & 5.020 & 5.015 & 5.003 & 11.34 & 14.81 & 11.52 \\ \hline
2 & 2 & 1 & 5 & 5.020 & 5.015 & 5.003 & 11.34 & 14.81 & 11.52 \\ \hline
2 & 2 & 2 & 6 & 6.022 & 6.016 & 6.003 & 13.28 & 17.48 & 13.47 \\ \hline
\end{tabular}
}
{For each row, the random seed that generates the simulation setup is fixed. The optimal design $\bT^*$ as suggested in Theorem~\ref{thm:OptimalDesign}, though solved from a minimax program, still yields the smallest risk for the outcome model in \eqref{eqn:simu:MoreGeneral}. A few rows are redundant because our switchback experiment, combining with the causal estimand \eqref{eqn:estimand}, is only able to measure the total additive treatment effect. We cannot distinguish the source of the additive treatment effects, i.e., we are unable to distinguish $\delta^{(1)}$, $\delta^{(2)}$, and $\delta^{(3)}$.}
\end{table}
Moreover, as $r(\eta_{\bT^\mathsf{H2}})$ is close to $r(\eta_{\bT^*})$ and both are much smaller than $r(\eta_{\bT^\mathsf{H1}})$, our results suggest that when $m$ is unknown, it is better to select $p$ to be slightly larger than the true $m$ as opposed to significantly smaller.

As the magnitude of treatment effects increase, the associated risk functions also increase.
The relative difference between risk functions of $r(\eta_{\bT^\mathsf{H1}})$ and $r(\eta_{\bT^*})$ increases, while the relative difference between $r(\eta_{\bT^\mathsf{H1}})$ and $r(\eta_{\bT^*})$ decreases.
This coincides with the intuitions discussed in Section~\ref{sec:MinimaxOptimization}.


\subsection{Asymptotic Normality}
\label{sec:simu:AsymptoticNormality}
\subsubsection{Simulation setup.}
\label{sec:simu:AsymptoticNormality:setup}
We run simulations based on the outcome model in \eqref{eqn:simu:MoreGeneral}, with $T=120$ and $m=2$.
We will consider three cases: (i) $m$ is correctly specified so $p=2$; (ii) $p=3$, and we estimate lag-$3$ causal estimand as in \eqref{eqn:estimand}; (iii) $p=1$, and we pretend as if we estimated the lag-$1$ causal estimand.
However, as the lag-$1$ causal estimand is not well defined, we instead estimate a different quantity, which we refer to as the ``$m$-misspecified lag-$p$ causal estimand'' (See details and definition in \eqref{eqn:estimandgeneral}).

For the outcome model, we consider $\mu = 0$, $\alpha_t = \log{(t)}$, and $\epsilon_t \sim N(0,1)$ are i.i.d. standard normal distributions.
For any $t >3$, let $\delta^{(t)} = 0$.
For simplicity, let $\delta^{(1)} = \delta^{(2)} = \delta^{(3)} = \delta$.
We vary $\delta \in \{1,2,3\}$ and conduct experiments under $3$ different scenarios.
We simulate one assignment path at a time, and conduct experiments following this assignment path.
Since the outcome model is prescribed, we calculate the observed outcomes based on the simulated assignment path.
Then we calculate the Horvitz-Thompson estimator, and the conservative estimator of the randomization variance (Corollary~\ref{coro:VarianceUpperBound}), based on the simulated assignment path and the simulated observed outcomes.
On the other hand, the lag-$p$ causal estimand is easy to calculate once the outcome model is prescribed.
Yet the $m$-misspecified lag-$p$ causal estimand has to be calculated in conjunction with the simulated assignment path.
By repeating the above procedure enough ($100000$) times we obtain a distribution of the estimator.

\subsubsection{Simulation results.}
\begin{figure}[!htb]
\centering
\includegraphics[width=0.7\textwidth]{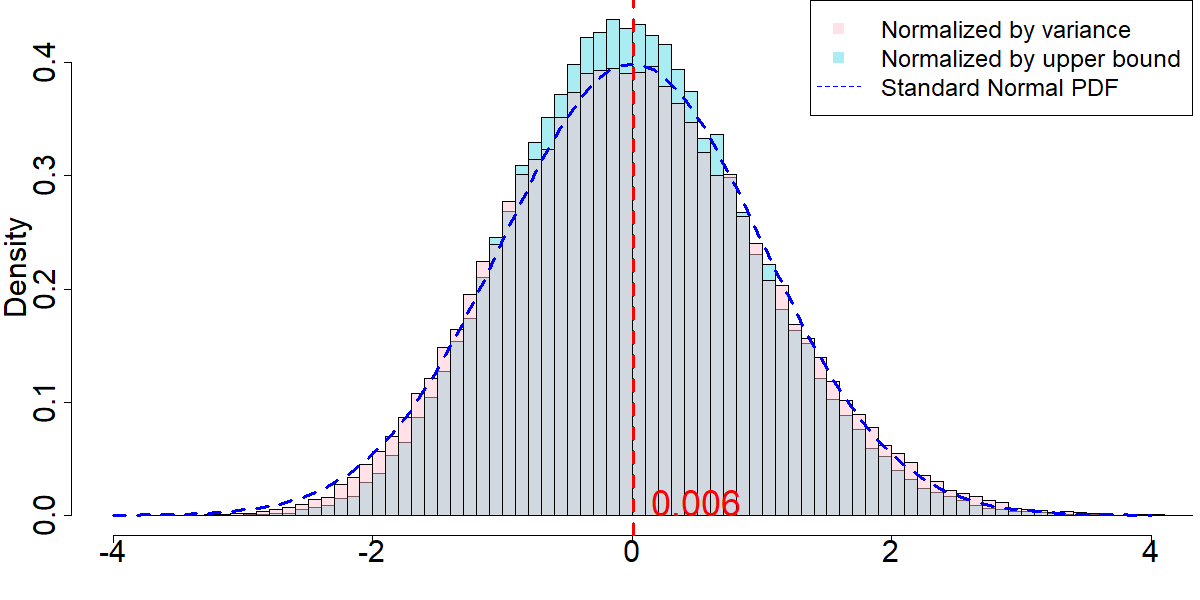}
\caption{Approximate normality of the randomization distribution when $m=2, p=2, \delta=3$.}
\label{fig:mCorrect:ApproximateNormality:delta=3}
\end{figure}

In Figure~\ref{fig:mCorrect:ApproximateNormality:delta=3}, the dotted dark blue line is the Probability Density Function of the standard normal distribution.
The pink histogram corresponds to the distribution induced by $\frac{\widehat{\tau}_p - \tau_p}{\sqrt{\var(\widehat{\tau}_p)}}$, which is the estimator (after re-centering at zero) normalized by the square root of the true randomization variance\footnote{We numerically find such variance $\var(\widehat{\tau}_p)$, and the expectation of the conservative upper bound $\bE[\widehat{\sigma}^2_{U}]$}.
Such a distribution, as suggested by Theorem~\ref{thm:AsymptoticNormality}, converges to a standard normal distribution when $T$ is large.
Comparing to the dotted dark blue line, Figure~\ref{fig:mCorrect:ApproximateNormality:delta=3} suggests that Theorem~\ref{thm:AsymptoticNormality} approximately holds for moderate values of $T$.
The light blue histogram corresponds to the distribution induced by $\frac{\widehat{\tau}_p - \tau_p}{\sqrt{\bE[\widehat{\sigma}^2_{U}]}}$, which is the estimator (after re-centering at zero) normalized by the expectation of the conservative upper bound of the randomization variance.
Since we replace the true variance by the conservative upper bound, the shape of the distribution is more concentrated around zero, as we see from the ``taller'' histogram.
The red vertical line is the expected value of the randomization distribution for the pink histogram.
The cases of $\delta=1$ and $\delta=2$ are similar, and the cases of overestimated $m$ and underestimated $m$ are also similar. We discuss them in Section~\ref{sec:simu:additional:AsymptoticNormality} in the Appendix.

For all the nine cases ($p \in \{1,2,3\}$ and $\delta\in\{1,2,3\}$), see Table~\ref{tbl:simu:RandomizationDistribution} for the expected values and the variances of the randomization distributions, as well as the conservative estimator of the randomization variances.
Note that the three cases all have the same underlying outcome model.
It is the different knowledge of $m$ that leads to three different designs of experiments.

\begin{table}[!htb]
\TABLE{Simulation results for the randomization distribution. \label{tbl:simu:RandomizationDistribution}}
{
\begin{tabular}{| p{3.6cm} | p{1.6cm} | p{1.6cm} | p{1.6cm} | p{1.6cm} | p{1.6cm} | p{1.6cm} |}
\hline
\multicolumn{2}{|l|}{}                              & $\tau_p$ & $\tau^{[m]}_p$ & $\bE[\widehat{\tau}_p]$ & $\var(\widehat{\tau}_p)$ & $\bE[\widehat{\sigma}^2_{\mathsf{U}}]$ \\ \hline
\multirow{3}{*}{$m=2, p=2$}         & $\delta=1$    & 3        & $-$            & 3.016          & 7.96                 & 8.48  \\ \cline{2-7} 
                                    & $\delta=2$    & 6        & $-$            & 6.022          & 13.28                & 15.16 \\ \cline{2-7} 
                                    & $\delta=3$    & 9        & $-$            & 9.028          & 20.10                & 24.25 \\ \hline
\multirow{3}{*}{$m=2, p=3$}         & $\delta=1$    & 3        & $-$            & 3.006          & 11.92                & 12.67 \\ \cline{2-7} 
                                    & $\delta=2$    & 6        & $-$            & 6.009          & 19.89                & 22.70 \\ \cline{2-7} 
                                    & $\delta=3$    & 9        & $-$            & 9.012          & 30.10                & 36.32 \\ \hline
\multirow{3}{*}{$m=2, p=1$}         & $\delta=1$    & $-$      & 2              & 2.016          & 4.00                 & 4.13  \\ \cline{2-7} 
                                    & $\delta=2$    & $-$      & 4              & 4.026          & 6.69                 & 7.06  \\ \cline{2-7} 
                                    & $\delta=3$    & $-$      & 6              & 6.037          & 10.14                & 10.92 \\ \hline
\end{tabular}
}
{The randomization distribution is unbiased in all 9 cases (when $p<m$ it is unbiased for the $m$-misspecified average lag-$1$ causal effect). The conservative estimation of the variance upper bound from Corollary~\ref{coro:VarianceUpperBound} is close to the true variance.}
\end{table}

From Table~\ref{tbl:simu:RandomizationDistribution}, we make the following two observations.
\textbf{(i) Unbiasedness of the Horvitz-Thompson estimator}.
When $m$ is correctly specified, $\bR[\widehat{\tau}_p]$ is very close to $\tau_p$, verifying the unbiasedness of the estimator.
When $m=2, p=3$, the estimand remains unchanged, and the estimator remains unbiased.
But the variance of the estimator is larger.
When $m=2, p=1$, the estimand is the $m$-misspecified estimand, and the estimator is unbiased for this $m$-misspecified estimand.
\textbf{(ii) Quality of Corollary~\ref{coro:VarianceUpperBound} and~\ref{coro:MisspecifiedmCLT}}.
As we increase $\delta$, the variances of the randomization distributions also increase.
The conservative estimators of the randomization variances are very close to the true variances, which suggests that Corollary~\ref{coro:VarianceUpperBound} and~\ref{coro:MisspecifiedmCLT} approximate the true variances quite well.

\subsubsection{Robustness Check.}
\label{sec:simu:Cauchy:AsymptoticNormality}
In this section we run simulations under almost the same setup as introduced in Section~\ref{sec:simu:AsymptoticNormality:setup}, with the only difference that we select each $\epsilon_t$ to be an i.i.d. Student's t-distribution with 1 degree of freedom.
The purpose of this section is to verify our theory when $\epsilon_t$ are drawn from heavy tailed distributions.

\begin{figure}[!htb]
\centering
\includegraphics[width=0.7\textwidth]{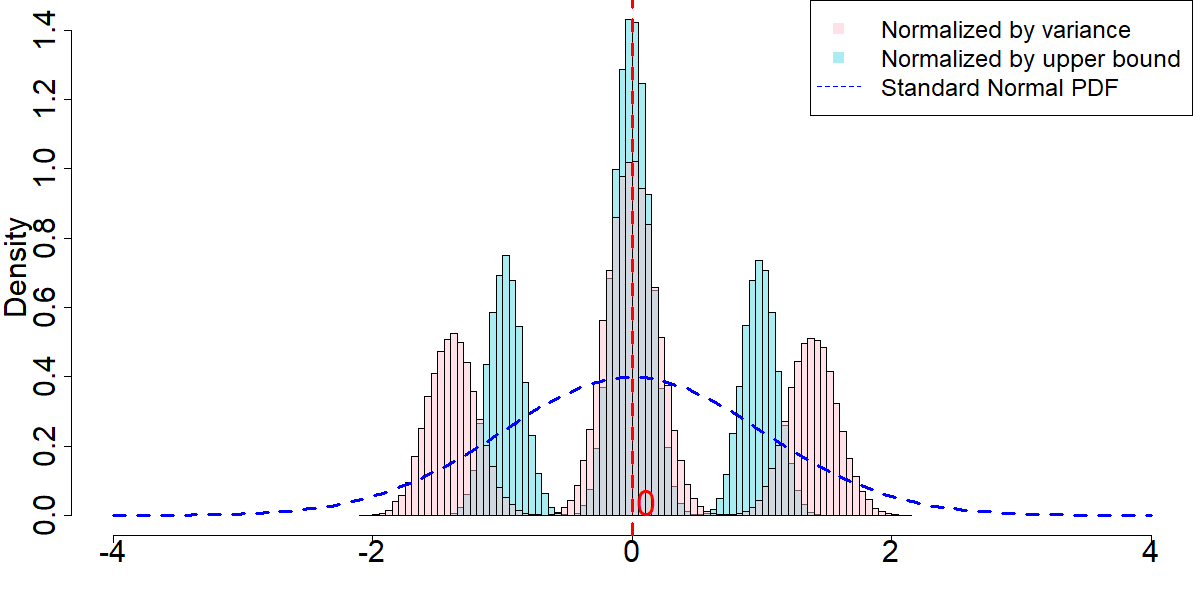}
\caption{Randomization distribution when random noises are Student's t-distributions, and when $m=2, p=2, \delta=1, T=120$.}
\label{fig:mCorrect:ApproximateNormality:delta=1;T=120}
\end{figure}

When $m=2, p=2, \delta=1$, as we can see from Figure~\ref{fig:mCorrect:ApproximateNormality:delta=1;T=120}, the randomization distribution is significantly different from a standard normal distribution.
This is because $T=120$ is too small.
Alternatively, we increase $T=1200$ to see that the randomization distribution behaves like a normal distribution.
In other words, when $\epsilon_t$ noises are heavy tailed, our Theorem~\ref{thm:AsymptoticNormality} has a slower convergence rate to a normal distribution.
We conduct extensive simulation study under other parameters, as we will show in Section~\ref{sec:simu:additional:AsymptoticNormality} in the Appendix.

\begin{figure}[!htb]
\centering
\includegraphics[width=0.7\textwidth]{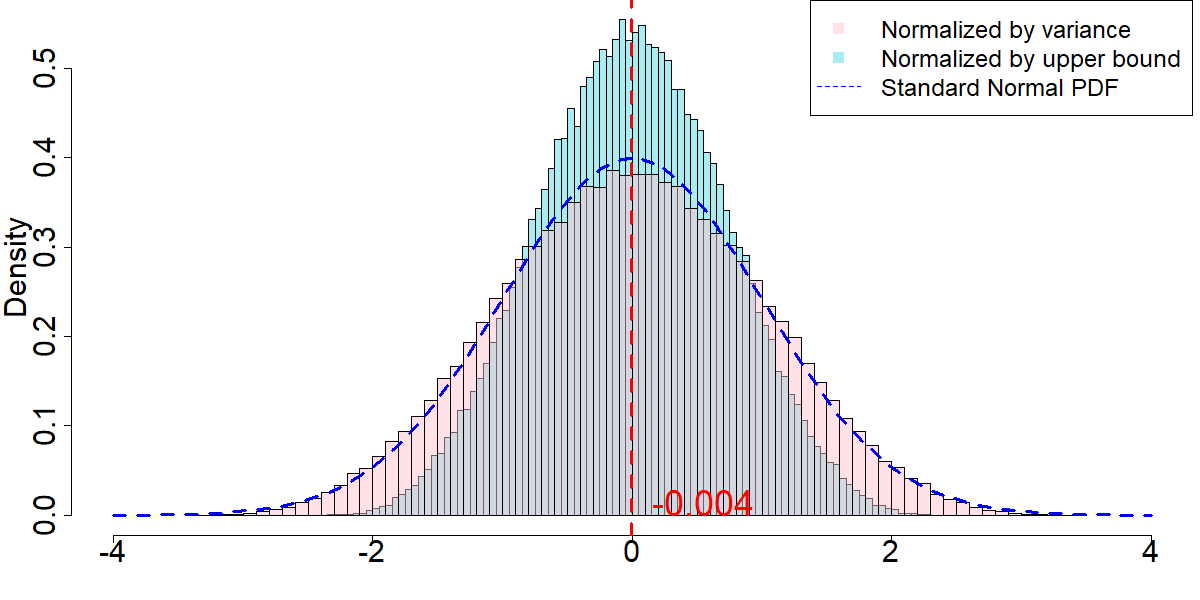}
\caption{Randomization distribution when random noises are Student's t-distributions, and when $m=2, p=2, \delta=1, T=1200$.}
\label{fig:mCorrect:ApproximateNormality:delta=1;T=1200}
\end{figure}

\subsection{Rejection Rates}
\label{sec:simu:RejectionRates}

\subsubsection{Simulation setup.}
We run simulations based on the outcome model as in \eqref{eqn:simu:MoreGeneral}.
We vary $T \in \{120, 240, ..., 1200\}$.
We consider $p=m=2$ where $m$ is correctly specified.
Similar to Section~\ref{sec:simu:AsymptoticNormality}, we consider the same parameterization and conduct experiments under $3$ different scenarios $\delta \in \{1,2,3\}$.

We simulate one assignment path at a time, and conduct experiments following this assignment path.
We first calculate the observed outcomes and the Horvitz-Thompson estimator.
Then we conduct the two inference methods as proposed in Section~\ref{sec:InferenceAndTesting}, and obtain two estimated $p$-values.
For the asymptotic inference method, we plug in $\widehat{\sigma}^2_{\mathsf{U}}$, the conservative upper bound of the variance.
We reject the corresponding null hypothesis when the $p$-value is smaller than $0.1$ (In Section~\ref{sec:simu:additional:RejectionRates} we run additional simulations by replacing such $0.1$ threshold by $0.05$ and $0.01$).
By repeating the above procedure enough (in this simulation, 1000) times we obtain the frequency of a null hypothesis being rejected, which we refer to as the rejection rate.

\subsubsection{Simulation Results.}
\begin{figure}[!htb]
\begin{subfigure}{.33\textwidth}
\centering
\includegraphics[width=\textwidth]{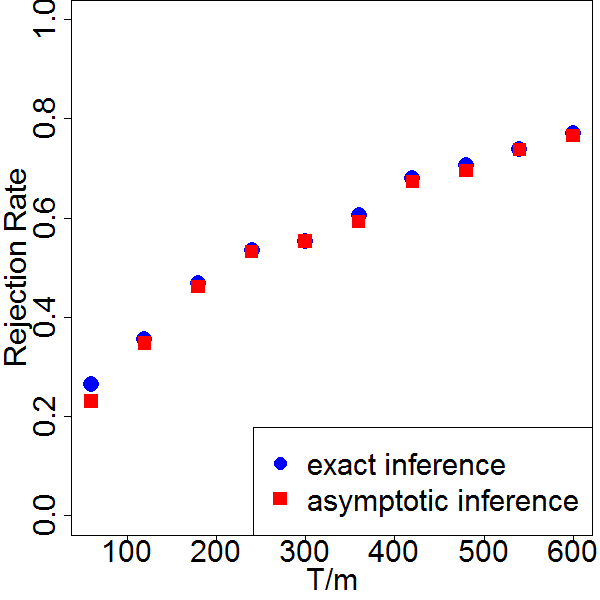}
\end{subfigure}\hfill
\begin{subfigure}{.33\textwidth}
\centering
\includegraphics[width=\textwidth]{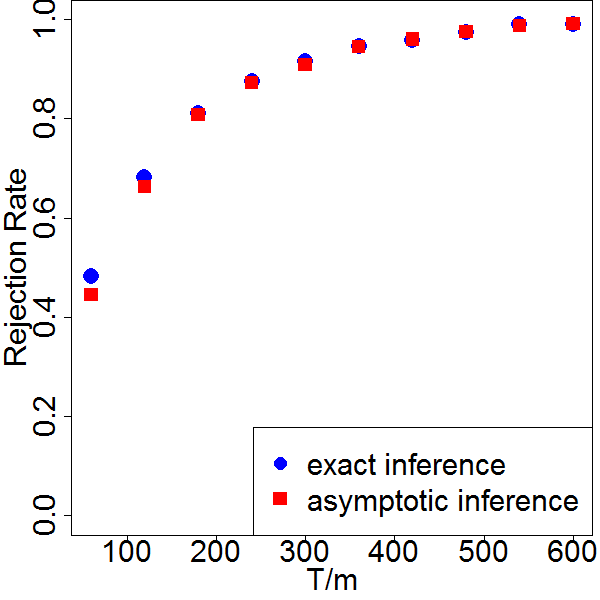}
\end{subfigure}\hfill
\begin{subfigure}{.33\textwidth}
\centering
\includegraphics[width=\textwidth]{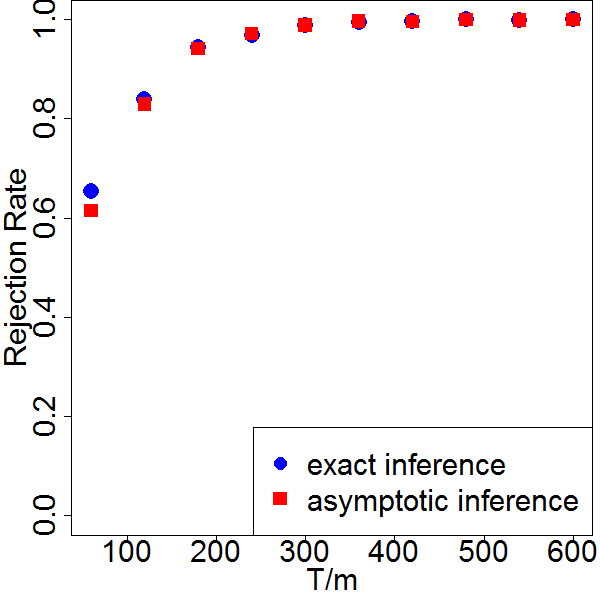}
\end{subfigure}
\caption{Rejection rates and their dependence on $T/m$. Left: $\delta=1$; Middle: $\delta=2$; Right: $\delta=3$}
\label{fig:RejectionRates}
\end{figure}

We calculate the rejection rates via simulations and then plot Figure~\ref{fig:RejectionRates}.
The blue dots are rejection rates under exact inference; the red dots are under asymptotic inference.
In all the simulations, $\delta \ne 0, \tau_p \ne 0$.
So, ideally, we would wish to reject both the Fisher's null hypothesis \eqref{Fisher null} and the Neyman's null hypothesis \eqref{Neyman null}.

From Figure~\ref{fig:RejectionRates} we make the following three observations. \textbf{(i) Dependence on $T/m$}. The rejection rates increase as the length of the horizon increases -- more specifically, as $T/m$ the total number of epochs increases.
In practice, when firms have to capability to choose the length of $T$, they can refer to Figure~\ref{fig:RejectionRates} to choose $T$ properly. Also see discussion in Section~\ref{sec:PracticalImplications}.
\textbf{(ii) Between two inference methods}. In all three cases, the rejection rate from testing a sharp null hypothesis \eqref{Fisher null} is slightly higher than that from testing the Neyman's null \eqref{Neyman null}.
This coincides with our intuition that a sharp null is more likely to be rejected.
We discuss this in Section~\ref{sec:simu:misspecification:Results} together with the associated $p$-values.
\textbf{(iii) Dependence on the signal-to-noise ratio}. The rejection rates all increase as $\delta$ increases from 1 to 3 (while holding the noise from the model fixed). This suggests that when the treatment effect is relatively larger, we do not require a long experimental horizon to achieve a desired rejection rate.

\subsection{Comparison of the Type I and Type II Errors for Different Designs}
\label{sec:simu:RejectionRates:DifferentDesigns}

\subsubsection{Simulation setup.}
We run simulations based on the outcome model as in \eqref{eqn:simu:MoreGeneral}.
We vary $T \in \{120, 240, ..., 1200\}$.
We consider $p=m=2$ where $m$ is correctly specified.
Similar to Section~\ref{sec:simu:AsymptoticNormality}, we consider the same parameterization and conduct experiments under $3$ different scenarios $\delta \in \{1,2,3\}$.
We compare three designs of experiments as described in Section~\ref{sec:simu:RiskFunctions}:
the optimal design $\bT^*=\{1,5,7,...,117\}$, which we refer to as \textit{Optimal Design} as in Figure~\ref{fig:TypeIandIIErrors};
the most commonly adopted heuristic $\bT^\mathsf{H1}=\{1,2,3,...,120\}$, which we refer to as \textit{Heuristic Design H1};
and the so-called intuitive design $\bT^\mathsf{H2}=\{1,4,7,...,118\}$, which we refer to as \textit{Heuristic Design H2}.

In this simulation, we first calculate the frequency of rejecting the Fisher's null hypothesis as in \eqref{Fisher null} out of a total of 1000 repetitions.
And then, we use the frequency to calculate the Type I and Type II errors.
Type I error is the probability of rejecting the null hypothesis when there is no treatment effect, which we simulate the frequency of rejection using $\delta=0$ when there is no treatment effect.
Type II error is the probability of not rejecting the null hypothesis when there is a treatment effect, which we simulate as $1$ minus the frequency of rejection using $\delta\in\{1,2,3\}$ when there is a non-negligible treatment effect.

\subsubsection{Simulation results.}
\begin{figure}[!htb]
\begin{subfigure}{0.475\textwidth}
\centering
\includegraphics[width=\textwidth]{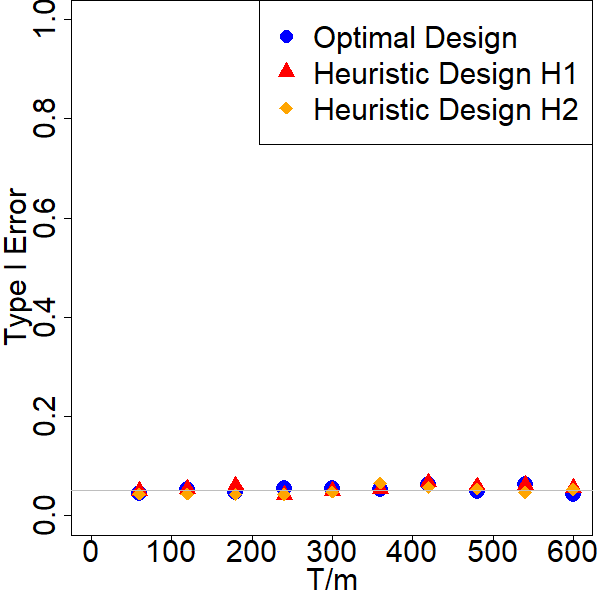}
\end{subfigure}
\hfill
\begin{subfigure}{0.475\textwidth}
\centering 
\includegraphics[width=\textwidth]{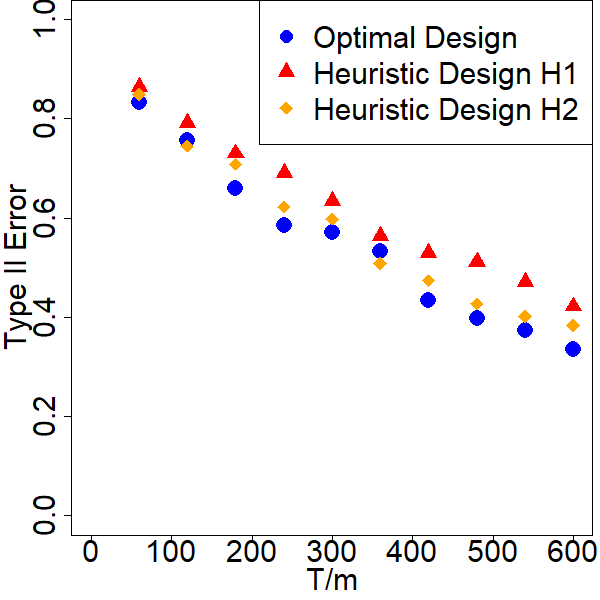}
\end{subfigure}
\vskip\baselineskip
\begin{subfigure}{0.475\textwidth}   
\centering 
\includegraphics[width=\textwidth]{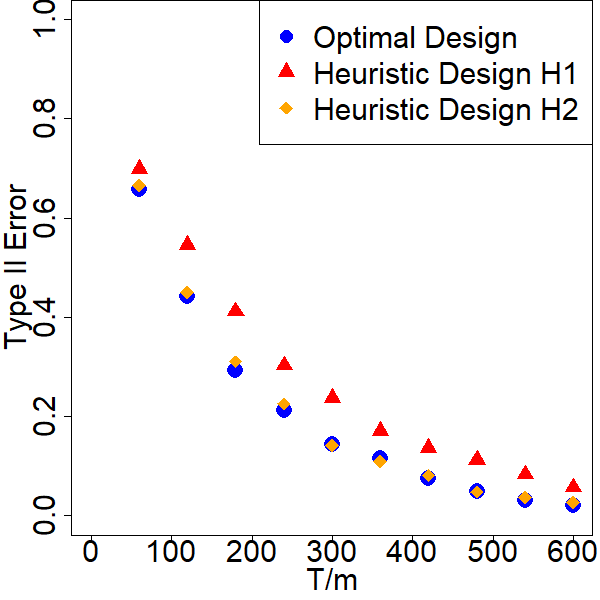}
\end{subfigure}
\hfill
\begin{subfigure}{0.475\textwidth}   
\centering 
\includegraphics[width=\textwidth]{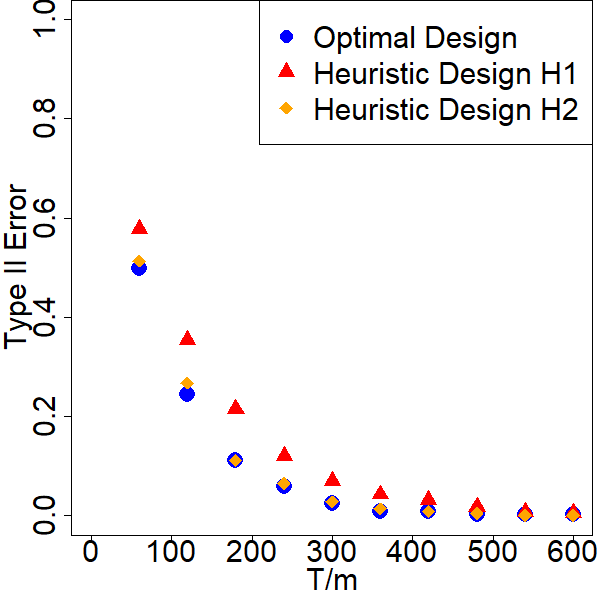}
\end{subfigure}
\caption{Type I and Type II errors. Upperleft: $\delta=0$, Type I error; Upperright: $\delta=1$, Type II error; Bottomleft: $\delta=2$, Type II error; Bottomright: $\delta=3$, Type II error.}
\label{fig:TypeIandIIErrors}
\end{figure}

The simulation results are summarized in Figure~\ref{fig:TypeIandIIErrors}.
The blue dots are the Type I and Type II errors of the optimal design; the red dots are the Type I and Type II errors of the heuristic design $H1$; the yellow dots are the Type I and Type II errors of the heuristic design $H2$.
The figure on the top-left corner reports the Type I error generated from $\delta=0$.
The grey horizontal line in the top-left figure represents the $0.05$ nominal level.
The other figures report the Type II errors generated from $\delta\in\{1,2,3\}$.

From Figure~\ref{fig:TypeIandIIErrors} we make the following observations.
First, for Type I error, all the three designs have similar performance --- all are very close to the $0.05$ nominal level.
Second, the optimal design almost always has the smallest Type II error.
This suggests that, even though we design our optimal experiment under the minimax criterion, the optimal design derived from this criterion outperforms the two heuristic benchmarks with respect to the Type II error.
The Type II error becomes smaller when $T/m$, the effective experimental periods, increases.
The gaps between the optimal design and the two heuristic designs also become smaller when $T/m$ increases.

\subsection{Estimation under a Misspecified $m$}
\label{sec:simu:misspecifiedm}

\subsubsection{Simulation setup.}
We run simulations whose setup are similar to Section~\ref{sec:simu:AsymptoticNormality:setup}; the only difference is that we only simulate one assignment path in this Section, and conduct hypothesis testing for this single run of the experiment.

The outcome model we consider is in \eqref{eqn:simu:MoreGeneral}, and we consider the same parameterization as in Section~\ref{sec:simu:AsymptoticNormality:setup}, and conduct experiments under $3$ different scenarios $\delta \in \{1,2,3\}$.
We consider three cases: (i) $m$ correctly specified so $p=2$; (ii) $p=3$, and we estimate the lag-$3$ causal estimand as in \eqref{eqn:estimand}; (iii) $p=1$, and we pretend as if we estimated the lag-$1$ causal estimand.
However, the lag-$1$ causal estimand is not well defined. Instead, we estimate the $2$-misspecified lag-$1$ causal estimand as in \eqref{eqn:estimandgeneral}.

We only simulate one assignment path.
Since the outcome model is prescribed, we calculate the observed outcomes.
There is only one time series of such observed outcomes.
We calculate the Horvitz-Thompson estimator based on the simulated assignment path and the simulated observed outcomes.
We calculate the lag-$p$ causal estimand directly, and also the $m$-misspecified lag-$p$ causal estimand in conjunction with the simulated assignment path.
Finally, we perform the two inference methods from Section~\ref{sec:InferenceAndTesting}, and report their associated estimated $p$-values.
For the asymptotic inference method we plug in $\widehat{\sigma}^2_{\mathsf{U}}$ the conservative upper bound of the variance.
We choose $I=100000$ to be the number of samples drawn in the exact inference method as shown in Algorithm~\ref{alg:FisherExact}.

\subsubsection{Simulation results.} \label{sec:simu:misspecification:Results}

Notice this is only one experiment under one simulated experimental setup from one simulated assignment path.
So the estimators $\widehat{\tau}_p$ we derive are different from $\tau_p$ (or $\tau_p^{(m)}$, which stands for the treatment effect when $m$ is misspecified; see Section~\ref{sec:FurtherMisspecified} in the Appendix for more details).
But they still follow the true causal effects which they estimate.
See Table~\ref{tbl:simu:SpecificInstances}.
\begin{table}[!htb]
\TABLE{Simulation results for correctly specified $m$ case, and two misspecified $m$ cases. \label{tbl:simu:SpecificInstances}}
{
\begin{tabular}{| p{3cm} | p{1.3cm} | p{1.3cm} | p{1.3cm} | p{1.3cm} | p{1.3cm} | p{1.3cm} | p{1.3cm} |}
\hline
\multicolumn{2}{|l|}{}                      & $\tau_p$ & $\tau^{(m)}_p$ & $\widehat{\tau}_p$ & $\widehat{\sigma}^2_{\mathsf{U}}$ & $\widehat{p}_\mathsf{F}$ & $\widehat{p}_\mathsf{N}$ \\ \hline      
\multirow{3}{*}{$m=2,p=2$}  & $\delta=1$    & 3        & $-$            & 1.35           & 8.81                          & 0.626                & 0.648                \\ \cline{2-8} 
                            & $\delta=2$    & 6        & $-$            & 4.30           & 15.16                         & 0.231                & 0.269                \\ \cline{2-8} 
                            & $\delta=3$    & 9        & $-$            & 7.25           & 23.88                         & 0.101                & 0.138                \\ \hline      
\multirow{3}{*}{$m=2,p=3$}  & $\delta=1$    & 3        & $-$            & 1.77           & 14.26                         & 0.606                & 0.639                \\ \cline{2-8} 
                            & $\delta=2$    & 6        & $-$            & 5.00           & 24.69                         & 0.262                & 0.314                \\ \cline{2-8} 
                            & $\delta=3$    & 9        & $-$            & 8.23           & 39.00                         & 0.136                & 0.188                \\ \hline      
\multirow{3}{*}{$m=2,p=1$}  & $\delta=1$    & $-$      & 2              & -1.03          & 3.87                          & 0.590                & 0.599                \\ \cline{2-8} 
                            & $\delta=2$    & $-$      & 4              & 0.41           & 6.28                          & 0.866                & 0.870                \\ \cline{2-8} 
                            & $\delta=3$    & $-$      & 6              & 1.86           & 9.47                          & 0.530                & 0.547                \\ \hline      
\end{tabular}
}
{The simulation setup for the three $\delta=1$ cases is the same; so are the $\delta=2$ cases and $\delta=3$ cases. The estimated $p$-values $\widehat{p}_{\mathsf{F}}$ derived from the exact inference are slightly smaller than the $p$-values $\widehat{p}_{\mathsf{N}}$ derived from the asymptotic inference.}
\end{table}

From Table~\ref{tbl:simu:SpecificInstances} we see that both our estimator and the estimated variance are well defined in all the cases when $p=m$, $p>m$, and $p<m$.
In each case, as $\delta$ increases from 1 to 3, the associated $p$-values exhibit decreasing trends, suggesting a stronger rejection rate against the null hypothesis.
Moreover, the $p$-values suggested by the exact inference are always slightly smaller than the $p$-values suggested by the asymptotic inference.
This coincides with our intuition that: (i) the exact inference method possesses a stronger null hypothesis \eqref{Fisher null} which implies the null hypothesis of \eqref{Neyman null}; (ii) in the asymptotic inference we replaced the true randomization variance by its conservative upper bound, which further leads to a larger $p$-value.

\subsection{Estimation of $m$}
\label{sec:simu:Estimatem}

We run simulations based on the outcome model as in \eqref{eqn:simu:MoreGeneral}, to test the performance of the procedure described in Section~\ref{sec:FigureOutm}.
In this section we only focus on $\delta = 3$.
Suppose we have narrowed down the range of the order of the carryover effect to be $m \leq 3$.
In the first round, we use our procedure to test a null hypothesis $m \leq 2$.
Then we would observe row 3 and 6 from Table~\ref{tbl:simu:SpecificInstances}, with $\widehat{\tau}_2 = 7.25, \widehat{\sigma}^2_{2} = 23.88; \widehat{\tau}_3 = 8.23, \widehat{\sigma}^2_{3} = 39.00$.
So the estimated $p$-value for the null hypothesis $m \leq 2$ is estimated to be $\widehat{p} = 0.902$, which is too large to reject the null hypothesis.
In the second round, we consult the procedure to test a null hypothesis $m \leq 1$.
Then we would observe row 3 and 9 from Table~\ref{tbl:simu:SpecificInstances}, with $\widehat{\tau}_1 = 1.86, \widehat{\sigma}^2_{3} = 9.47; \widehat{\tau}_2 = 7.25, \widehat{\sigma}^2_{2} = 23.88$.
The estimated $p$-value for the null hypothesis $m \leq 1$ is estimated to be $\widehat{p} = 0.350$.
This is still rather large, yet a significant difference from $0.902$.

We conduct a few more numerical simulations with different time periods.
The setup is the same as in Section~\ref{sec:simu:misspecifiedm}, except that $T$ takes values in $T \in \{210, 1020, 2010\}$\footnote{The values of $T$ were selected such that they were both divisible by both 2 and 3, the possible values of the carryover effect.}.
When $T=210$,
in the first round the estimated $p$-value for the null hypothesis $m \leq 2$ is estimated to be $\widehat{p} = 0.956$;
in the second round the estimated $p$-value for the null hypothesis $m \leq 1$ is estimated to be $\widehat{p} = 0.182$.
When $T=1020$,
in the first round the estimated $p$-value for the null hypothesis $m \leq 2$ is estimated to be $\widehat{p} = 0.869$;
in the second round the estimated $p$-value for the null hypothesis $m \leq 1$ is estimated to be $\widehat{p} = 0.163$.
When $T=2010$,
in the first round the estimated $p$-value for the null hypothesis $m \leq 2$ is estimated to be $\widehat{p} = 0.760$;
in the second round the estimated $p$-value for the null hypothesis $m \leq 1$ is estimated to be $\widehat{p} = 0.037$.
In practice, we suggest increasing the horizon's length to a degree such that $T/p > 100$.

\section{Practical Implications, Limitations, and Concluding Remarks} \label{sec:PracticalImplications}

When a firm decides to use a switchback experiment to evaluate  a new product or initiative, they have to make multiple decisions to ensure that the results are reliable, practical, and replicable. First, the firm must determine an appropriate outcome(s) that adequately captures the relative effectiveness of the change. In practice, this requires substantive domain knowledge combined with an understanding of the likely impact of the change; see \cite{kohavi2020trustworthy} for an in-depth discussion of metric definition strategies.

Second, as part of the design of the experiment, the firm often has control over the granularity of one single experimental period.
As we have shown in Example~\ref{exa:TimeGranularity}, as long as each time period is smaller than the length of the carryover effect and the length of the carryover effect is divisible by the length of one time unit, the selection of granularity makes no difference to the optimal design and analysis of switchback experiments.
On the other hand, setting each period's length longer than the carryover effect will lead to a loss in precisions. Consider an extreme case where the carryover effect is 1 minute, while each period is selected to be an hour. Had we set each period to be a minute, we would have collected order of magnitude more useful data. Hence, we suggest that each period's length be smaller than the carryover effect duration.


Third, the firm must use prior knowledge to decide an appropriate value $p$ for the order of the carryover effect $m$. When a firm lacks such knowledge, we propose using the procedure outline in Section~\ref{sec:FigureOutm} to select an appropriate value of the order of the carryover effect. Practically, researchers should try to narrow down the set of possible values of $m$ as, when $m$ is relatively large compared to $T$, our procedure could fail to reject the null hypothesis simply due to insufficient statistical power. Also, it is important to keep in mind that each hypothesis test to identify \eqref{SubRoutineNull} needs to consume experimental resources at the scale of $T/m > 100$ to distinguish two candidate values, which could be over burdensome when the resource is scarce.

Fourth, when the firm has control over the experiment's horizon, the firm should set $p=m$ and control the overall duration of the experiment $n = T/p = T/m$.  We suggest choosing $n$ by referring to the rejection rate curve, as shown in Section~\ref{sec:simu:RejectionRates}; intuitively, this procedure resembles a typical power analysis. We begin with selecting our inference method, as described in Section~\ref{sec:InferenceAndTesting}. We then use our domain knowledge to estimate the expected signal-to-noise ratio; this could be done by looking at historical experiments or through dummy experiments. Then, we choose the desired rejection rate and find out the length of the horizon required.

Finally, using the previous four points, the firm decides the randomization points and samples the assignment path from the appropriate randomization distribution.  This final step has already been discussed at length, as we showed in Section~\ref{sec:MinimaxOptimization} the optimal design is obtained from Theorems~\ref{thm:FairCoinOptimal} and~\ref{thm:OptimalDesign}.
In cases when the time horizon is pre-determined and when $T/p$ is not an integer, our optimization formulation as shown in Theorem~\ref{thm:OptimalDesign} can always be used to find an optimal solution without discarding any periods. Just in the ``imperfect cases,'' we do not have closed-form solutions. Our suggestion is that if the experimental designer wishes not to discard any periods, then solve the optimal solution (using any commercial software); if the experimental designer wishes not to solve an optimization problem, then discard a few periods and consult the explicit solution suggested in Theorem~\ref{thm:OptimalDesign}.

After designing the experiment, the firm can use the data collected from the test to draw causal conclusions about the new innovation's performance using the two inferential methods as discussed in Section~\ref{sec:InferenceAndTesting}.
As a more practical consideration, when the firm have the capability to run multiple experiments on multiple experimental units, we suggest the firm to run the optimal design on each of the experimental units and then combine them to increase both precision and power. See \citet{bojinov2019time} for detailed discussions.

We point out three limitations of our paper.
First, when $m$, the order of the carryover effect is as large as comparable to $T$ the horizon's length, our method, though still unbiased in theory, incurs a large variance that typically prohibits the firm from making meaningful inference. This is because our method is general and requires the minimum amount of modeling assumptions. If we have strong domain knowledge about the outcome model, we can incorporate it to improve the design.
Second, our method only considers flipping independent coins before the experiment even begins. We do not consider adaptively changing the coin flip probabilities, as it requires further assumptions about the outcome model, \emph{e.g.}, some time-homogeneity of the data generating process.
Third, in this paper, we have only considered the estimand as in \eqref{eqn:estimand}, which is motivated when firms want to decide whether to permanently adopt a policy. If the primary focus is on some other general causal estimands, our results do not directly apply. It remains open to derive new results for other estimands, using a similar strategy that we have employed.

\section*{Acknowledgment}
The authors thank the department editor George Shanthikumar, the anonymous associate editor, and three anonymous referees whose comments improved the manuscript.
The authors also thank the MIT-IBM partnership in AI and the MIT Data Science Laboratory for support.

\bibliographystyle{informs2014} 
\bibliography{bibliography,reference} 

\ECSwitch


\ECHead{E-Companion}

Within this paper, let $\bN, \bN_0$ be the set of positive integers and non-negative integers, respectively.
For any $T \in \bN$, let $[T] = \{1,...,T\}$ be the set of positive integers no larger than $T$.
For any $t < t' \in \bN$, let $\{t:t'\} = \{t,t+1,...,t'\}$ be the set of integers between (including) $t$ and $t'$.
For any $m \in \bN$, let $\bm{1}_{m} = (1,1,...,1)$ be a vector of $m$ ones; let $\bm{0}_{m} = (0,0,...,0)$ be a vector of $m$ zeros.
We use parentheses for probabilities, i.e., $\Pr(\cdot)$; brackets for expectations, i.e., $\bE[\cdot]$; and curly brackets for indicators, i.e., $\bI\{\cdot\}$.
For any $a \in \bR$, let $(a)^+ = \max\{a,0\}$.

\section{Theorems Used}
We summarize here the results that we have directly used in our proofs.

\begin{definition}
[$\phi$-Dependent Random Variables, \citet{hoeffding1948central}]
\label{defn:MDependentRandomVariables}
For any sequence $\{X_1, X_2, ...\}$, if there exists $\phi$ such that for any $s-r > \phi$, the two sets $$(X_1, X_2, ..., X_r), \quad (X_s, X_{s+1}, ..., X_n)$$
are independent, then the sequence is said to be $\phi$-dependent.
\end{definition}

\begin{lemma}
[\citet{romano2000more}, Theorem 2.1]
\label{lem:RomanoWolfCLT}
Let $\left\{X_{n,i}\right\}$ be a triangular array of zero-mean random variables.
Let $\phi \in \bN$ be a fixed constant.
For each $n=1,2,...$, let $d=d_n$, and suppose that $X_{n,1}, X_{n,2}, ..., X_{n,d}$ is an $\phi$-dependent sequence of random variables. Define
\begin{align*}
B^2_{n,k,a} = \var\left( \sum_{i=a}^{a+k-1} X_{n,i} \right), & & B^2_n = B^2_{n,d,1} = \var\left( \sum_{i=1}^{d} X_{n,i} \right)
\end{align*}
For some $\delta>0$ and $-1 \leq \gamma \leq 1$, if the following conditions hold:
\begin{enumerate}
\item $\bE\left| X_{n,i} \right|^{2+\delta} \leq \Delta_n$, for all $i$;
\item $B^2_{n,k,a}/k^{1+\gamma} \leq K_n$, for all $a$ and $k \geq \phi$;
\item $B^2_n / (d \phi^\gamma) \geq L_n$;
\item $K_n / L_n = O(1)$;
\item $\Delta / L_n^{(2+\delta)/2} = O(1)$,
\end{enumerate}
then
\begin{align*}
\frac{\sum_{i=1}^d X_{n,i}}{B_n} \xrightarrow{D} \mathcal{N}(0,1).
\end{align*}
\end{lemma}

We explain Lemma~\ref{lem:RomanoWolfCLT}.
The $\xrightarrow{D}$ notation stands for convergence in distribution.
The definition of a sequence of $\phi$-dependent random variables is given in Definition~\ref{defn:MDependentRandomVariables}.
To check if the conditions in Lemma~\ref{lem:RomanoWolfCLT} hold, we will first calculate $B^2_{n,k,a}$ for any $k$ and $a$, and then construct some proper $\Delta_n, K_n,$ and $L_n$.

\begin{lemma}
\label{lem:ElegantLemma}
For any $n \in \bN$ and $q_1, ..., q_n \in (0,1)$, define $$f(q_1, ...,q_n) = \frac{1}{\prod_{i=1}^n q_i} + \frac{1}{\prod_{i=1}^n (1-q_i)}.$$ Then $$f(q_1, ...,q_n) \geq 2^{n+1},$$ where equality holds if and only if $q_1=q_2=...=q_n=1/2$.
\end{lemma}

The proof of Lemma~\ref{lem:ElegantLemma} is elegant and is of its own interests. We prove Lemma~\ref{lem:ElegantLemma} below.

\proof{Proof of Lemma~\ref{lem:ElegantLemma}.}
For all $i\in[n]$ denote $\bq_i = 1-q_i$. We re-write our objective, such that we wish to find the minimum for $$\frac{1}{\prod_{i=1}^n q_i} + \frac{1}{\prod_{i=1}^n \bq_i},$$ under the constraints that $q_i+\bq_i=1$ for all $i\in[n]$.
Note that $\prod_{i=1}^n (q_i+\bq_i) = 1$. By expanding expand the product term and we have
\begin{align*}
\frac{1}{\prod_{i=1}^n q_i} = \frac{\prod_{i=1}^n (q_i+\bq_i)}{\prod_{i=1}^n q_i} = 1 + \left( \frac{\bq_1}{q_1} + \frac{\bq_2}{q_2} + \ldots + \frac{\bq_n}{q_n} \right) + \left( \frac{\bq_1\bq_2}{q_1q_2} + \frac{\bq_1\bq_3}{q_1q_3} + \ldots + \frac{\bq_{n-1}\bq_n}{q_{n-1}q_n} \right) + \ldots + \frac{\prod_{i=1}^n \bq_i}{\prod_{i=1}^n q_i}
\end{align*}
And similarly we can expand the product term for the second fractional expression. Putting them together we have:
\begin{align*}
\frac{1}{\prod_{i=1}^n q_i} + \frac{1}{\prod_{i=1}^n \bq_i} = & 1 + \left( \frac{\bq_1}{q_1} + \frac{\bq_2}{q_2} + \ldots + \frac{\bq_n}{q_n} \right) + \left( \frac{\bq_1\bq_2}{q_1q_2} + \frac{\bq_1\bq_3}{q_1q_3} + \ldots + \frac{\bq_{n-1}\bq_n}{q_{n-1}q_n} \right) + \ldots + \frac{\prod_{i=1}^n \bq_i}{\prod_{i=1}^n q_i} \\
& \ + 1 + \left( \frac{q_1}{\bq_1} + \frac{q_2}{\bq_2} + \ldots + \frac{q_n}{\bq_n} \right) + \left( \frac{q_1q_2}{\bq_1\bq_2} + \frac{q_1q_3}{\bq_1\bq_3} + \ldots + \frac{q_{n-1}q_n}{\bq_{n-1}\bq_n} \right) + \ldots + \frac{\prod_{i=1}^n q_i}{\prod_{i=1}^n \bq_i}
\end{align*}
Now focus on the right hand side.
There are a total of $2^{n+1}$ terms, and we match them into $2^n$ pairs.
We match the first term in the first line with the first term in the second line, the second term in the first line with the second term in the second line, ..., the last term in the first line with the last term in the second line.
For each pair indexed by subset $I \subseteq [T]$, we have that $$\frac{\prod_{i \in I \subseteq [T]} \bq_i}{\prod_{i \in I \subseteq [T]} q_i} + \frac{\prod_{i \in I \subseteq [T]} q_i}{\prod_{i \in I \subseteq [T]} \bq_i} \geq 2,$$
where equality holds if and only if $\prod_{i \in I \subseteq [T]} q_i = \prod_{i \in I \subseteq [T]} \bq_i$. Putting all the $2^n$ pairs together we finish the proof.
\Halmos \endproof

\section{Proof from Section~\ref{sec:Definitions}}
\label{sec:proof:thm:HTUnbiased}

The only proof from Section~\ref{sec:Definitions} is the unbiasedness of the Horvitz-Thompson estimator.
We prove by checking the expectations.

\proof{Proof of Proposition~\ref{thm:HTUnbiased}.}
First observe 
that for regular switchback experiments, both $0 < \Pr(\bm{W}_{t-p:t} = \bm{1}_{p+1}), \Pr(\bm{W}_{t-p:t} = \bm{0}_{p+1}) < 1$.
So for any $t \in \{p+1:T\}$, with probability $\Pr(\bm{W}_{t-p:t} = \bm{1}_{p+1}) \ne 0$, $\bI{\{\bm{W}_{t-p:t} = \bm{1}_{p+1}\}} = 1$, and $Y_t^{\obs} = Y_t(\bm{1}_{p+1})$.
So $\bE\left[ Y_t^{\obs} \frac{\bI{\{\bm{W}_{t-p:t} = \bm{1}_{p+1}\}}}{\Pr(\bm{W}_{t-p:t} = \bm{1}_{p+1})} \right] = Y_t(\bm{1}_{m+1})$.
Similarly $\bE\left[ Y_t^{\obs} \frac{\bI{\{\bm{W}_{t-p:t} = \bm{0}_{p+1}\}}}{\Pr(\bm{W}_{t-p:t} = \bm{0}_{p+1})} \right] = Y_t(\bm{0}_{p+1})$.
Sum them up for any $t \in \{p+1:T\}$ we finish the proof.
\Halmos \endproof

\section{Proofs and Discussions from Section~\ref{sec:MinimaxOptimization}}
\label{sec:ProofsDiscussions:APD}
In Section~\ref{sec:MinimaxOptimization} we focus on the case when $p=m$.
Throughout this section in the appendix, we use only $m$ instead of $p$.

\subsection{Extra Notations Used in the Proofs from Section~\ref{sec:MinimaxOptimization}}
\label{sec:NotationsDiscussion}

Recall that any regular switchback experiment can be represented by $\bT = \{t_0, t_1, ..., t_K\} \subseteq [T]$ and $\bQ = (q_0, q_1, ..., q_K) \in (0,1)^{K+1}$.
We first focus on the dependence on $\bT$, the randomization points.
Define $f_\bT: [T] \to \bT$ to be the ``determining randomization point of period $t$'', i.e.,
$$f_\bT(t) = \max \left\{ j \left| j \in \bT, j \leq t \right. \right\}$$
such that the coin flip in period $f_\bT(t)$ uniquely determines the distribution of $W_t$, i.e., $W_t = W_{f_\bT(t)}$.
When $\bT$ is clear from the context we also omit the subscript and use $f(t)$ for $f_\bT(t)$.

Similarly, we define $f^m_\bT(t): [T] \to \{0,1\}^\bT$, which maps a time period to a subset of $\bT$, to be the ``determining randomization points of periods $\{t-m, t-m+1, ..., t\}$'', i.e.
$$f^m_\bT(t) = \left\{ j \left| \exists i \in \{t-m, ..., t\}, s.t. \ j = f_\bT(i) \right. \right\}$$
such that $f^m_\bT(t) \subseteq \bT \subseteq [T]$.
And $f^m_\bT(t)$ contains all the time periods whose coin flips uniquely determine the distributions of $W_{t-m}, W_{t-m+1}, ..., W_t$.
Denote $\left| f^m_\bT(t) \right| = J$, the cardinality of set $f^m_\bT(t)$.
We keep in mind that $J$ depends on $m, t$ and $\bT$, yet they are all omitted for brevity.
Since the treatment assignments $\bm{W}_{t-m:t}$ are determined by at least one randomization point $f(t-m)$, we know that $f^m_\bT(t) \ne \emptyset$ is non-empty, i.e.,
\begin{align}
\label{eqn:nonemptyJ}
\left| f^m_\bT(t) \right| = J \geq 1.
\end{align}
Let the elements be $f^m_\bT(t) = \{u_1, u_2, ..., u_J\}$, and let $u_1 < u_2 < ... < u_J$.

Finally, define ``overlapping randomization points of periods $\{t-m, t-m+1, ..., t\}$ and $\{t'-m, t'-m+1, ..., t'\}$'' to be
$$O_\bT(t, t') = f^m_\bT(t) \cap f^m_\bT(t')$$
Denote $\left| O_\bT(t, t') \right| = J^\so$.
We keep in mind that $J^\so$ depends on $m, t, t'$ and $\bT$, yet they are all omitted for brevity.

Now we introduce an important short-hand notation. Recall that for any randomization point $t_k$, the associated $q_k$ is the probability that $W_{t_k}$ receives treatment, i.e., $q_k= \Pr(W_{t_k} = 1)$. And recall that $\bq_k = 1 - q_k$. Now define for any $t \in \{m+1:T\}$,
\begin{multline}
\label{eqn:defn:bI}
\bI_t(\bT, \bQ, \bY) = Y_t(\bm{1}_{m+1}) \left[ \bI\{\bm{W}_{t-m:t} = \bm{1}_{m+1}\} \prod_{j=1}^J {\frac{1}{q_{u_j}}} - 1 \right] \\
- Y_t(\bm{0}_{m+1}) \left[ \bI\{\bm{W}_{t-m:t} = \bm{0}_{m+1}\} \prod_{j=1}^J {\frac{1}{\bq_{u_j}}} - 1 \right]
\end{multline}
where we use $\prod_{j=1}^J (1 / q_{u_j})$ and $\prod_{j=1}^J (1 / \bq_{u_j})$ to calculate the inverse propensity scores.
When $\bT, \bQ$ and $\bY$ are clear from the context we omit them and use $\bI_t$ for $\bI_t(\bT, \bQ, \bY)$.

Using the above notation, we could re-write
\begin{align*}
\widehat{\tau}_m - \tau_m = \frac{1}{T-m} \sum_{t=m+1}^T \bI_t
\end{align*}
Similar to Proposition~\ref{thm:HTUnbiased}, we can check the expectation of $\bI_t$ by expanding the probability governing $\bm{W}$ (the only source of randomness is our assignment path $\bm{W}$).
For any $t \in \{m+1, m+2, ..., T\}$,
\begin{align}
\label{eqn:UnbiasedEachPeriod}
\bE[\bI_t] = 0.
\end{align}

\subsection{Preliminary Results}
In this section we introduce two Lemmas for the proof of Theorem~\ref{thm:FairCoinOptimal} and proof of Lemma~\ref{lem:AdversaryStrategy}.

\begin{lemma}
\label{lem:square-terms}
Under Assumptions~\ref{asp:nonanticipating}--\ref{asp:nomcarryover}, for any $t \in [T]$, let $\left| f^m_\bT(t) \right| = J$.
\begin{align}
\bE[\bI_t^2] = & \left(\prod_{j=1}^J \frac{1}{q_{u_j}} - 1\right) Y_t(\bm{1}_{m+1})^2 + 2 Y_t(\bm{1}_{m+1}) Y_t (\bm{0}_{m+1}) + \left(\prod_{j=1}^J \frac{1}{\bq_{u_j}} - 1\right) Y_t(\bm{0}_{m+1})^2. \label{eqn:square-terms}
\end{align}
\end{lemma}

\proof{Proof of Lemma~\ref{lem:square-terms}.}

Denote $\left| f^m_\bT(t) \right| = J$.
Let the elements be $f^m_\bT(t) = \{u_1, u_2, ..., u_J\}$.
Let $u_1 < u_2 < ... < u_J$.

Using the notations defined earlier in Section~\ref{sec:NotationsDiscussion} and, in particular, the definition of \eqref{eqn:defn:bI}, we can directly calculate the squared terms of $\bE[\bI_t^2]$ by consulting the law of total expectation.
\begin{align*}
\bE[\bI_t^2] = & \Pr\left( \bm{W}_{t-m:t} = \bm{1}_{m+1} \right) \cdot \bE[\bI_t^2 \left| \bm{W}_{t-m:t} = \bm{1}_{m+1} \right.] \\
& + \Pr\left( \bm{W}_{t-m:t} = \bm{1}_{m+1} \right) \cdot \bE[\bI_t^2 \left| \bm{W}_{t-m:t} = \bm{1}_{m+1} \right.] \\
& + \Pr\left( \bm{W}_{t-m:t} = \bm{1}_{m+1} \right) \cdot \bE[\bI_t^2 \left| \bm{W}_{t-m:t} = \bm{1}_{m+1} \right.] \\
= & \Pr\left( \bm{W}_{t-m:t} = \bm{1}_{m+1} \right) \cdot \left\{ Y_t(\bm{1}_{m+1}) \left(\prod_{j=1}^J\frac{1}{q_{u_j}} - 1\right) - Y_t(\bm{0}_{m+1}) (0 - 1)\right\}^2 \\
& + \Pr\left( \bm{W}_{t-m:t} = \bm{0}_{m+1} \right) \cdot \left\{ Y_t(\bm{1}_{m+1}) (0 - 1) - Y_t(\bm{0}_{m+1}) \left(\prod_{j=1}^J\frac{1}{\bq_{u_j}} - 1\right) \right\}^2 \\
& + \Pr\left( \bm{W}_{t-m:t} \ne \bm{1}_{m+1} \text{ \ or \ } \bm{0}_{m+1} \right) \cdot \left\{ Y_t(\bm{1}_{m+1}) (0 - 1) - Y_t(\bm{0}_{m+1}) (0 - 1) \right\}^2 \\
= & \Pr\left( (W_{u_1},...,W_{u_J}) = \bm{1}_J \right) \cdot \left\{ \left(\prod_{j=1}^J\frac{1}{q_{u_j}} - 1\right) Y_t(\bm{1}_{m+1}) + Y_t(\bm{0}_{m+1}) \right\}^2 \\
& + \Pr\left( (W_{u_1},...,W_{u_J}) = \bm{0}_J \right) \cdot \left\{ - Y_t(\bm{1}_{m+1}) - \left(\prod_{j=1}^J\frac{1}{\bq_{u_j}} - 1\right) Y_t(\bm{0}_{m+1}) \right\}^2 \\
& + \Pr\left( (W_{u_1},...,W_{u_J}) \ne \bm{1}_J \text{ \ or \ } \bm{0}_J \right) \cdot \left\{ - Y_t(\bm{1}_{m+1}) + Y_t(\bm{0}_{m+1}) \right\}^2 \\
= & \prod_{j=1}^J q_{u_j} \cdot \left\{ \prod_{j=1}^J\frac{1}{q_{u_j}} \cdot Y_t(\bm{1}_{m+1}) - Y_t(\bm{1}_{m+1}) + Y_t(\bm{0}_{m+1}) \right\}^2 \\
& + \prod_{j=1}^J \bq_{u_j} \cdot \left\{ - \prod_{j=1}^J\frac{1}{\bq_{u_j}} \cdot Y_t(\bm{0}_{m+1}) - Y_t(\bm{1}_{m+1}) + Y_t(\bm{0}_{m+1}) \right\}^2 \\
& + \left( 1 - \prod_{j=1}^J q_{u_j} - \prod_{j=1}^J \bq_{u_j} \right) \cdot \left\{ - Y_t(\bm{1}_{m+1}) + Y_t(\bm{0}_{m+1}) \right\}^2 \\
= & \left(\prod_{j=1}^J \frac{1}{q_{u_j}} - 1\right) Y_t(\bm{1}_{m+1})^2 + 2 Y_t(\bm{1}_{m+1}) Y_t (\bm{0}_{m+1}) + \left(\prod_{j=1}^J \frac{1}{\bq_{u_j}} - 1\right) Y_t(\bm{0}_{m+1})^2
\end{align*}
which finishes the proof.
\Halmos \endproof

\begin{lemma}
\label{lem:cross-products}
Under Assumptions~\ref{asp:nonanticipating}--\ref{asp:nomcarryover}, for any $t < t' \in [T]$, when $\left| O_\bT(t, t') \right| = J^\so = 0$,
\begin{align}
\bE[\bI_t \bI_{t'}] = & 0. \label{eqn:crossIndependent}
\end{align}
When $\left| O_\bT(t, t') \right| = J^\so \geq 1$,
\begin{align}
\bE[\bI_t \bI_{t'}] = & (\prod_{j=1}^{J^\so} \frac{1}{q_{u^\so_j}} - 1) Y_t(\bm{1}_{m+1}) Y_{t'}(\bm{1}_{m+1}) + Y_t(\bm{1}_{m+1}) Y_{t'}(\bm{0}_{m+1}) \nonumber \\
& + Y_t(\bm{0}_{m+1}) Y_{t'}(\bm{1}_{m+1}) + (\prod_{j=1}^{J^\so} \frac{1}{\bq_{u^\so_j}} - 1) Y_t(\bm{0}_{m+1}) Y_{t'}(\bm{0}_{m+1}). \label{eqn:cross-products}
\end{align}
\end{lemma}

\proof{Proof of Lemma~\ref{lem:cross-products}.}
Denote $\left| f^m_\bT(t) \right| = J$, $\left| f^m_\bT(t') \right| = J'$, and $\left| O_\bT(t, t') \right| = J^\so$.
Let the elements be $f^m_\bT(t) = \{u_1, u_2, ..., u_J\}$, $f^m_\bT(t') = \{u'_1, u'_2, ..., u'_{J'}\}$, and $O_\bT(t, t') = \{u^\so_1, u^\so_2, ..., u^\so_{J^\so}\}$.
Let $u_1 < u_2 < ... < u_J$, $u'_1 < u'_2 < ... < u'_{J'}$, and $u^\so_1 < u^\so_2 < ... < u^\so_{J^\so}$.

One time period could have different numberings in $f_{\bT}^m(t)$, $f_{\bT}^m(t')$, and $O_{\bT}(t,t')$.
For example, $u_{J-J^\so+1} = u'_1 = u^\so_1$, and $u_{J} = u'_{J^\so} = u^\so_{J^\so}$.
See Table~\ref{tbl:IllustratorJs} for an illustrator of the determining randomization points and the overlapping randomization points.

\begin{table}[!htb]
\TABLE{Illustrator of the determining randomization points and the overlapping randomization points. \label{tbl:IllustratorJs}}
{
\begin{tabular}{| p{1.3cm} | p{1.3cm} | p{1.3cm} | p{1.3cm} | p{1.3cm} | p{1.3cm} | p{1.3cm} | p{1.3cm} | p{1.3cm} |}
\hline
$u_1$ & $u_2$ & ... & $u_{J-J^\so+1}$ & ... & $u_J$                 &                         &     &  \\
\hline
          &           &     & $u^\so_1$          & ... & $u^\so_{J^\so}$ &                         &     &  \\
\hline
          &           &     & $u'_1$                & ... & $u'_{J^\so}$       & $u'_{J^\so+1}$ & ... & $u'_{J'}$ \\
\hline
\end{tabular}
}
{Each columns stands for one time period. The first row stands for the determining randomization points of $f_{\bT}^m(t)$; the second row for the overlapping randomization points of $O_{\bT}(t,t')$; and the third row for the determining randomization points of $f_{\bT}^m(t')$.}
\end{table}

First, when $J^\so = 0$, this implies that $\bI_t$ and $\bI_{t'}$ are independent.
Then $\bE[\bI_t \bI_{t'}] = \bE[\bI_t] \bE[\bI_{t'}] = 0$, where the second equality is due to \eqref{eqn:UnbiasedEachPeriod}.

When $J^\so \geq 1$, this implies that $\bI_t$ and $\bI_{t'}$ are correlated.
Using the notations defined above,
\begin{align}
\bE[\bI_{t} \bI_{t'}] = & \bE_{W_{u_1^\so}, ..., W_{u_{J^\so}^\so}} \left[ \bE\left[ \bI_{t} \bI_{t'} \left| W_{u_1^\so}, ..., W_{u_{J^\so}^\so} \right. \right] \right] \label{eqn:cross-productCalculation} \\
= & \Pr\left( (W_{u_1^\so}, ..., W_{u_{J^\so}^\so}) = \bm{1}_{J^\so} \right) \bE\left[ \bI_{t} \bI_{t'} \left| (W_{u_1^\so}, ..., W_{u_{J^\so}^\so}) = \bm{1}_{J^\so} \right. \right] \nonumber \\
& + \Pr\left( (W_{u_1^\so}, ..., W_{u_{J^\so}^\so}) = \bm{0}_{J^\so} \right) \bE\left[ \bI_{t} \bI_{t'} \left| (W_{u_1^\so}, ..., W_{u_{J^\so}^\so}) = \bm{0}_{J^\so} \right. \right] \nonumber \\
& + \Pr\left( (W_{u_1^\so}, ..., W_{u_{J^\so}^\so}) \ne \bm{1}_{J^\so} \text{\ or \ } \bm{0}_{J^\so} \right) \bE\left[ \bI_{t} \bI_{t'} \left| (W_{u_1^\so}, ..., W_{u_{J^\so}^\so}) \ne \bm{1}_{J^\so} \text{\ or \ } \bm{0}_{J^\so} \right. \right] \nonumber
\end{align}

Next we go over the three cases of $(W_{u_1^\so}, ..., W_{u_{J^\so}^\so})$ as decomposed above.
Note that conditional on $(W_{u_1^\so}, ..., W_{u_{J^\so}^\so})$, $\bI_{t}$ and $\bI_{t'}$ are independent, i.e.,
\begin{align*}
\bE\left[\bI_t \bI_{t'} \left| W_{u_1^\so}, ..., W_{u_{J^\so}^\so} \right. \right] = \bE\left[\bI_t \left| W_{u_1^\so}, ..., W_{u_{J^\so}^\so} \right. \right] \bE\left[\bI_{t'} \left| W_{u_1^\so}, ..., W_{u_{J^\so}^\so} \right. \right]
\end{align*}

\noindent \textbf{(1)} With probability $\prod_{j=1}^{J^\so} q_{u^\so_j}$, $(W_{u_1^\so}, ..., W_{u_{J^\so}^\so}) = \bm{1}_{J^\so}$. In this case
\begin{align*}
\bE\left[\bI_t \left| W_{u_1^\so}, ..., W_{u_{J^\so}^\so} \right. \right] = & \Pr\left( \bm{W}_{t-m:t} = \bm{1}_{m+1} \right) \cdot \left\{ Y_t(\bm{1}_{m+1}) (\prod_{j=1}^J \frac{1}{q_{u_j}} - 1) + Y_t(\bm{0}_{m+1})\right\} \\
& + \Pr\left( \bm{W}_{t-m:t} \ne \bm{1}_{m+1} \right) \cdot \left\{ Y_t(\bm{1}_{m+1}) (0 - 1) + Y_t(\bm{0}_{m+1})\right\} \\
= & \Pr\left( (W_{u_1}, W_{u_2}, ..., W_{u_{J-J^\so}}) = \bm{1}_{J-J^\so} \right) \cdot \left\{ Y_t(\bm{1}_{m+1}) (\prod_{j=1}^J \frac{1}{q_{u_j}} - 1) + Y_t(\bm{0}_{m+1})\right\} \\
& + \Pr\left( (W_{u_1}, W_{u_2}, ..., W_{u_{J-J^\so}}) \ne \bm{1}_{J-J^\so} \right) \cdot \left\{ - Y_t(\bm{1}_{m+1}) + Y_t(\bm{0}_{m+1})\right\} \\
= & \prod_{j=1}^{J-J^\so} q_{u_j} \cdot \left\{ Y_t(\bm{1}_{m+1}) (\prod_{j=1}^J \frac{1}{q_{u_j}} - 1) + Y_t(\bm{0}_{m+1})\right\} \\
& + (1 - \prod_{j=1}^{J-J^\so} q_{u_j}) \cdot \left\{ - Y_t(\bm{1}_{m+1}) + Y_t(\bm{0}_{m+1})\right\} \\
= & (\prod_{j=1}^{J^\so} \frac{1}{q_{u^\so_j}} - 1) Y_t(\bm{1}_{m+1}) + Y_t(\bm{0}_{m+1})
\end{align*}
where the third equality is due to \eqref{eqn:CoinFlip}.
Similarly, 
\begin{align*}
\bE\left[\bI_{t'} \left| W_{u_1^\so}, ..., W_{u_{J^\so}^\so} \right. \right] = & \Pr\left( \bm{W}_{t'-m:t'} = \bm{1}_{m+1} \right) \cdot \left\{ Y_{t'}(\bm{1}_{m+1}) (\prod_{j=1}^{J'} \frac {1}{q_{u_j'}} - 1) + Y_{t'}(\bm{0}_{m+1})\right\} \\
& + \Pr\left( \bm{W}_{t'-m:t'} \ne \bm{1}_{m+1} \right) \cdot \left\{ Y_{t'}(\bm{1}_{m+1}) (0 - 1) + Y_{t'}(\bm{0}_{m+1})\right\} \\
= & \Pr\left( (W_{u'_{J^\so + 1}}, W_{u'_{J^\so + 2}}, ..., W_{u'_{J'}}) = \bm{1}_{J'-J^\so} \right) \cdot \left\{ Y_{t'}(\bm{1}_{m+1}) (\prod_{j=1}^{J'} \frac {1}{q_{u'_j}} - 1) + Y_{t'}(\bm{0}_{m+1})\right\} \\
& + \Pr\left( (W_{u'_{J^\so + 1}}, W_{u'_{J^\so + 2}}, ..., W_{u'_{J'}}) \ne \bm{1}_{J'-J^\so} \right) \cdot \left\{ - Y_{t'}(\bm{1}_{m+1}) + Y_{t'}(\bm{0}_{m+1})\right\} \\
= & \prod_{j=J^\so+1}^{J'} q_{u'_j} \cdot \left\{ Y_{t'}(\bm{1}_{m+1}) (\prod_{j=1}^{J'} \frac {1}{q_{u'_j}} - 1) + Y_{t'}(\bm{0}_{m+1})\right\} \\
& + (1 - \prod_{j = J^\so + 1}^{J'} q_{u'_j}) \cdot \left\{ - Y_{t'}(\bm{1}_{m+1}) + Y_{t'}(\bm{0}_{m+1})\right\} \\
= & (\prod_{j=1}^{J^\so} \frac{1}{q_{u^\so_j}} - 1) Y_{t'}(\bm{1}_{m+1}) + Y_{t'}(\bm{0}_{m+1})
\end{align*}

\noindent \textbf{(2)} With probability $\prod_{j=1}^{J^\so} \bq_{u^\so_j}$, $(W_{u_1^\so}, ..., W_{u_{J^\so}^\so}) = \bm{0}_{J^\so}$. This case is similar to Case (1), and we can calculate the expectation similarly.
\begin{align*}
\bE\left[\bI_t \left| W_{u_1^\so}, ..., W_{u_{J^\so}^\so} \right. \right] = & \Pr\left( \bm{W}_{t-m:t} = \bm{0}_{m+1} \right) \cdot \left\{ -Y_t(\bm{1}_{m+1}) - Y_t(\bm{0}_{m+1}) (\prod_{j=1}^J \frac{1}{\bq_{u_j}} - 1) \right\} \\
& + \Pr\left( \bm{W}_{t-m:t} \ne \bm{0}_{m+1} \right) \cdot \left\{ - Y_t(\bm{1}_{m+1}) - Y_t(\bm{0}_{m+1}) (0 - 1) \right\} \\
= & \prod_{j=1}^{J-J^\so} \bq_{u_j} \cdot \left\{ - Y_t(\bm{1}_{m+1}) - Y_t(\bm{0}_{m+1}) (\prod_{j=1}^J \frac{1}{\bq_{u_j}} - 1) \right\} \\
& + (1 - \prod_{j=1}^{J-J^\so} \bq_{u_j}) \cdot \left\{ - Y_t(\bm{1}_{m+1}) + Y_t(\bm{0}_{m+1})\right\} \\
= & - Y_t(\bm{1}_{m+1}) - (\prod_{j=1}^{J^\so} \frac{1}{\bq_{u^\so_j}} - 1) Y_t(\bm{0}_{m+1})
\end{align*}
and again, similarly,
\begin{align*}
\bE\left[\bI_{t'} \left| W_{u_1^\so}, ..., W_{u_{J^\so}^\so} \right. \right] = & \Pr\left( \bm{W}_{t'-m:t'} = \bm{0}_{m+1} \right) \cdot \left\{ -Y_{t'}(\bm{1}_{m+1}) - Y_{t'}(\bm{0}_{m+1}) (\prod_{j=1}^{J'}\frac{1}{\bq_{u'_j}} - 1) \right\} \\
& + \Pr\left( \bm{W}_{t'-m:t'} \ne \bm{0}_{m+1} \right) \cdot \left\{ - Y_{t'}(\bm{1}_{m+1}) - Y_{t'}(\bm{0}_{m+1}) (0 - 1) \right\} \\
= & \prod_{j=J^\so+1}^{J'} \bq_{u'_j} \cdot \left\{ - Y_{t'}(\bm{1}_{m+1}) - Y_{t'}(\bm{0}_{m+1}) (\prod_{j=1}^{J'}\frac{1}{\bq_{u'_j}} - 1) \right\} \\
& + (1 - \prod_{j=J^\so+1}^{J'} \bq_{u'_j}) \cdot \left\{ - Y_{t'}(\bm{1}_{m+1}) + Y_{t'}(\bm{0}_{m+1})\right\} \\
= & - Y_{t'}(\bm{1}_{m+1}) - (\prod_{j=1}^{J^\so}\frac{1}{\bq_{u^\so_j}} - 1) Y_{t'}(\bm{0}_{m+1})
\end{align*}

\noindent \textbf{(3)} With probability $1 - 2 \cdot (1 / 2^{J^\so})$, $(W_{u_1^\so}, ..., W_{u_{J^\so}^\so}) \ne \bm{1}_{J^\so}$ or $\bm{0}_{J^\so}$. In this case
\begin{align*}
\bE\left[\bI_t \left| W_{u_1^\so}, ..., W_{u_{J^\so}^\so} \right. \right] = & - Y_{t}(\bm{1}_{m+1}) + Y_{t}(\bm{0}_{m+1}) \\
\bE\left[\bI_{t'} \left| W_{u_1^\so}, ..., W_{u_{J^\so}^\so} \right. \right] = & - Y_{t'}(\bm{1}_{m+1}) + Y_{t'}(\bm{0}_{m+1})
\end{align*}

Finally, putting all above together into \eqref{eqn:cross-productCalculation}, we have
\begin{align*}
\bE[\bI_{t} \bI_{t'}] = & \prod_{j=1}^{J^\so} q_{u^\so_j} \cdot \left\{ (\prod_{j=1}^{J^\so} \frac{1}{q_{u^\so_j}} - 1) Y_t(\bm{1}_{m+1}) + Y_t(\bm{0}_{m+1}) \right\} \cdot \left\{ (\prod_{j=1}^{J^\so} \frac{1}{q_{u^\so_j}} - 1) Y_{t'}(\bm{1}_{m+1}) + Y_{t'}(\bm{0}_{m+1}) \right\} \\
& + \prod_{j=1}^{J^\so} \bq_{u^\so_j} \cdot \left\{ - Y_t(\bm{1}_{m+1}) - (\prod_{j=1}^{J^\so} \frac{1}{\bq_{u^\so_j}} - 1) Y_t(\bm{0}_{m+1}) \right\} \cdot \left\{ - Y_{t'}(\bm{1}_{m+1}) - (\prod_{j=1}^{J^\so}\frac{1}{\bq_{u^\so_j}} - 1) Y_{t'}(\bm{0}_{m+1}) \right\} \\
& + \left\{ 1- \prod_{j=1}^{J^\so} q_{u^\so_j} - \prod_{j=1}^{J^\so} \bq_{u^\so_j} \right\} \cdot \left\{ - Y_{t}(\bm{1}_{m+1}) + Y_{t}(\bm{0}_{m+1}) \right\} \cdot \left\{ - Y_{t'}(\bm{1}_{m+1}) + Y_{t'}(\bm{0}_{m+1}) \right\} \\
= & (\prod_{j=1}^{J^\so} \frac{1}{q_{u^\so_j}} - 1) Y_t(\bm{1}_{m+1}) Y_{t'}(\bm{1}_{m+1}) + Y_t(\bm{1}_{m+1}) Y_{t'}(\bm{0}_{m+1}) \\
& + Y_t(\bm{0}_{m+1}) Y_{t'}(\bm{1}_{m+1}) + (\prod_{j=1}^{J^\so} \frac{1}{\bq_{u^\so_j}} - 1) Y_t(\bm{0}_{m+1}) Y_{t'}(\bm{0}_{m+1})
\end{align*}
which finishes the proof.
\Halmos \endproof

\subsection{Lemma~\ref{lem:AdversaryStrategy}: Adversarial Selection of Potential Outcomes}
\label{sec:AdversarialPolicyDiscussion}
In this section, we first prove Lemma~\ref{lem:AdversaryStrategy}, and then discuss the implications of Lemma~\ref{lem:AdversaryStrategy}.

\subsubsection{Proof of Lemma~\ref{lem:AdversaryStrategy}.}
\label{sec:Proof:lem:AdversaryStrategy}
The proof of Lemma~\ref{lem:AdversaryStrategy} is through careful expansion of the risk function, the expected square loss.

\proof{Proof of Lemma~\ref{lem:AdversaryStrategy}.}
From Lemma~\ref{lem:square-terms} and Lemma~\ref{lem:cross-products}, all the terms are quadratic, and all the coefficients are non-negative.
After multiplying the constant $(T-m)^2$, we can expand, for any design of experiment $(\bT, \bQ)$ and any potential outcomes $\bY \in \cY$, the following terms:
\begin{align*}
& (T-m)^2 \cdot \bE\left[ \left( \widehat{\tau}_m - \tau_m \right)^2 \right] \\
= & \sum_{t=m+1}^T \left\{ \left(\prod_{j=1}^J \frac{1}{q_{u_j}} - 1\right) Y_t(\bm{1}_{m+1})^2 + 2 Y_t(\bm{1}_{m+1}) Y_t (\bm{0}_{m+1}) + \left(\prod_{j=1}^J \frac{1}{\bq_{u_j}} - 1\right) Y_t(\bm{0}_{m+1})^2 \right\} \\
& + \sum_{\substack{m+1 \leq t<t' \leq T \\ \left| O_\bT(t,t') \right| \geq 1}} \left\{\left(\prod_{j=1}^{J^\so} \frac{1}{q_{u^\so_j}} - 1\right) Y_t(\bm{1}_{m+1}) Y_{t'}(\bm{1}_{m+1}) + Y_t(\bm{1}_{m+1}) Y_{t'}(\bm{0}_{m+1}) \right. \\
& \left.\hphantom{+ \sum_{\substack{m+1 \leq t<t' \leq T \\ \left| O_\bT(t,t') \right| \geq 1}}} + Y_t(\bm{0}_{m+1}) Y_{t'}(\bm{1}_{m+1}) + \left(\prod_{j=1}^{J^\so} \frac{1}{\bq_{u^\so_j}} - 1\right) Y_t(\bm{0}_{m+1}) Y_{t'}(\bm{0}_{m+1}) \right\}
\end{align*}
where the equality is due to Lemma~\ref{lem:square-terms} and Lemma~\ref{lem:cross-products}.
Notice that in the first summation, all the coefficients in the front of $Y_t(\bm{1}_{m+1})^2$, $Y_t(\bm{1}_{m+1}) Y_t (\bm{0}_{m+1})$, and $Y_t(\bm{0}_{m+1})^2$ are strictly positive, because $q_{u_j}$ are strictly between $(0,1)$.
In the second summation, for those periods such that $\left| O_\bT(t,t') \right| \geq 1$, all the coefficients in the front of $Y_t(\bm{1}_{m+1}) Y_{t'}(\bm{1}_{m+1})$, $Y_t(\bm{1}_{m+1}) Y_{t'}(\bm{0}_{m+1})$, $Y_t(\bm{0}_{m+1}) Y_{t'}(\bm{1}_{m+1})$, and $Y_t(\bm{0}_{m+1}) Y_{t'}(\bm{0}_{m+1})$ are strictly positive as well, because $q_{u_j}$ are strictly between $(0,1)$.

For the squared terms in the above expression, $Y_t(\bm{1}_{m+1})^2 \leq B^2, Y_t(\bm{0}_{m+1})^2 \leq B^2$ for any $t \in \{m+1:T\}$.
This is because $f(y)=y^2$ attains maximum at the end points of the interval $[-B, B]$.
For the cross-product terms in the above expression, no matter if $(y_1, y_2)$ takes $(Y_t(\bm{1}_{m+1}),Y_t(\bm{0}_{m+1}))$, $(Y_t(\bm{1}_{m+1}),Y_{t'}(\bm{1}_{m+1}))$, $(Y_t(\bm{1}_{m+1}),Y_{t'}(\bm{0}_{m+1}))$, $(Y_t(\bm{0}_{m+1}),Y_{t'}(\bm{1}_{m+1}))$, or $(Y_t(\bm{0}_{m+1}),Y_{t'}(\bm{0}_{m+1}))\}$, we have that $y_1 \cdot y_2 \leq (y_1^2 + y_2^2)/2 \leq B^2$
where the first inequality is due to Cauchy-Schwarz, and the second inequality is due to convexity.
Combining that fact that all coefficients are positive, $r(\eta_{\bT, \bQ}, \bY) \leq r(\eta_{\bT, \bQ}, \bY^{+}) = r(\eta_{\bT, \bQ}, \bY^{-})$.

Moreover, for any $\bY \in \cY$ such that $\bY \ne \bY^{+}$ or $\bY^{-}$, if $\exists \ t \in \{m+1, ..., T\}$ such that $-B < Y_t(\bm{1}_{m+1}) < B$.
Then from inequality \eqref{eqn:nonemptyJ}, $\prod_{j=1}^J \frac{1}{q_{u_j}} - 1 > 0$, so the inequality is strict.
Similarly, if $\exists t \in \{m+1, ..., T\}$ such that $-B < Y_t(\bm{0}_{m+1}) < B$, then combine $\prod_{j=1}^J \frac{1}{\bq_{u_j}} - 1 > 0$, so the inequality is strict.
\Halmos \endproof

\subsubsection{Implications of Lemma~\ref{lem:AdversaryStrategy}.}
\label{sec:lem:HandyExpressionGeneralCoin}

Lemma~\ref{lem:AdversaryStrategy} simplifies the minimax problem in \eqref{eqn:TheProblem}.
Instead of thinking it as a minimax problem, we can now replace $\bY$ by either $\bY^+$ or $\bY^-$, and solve only a minimization problem.

Here we state Lemma~\ref{lem:HandyExpressionGeneralCoin} that is a direct implication of Lemma~\ref{lem:AdversaryStrategy}.
It will be frequently used later on.
\begin{lemma}
\label{lem:HandyExpressionGeneralCoin}
When $\bY = \bY^{+}$ or $\bY = \bY^{-}$, under Assumptions~\ref{asp:nonanticipating}--\ref{asp:BoundedPO}, for any $t \in [T]$,
\begin{align*}
\bE[\bI_t^2] = & \left( \frac{1}{\prod_{j=1}^J q_{u_j}} + \frac{1}{\prod_{j=1}^J \bq_{u_j}} \right) B^2.
\end{align*}
For any $t < t' \in [T]$, when $\left| O_\bT(t, t') \right| = J^\so = 0$,
\begin{align*}
\bE[\bI_t \bI_{t'}] = & 0
\end{align*}
When $\left| O_\bT(t, t') \right| = J^\so \geq 1$,
\begin{align*}
\bE[\bI_t \bI_{t'}] = & \left( \frac{1}{\prod_{j=1}^{J^\so} q_{u^\so_j}} + \frac{1}{\prod_{j=1}^{J^\so} \bq_{u^\so_j}} \right) B^2
\end{align*}
\end{lemma}

\proof{Proof of Lemma~\ref{lem:HandyExpressionGeneralCoin}.}
Replace $Y_t(\bm{1}_{m+1}) = Y_t(\bm{0}_{m+1})$ by $B$ or $-B$ into the expressions in Lemmas~\ref{lem:square-terms} and~\ref{lem:cross-products}.
\Halmos \endproof

\subsection{Theorem~\ref{thm:FairCoinOptimal}: Optimality of Fair Coin Flipping}
In this section, we first prove Theorem~\ref{thm:FairCoinOptimal}, and then discuss the implications of Theorem~\ref{thm:FairCoinOptimal}.

\subsubsection{Proof of Theorem~\ref{thm:FairCoinOptimal}.}
\label{sec:proof:thm:FairCoinOptimal}
The proof of Theorem~\ref{thm:FairCoinOptimal} is through an elegant inequality that highlights the balance between treatment probabilities and control probabilities.

\proof{Proof of Theorem~\ref{thm:FairCoinOptimal}.}
Similar to the proof of Lemma~\ref{lem:AdversaryStrategy}, we expand the quadratic terms using Lemma~\ref{lem:HandyExpressionGeneralCoin}.
After multiplying the constant $(T-m)^2$, we can expand, for any design of experiment $(\bT, \bQ)$ and any potential outcomes $\bY \in \cY$, the following terms:
\begin{align*}
(T-m)^2 \cdot \bE\left[ \left( \widehat{\tau}_m - \tau_m \right)^2 \right] = & \sum_{t=m+1}^T \left( \prod_{j=1}^J \frac{1}{q_{u_j}} + \prod_{j=1}^J \frac{1}{\bq_{u_j}} \right) \cdot B^2 + \sum_{\substack{m+1 \leq t<t' \leq T \\ \left| O_\bT(t,t') \right| \geq 1}} \left(\prod_{j=1}^{J^\so} \frac{1}{q_{u^\so_j}} + \prod_{j=1}^{J^\so} \frac{1}{\bq_{u^\so_j}} \right) \cdot B^2
\end{align*}
For each of them, due to Lemma~\ref{lem:ElegantLemma}, the minimum is obtained at $q_0=q_1=...=q_K=1/2$.
\Halmos \endproof

\subsubsection{Implications of Theorem~\ref{thm:FairCoinOptimal}.}
\label{sec:lem:HandyExpression}

Theorem~\ref{thm:FairCoinOptimal} further simplifies the minimax problem in \eqref{eqn:TheProblem}.
Now that we have identified the optimal randomization probabilities, we can directly plug in the optimal probabilities being $1/2$.
Here we state Lemma~\ref{lem:HandyExpression} that is a combination of Lemma~\ref{lem:HandyExpressionGeneralCoin} and Theorem~\ref{thm:FairCoinOptimal}.
It will be frequently used later on.
\begin{lemma}
\label{lem:HandyExpression}
Under Assumptions~\ref{asp:nonanticipating}--\ref{asp:BoundedPO}, when $\bY = \bY^{+}$ or $\bY = \bY^{-}$, and when $q_0=q_1=...=q_K=1/2$, for any $t \in [T]$,
\begin{align*}
\bE[\bI_t^2] = & 2^{J+1} B^2.
\end{align*}
For any $t < t' \in [T]$, when $\left| O_\bT(t, t') \right| = J^\so = 0$,
\begin{align*}
\bE[\bI_t \bI_{t'}] = & 0
\end{align*}
When $\left| O_\bT(t, t') \right| = J^\so \geq 1$,
\begin{align*}
\bE[\bI_t \bI_{t'}] = & 2^{J^\so+1} B^2
\end{align*}
\end{lemma}

\proof{Proof of Lemma~\ref{lem:HandyExpression}.}
Simply replace $q_0=q_1=...=q_K=1/2$ into Lemma~\ref{lem:HandyExpressionGeneralCoin}.
\Halmos \endproof

\subsection{Structural Results of the Optimal Design}

Using Lemma~\ref{lem:AdversaryStrategy}, we now establish two structural results that further characterize the class of optimal designs of regular switchback experiments.
Lemma~\ref{lem:StartEndStructure} states the optimal starting and ending structure;
Lemma~\ref{lem:MiddleStructure} states the optimal middle-case structure.
The proofs to Lemma~\ref{lem:StartEndStructure} and Lemma~\ref{lem:MiddleStructure} are deferred to Sections~\ref{sec:proof:lem:StartEndStructure} and~\ref{sec:proof:lem:MiddleStructure}, respectively.

\begin{lemma}
\label{lem:StartEndStructure}
When $\bY = \bY^{+}$ or $\bY = \bY^{-}$, under Assumptions~\ref{asp:nonanticipating}--\ref{asp:BoundedPO}, any optimal design of experiment $\bT$ must satisfy
\begin{align*}
t_1 \geq m+2, & & \text{and} & & t_K \leq T-m.
\end{align*}
\end{lemma}

Lemma~\ref{lem:StartEndStructure} states that the first randomization point on period $1$ should be followed by at least $m$ periods that do not flip a coin, and that the last randomization point should be followed by at least $m$ periods that do not flip a coin.
This guarantees that the assignments during $\{1:m+1\}$ and during $\{T-m:T\}$ both produce observed data that can be used to estimate the lag-$m$ effect.

\begin{lemma}
\label{lem:MiddleStructure}
When $\bY = \bY^{+}$ or $\bY = \bY^{-}$, under Assumptions~\ref{asp:nonanticipating}--\ref{asp:BoundedPO}, any optimal design of experiment $\bT$ must satisfy
\begin{align*}
t_{k+1} - t_{k-1} \geq m, \ \forall k \in [K].
\end{align*}
\end{lemma}

Lemma~\ref{lem:MiddleStructure} suggests that in every consecutive $m+1$ periods, there could be at most $3$ randomization points.
Intuitively, too many randomization points in every consecutive $m+1$ periods decreases the chance of observing a useful assignment path of $\bm{1}_{m+1}$ or $\bm{0}_{m+1}$.
Lemma~\ref{lem:MiddleStructure} formalizes such intuition, and suggests that as the persistence of the carryover effect increases, the optimal design randomizes less often.

Lemmas~\ref{lem:StartEndStructure} and \ref{lem:MiddleStructure} restrict the space of possible optimal regular switchback experiment to a smaller class of switchback experiments.
Under such a smaller class of switchback experiments, we can explicitly express the risk function in closed form, which we define below.

\begin{lemma}
[Risk Function]
\label{lem:RiskFunctionExplicit}
When $\bY = \bY^{+}$ or $\bY = \bY^{-}$, under Assumptions~\ref{asp:nonanticipating}--\ref{asp:BoundedPO}, as long as the following three conditions are satisfied,
\begin{align*}
t_1 \geq m+2; & & t_K \leq T-m; & & t_{k+1} - t_{k-1} \geq m, \ \forall k \in [K],
\end{align*}
the risk function for any switchback experiment is given by
\begin{multline}
r(\eta_{\bT, \bQ}, \bY) = \frac{1}{(T-m)^2} \left\{ 4 \sum_{k=1}^{K+1} (t_{k} - t_{k-1})^2 + 8 m (t_K - t_1) + 4 m^2 K - 4 m^2 \right.\\
\left. + 4 \sum_{k=2}^{K} [(m-t_k+t_{k-1})^+]^2\right\} B^2 \label{eqn:StructuredSE:RiskExplicit}
\end{multline}
\end{lemma}

Lemma~\ref{lem:RiskFunctionExplicit} explicitly describes the risk function of any optimal design of regular switchback experiments, which lies in the optimal sub-class of switchback experiments.
The proof of Lemma~\ref{lem:RiskFunctionExplicit} is deferred to Section~\ref{sec:Proof:lem:RiskFunctionExplicit} in the appendix.

To understand the risk function in Lemma~\ref{lem:RiskFunctionExplicit}, we separately examine each term in \eqref{eqn:StructuredSE:RiskExplicit}. The first summation of the squares $\sum_{k=1}^{K+1} (t_{k} - t_{k-1})^2$ suggests that the gap between two consecutive randomization points should not be too large.
The middle term $8m(t_K-t_1)$ formalizes Lemma~\ref{lem:StartEndStructure}, suggesting that the second randomization point on period $t_1$ should not be too early and the last randomization point on period $t_K$ should not be too late.
The last summation of the squares $\sum_{k=2}^{K} [(m-t_k+t_{k-1})^+]^2$ suggests that the gap should not be too small.
Equation~\ref{eqn:StructuredSE:RiskExplicit} formalizes the trade-off that we have described earlier in this section.
First note that when we focus on the optimal design, we treat $T$ and $m$ both as constants. So the constant of $1/(T-m)$ in the expression of the risk function does not affect the optimal design.

\subsubsection{Proof of Lemma~\ref{lem:StartEndStructure}.}
\label{sec:proof:lem:StartEndStructure}

\proof{Proof of Lemma~\ref{lem:StartEndStructure}.}

We prove the two parts separately, both by contradiction.

\textbf{(1)} Suppose there exists an optimal design $\bT = \{t_0=1, t_1, t_2, ..., t_K\}$ such that $t_1 \leq m+1$.
Then we try to construct another design $\tbT$, such that $\left| \tbT \right| = K = \left| \bT \right| -1$.
And the $K$ elements are $\tbT = \{\ttt_0=1 , \ttt_1 = t_2 , \ttt_2 = t_3 , ... , \ttt_{K-1} = t_K\}$.

\begin{table}[!htb]
\TABLE{An example of two regular switchback experiments $\bT$ and $\tbT$ when $m=4$ and $t_1=3$. \label{tbl:StartStructure:TwoTs}}
{
\begin{tabular}{p{1.3cm} | p{1.3cm} | p{1.3cm} | p{1.3cm} | p{1.3cm} | p{1.3cm} | p{1.3cm} | p{1.3cm} }
\hline
& 1 & 2 & 3 & 4 & 5 & 6 & ... \\
\hline
$\bT$ & $\checkmark$ & $-$ & $\checkmark$ & $-$ & $-$ & $\checkmark$ & ... \\
$\tbT$ & $\checkmark$ & $-$ & $-$ & $-$ & $-$ & $\checkmark$ & ... \\
\hline
\end{tabular}
}
{Each checkmark beneath a number indicates that this number is within that set; and each dash beneath a number indicates that this number is not within that set. For example, the checkmark $\checkmark$ beneath number $3$ indicates that $3 \in \bT$; and the dash $-$ beneath number $3$ indicates that $3 \ne \tbT$.}
\end{table}

Next we argue that when $\bY = \bY^{+}$ or $\bY = \bY^{-}$,
$$r(\bT, \bY) > r(\tbT, \bY),$$
which suggests that $\bT$ is not the optimal design.

First, focus on the squared terms.
For any $m+1 \leq t \leq t_1+m-1$, $t_1 \in f^m_{\bT}(t), t_1 \ne f^m_{\tbT}(t)$.
Moreover, $t-m \leq t_1 - 1$, so that $t_0 \in f^m_{\tbT}(t)$.
So $f^m_{\bT}(t) - \{t_1\} = f^m_{\tbT}(t)$, and $\left| f^m_{\tbT}(t) \right| \geq 1$.
As a result,
$$\bE[\bI_t(\bT)^2] - \bE[\bI_t(\tbT)^2] \geq (2^{2+1} - 2^{1+1}) B^2 = 4B^2.$$
For any $t \geq t_1 + m$, either (i) $f_{\bT}(t-m) = t_1$, in which case $f_{\tbT}(t-m) = t_0$.
This is the only difference between $f^m_{\bT}(t)$ and $f^m_{\tbT}(t)$, i.e., $f^m_{\bT}(t) - \{t_1\} = f^m_{\tbT}(t) - \{t_0\}$.
So $\left| f^m_{\bT}(t) \right| = \left| f^m_{\tbT}(t) \right|$.
The second case is (ii) $f_{\bT}(t-m) \geq t_2$, in which case $f^m_{\bT}(t) = f^m_{\tbT}(t)$.
Both cases suggest that 
$$\bE[\bI_t(\bT)^2] - \bE[\bI_t(\tbT)^2] = 0.$$

So we have
\begin{align*}
\sum_{t=m+1}^T \bE\left[ \bI_t(\bT)^2 \right] - \sum_{t=m+1}^T \bE\left[ \bI_t(\tbT)^2 \right] 
= & \sum_{t=m+1}^{t_1+m-1} \left( \bE\left[ \bI_t(\bT)^2 \right] - \bE\left[ \bI_t(\tbT)^2 \right] \right) + \sum_{t=t_1+m}^{T} \left( \bE\left[ \bI_t(\bT)^2 \right] - \bE\left[ \bI_t(\tbT)^2 \right] \right) \\
\geq & \sum_{t=m+1}^{t_1+m-1} (4B^2) + 0 \\
= & 4(t_1-1)B^2 \\
> & 0
\end{align*}

Second, focus on the cross product terms.
For any $t$ and $t'$ such that $m+1 \leq t < t' \leq t_1+m-1$, $t_1 \in O_{\bT}(t,t'), t_1 \ne O_{\tbT}(t,t')$.
Moreover, $t-m \leq t_1 - 1$, so that $t_0 \in O_{\bT}(t,t')$.
So $O_{\bT}(t,t') - \{t_1\} = O_{\tbT}(t,t')$, and $\left| O_{\tbT}(t,t') \right| \geq 1$.
As a result,
$$\bE[\bI_t(\bT)\bI_{t'}(\bT)] - \bE[\bI_t(\tbT)\bI_{t'}(\tbT)] \geq (2^{2+1} - 2^{1+1}) B^2 = 4B^2 > 0.$$
For any $m+1 \leq t < t' \leq T$ such that $t' \geq t_1 + m$, either (i) $f_{\bT}(t'-m) = t_1$, in which case $f_{\tbT}(t'-m) = t_0$.
So $O_{\bT}(t,t') - \{t_1\} = O_{\tbT}(t,t') - \{t_0\}$.
So $\left| O_{\bT}(t,t') \right| = \left| O_{\tbT}(t,t') \right|$.
The second case is (ii) $f_{\bT}(t'-m) \geq t_2$, in which case $O_{\bT}(t,t') = O_{\tbT}(t,t')$.
Both cases suggest that 
$$\bE[\bI_t(\bT)\bI_{t'}(\bT)] - \bE[\bI_t(\tbT)\bI_{t'}(\tbT)] = 0.$$

So we have
\begin{align*}
& \sum_{m+1 \leq t < t' \leq T} \bE\left[ \bI_t(\bT)\bI_{t'}(\bT) \right] - \sum_{m+1 \leq t < t' \leq T} \bE\left[ \bI_t(\tbT)\bI_{t'}(\tbT) \right] \\
= & \sum_{m+1 \leq t < t' \leq t_1+m-1} \left( \bE\left[ \bI_t(\bT)\bI_{t'}(\bT) \right] - \bE\left[ \bI_t(\tbT)\bI_{t'}(\tbT) \right] \right) + \sum_{\substack{m+1 \leq t < t' \leq T \\ t' \geq t_1+m}} \left( \bE\left[ \bI_t(\bT)\bI_{t'}(\bT) \right] - \bE\left[ \bI_t(\tbT)\bI_{t'}(\tbT) \right] \right) \\
\geq & 0
\end{align*}

Combine both square terms and cross-product terms we know that $$r(\bT, \bY) > r(\tbT, \bY).$$

\textbf{(2)} Suppose there exists an optimal design $\bT = \{t_0=1, t_1, t_2, ..., t_K\}$ such that $t_K \geq T-m+1$.
Then we try to construct another design $\tbT$, such that $\left| \tbT \right| = K = \left| \bT \right| - 1$.
And the $K$ elements are $\tbT = \{\ttt_0=1 , \ttt_1 = t_1 , \ttt_2 = t_2, ... , \ttt_{K-1} = t_{K-1}\}$.

\begin{table}[!htb]
\TABLE{An example of two regular switchback experiments $\bT$ and $\tbT$ when $m=4$ and $t_K=T-2$. \label{tbl:EndStructure:TwoTs}}
{
\begin{tabular}{| p{1.3cm} | p{1.3cm} | p{1.3cm} | p{1.3cm} | p{1.3cm} | p{1.3cm} | p{1.3cm} | p{1.3cm} |}
\hline
 & ... & $T-5$ & $T-4$ & $T-3$ & $T-2$ & $T-1$ & $T$\\
\hline
$\bT$ & ... & $\checkmark$ & $-$ & $\checkmark$ & $\checkmark$ & $-$ & $-$ \\
$\tbT$ & ... & $\checkmark$ & $-$ & $\checkmark$ & $-$ & $-$ & $-$ \\
\hline
\end{tabular}
}
{Each checkmark beneath a number indicates that this number is within that set; and each dash beneath a number indicates that this number is not within that set. For example, the checkmark $\checkmark$ beneath number $T-2$ indicates that $T-2 \in \bT$; and the dash $-$ beneath number $T-2$ indicates that $T-2 \ne \tbT$.}
\end{table}

Next we argue that when $\bY = \bY^{+}$ or $\bY = \bY^{-}$,
$$r(\bT, \bY) > r(\tbT, \bY),$$
which suggests that $\bT$ is not the optimal design.

First focus on the squared terms.
For any $m+1 \leq t \leq t_K-1$, $f^m_{\bT}(t) = f^m_{\tbT}(t)$ is totally unchanged.
$$\bE[\bI_t(\bT)^2] - \bE[\bI_t(\tbT)^2] = 0.$$
For any $t_K \leq t \leq T$, $t_K \notin f^m_{\tbT}(t), t_K \in f^m_{\bT}(t)$.
And all the other determining randomization points are unchanged. So $f^m_{\tbT}(t) \subset f^m_{\bT}(t)$ and $f^m_{\bT}(t) - \{t_K\} = f^m_{\tbT}(t)$ and $\left| f^m_{\tbT}(t) \right| \geq 1$.
$$\bE[\bI_t(\bT)^2] - \bE[\bI_t(\tbT)^2] \geq (2^{2+1} - 2^{1+1}) B^2 = 4B^2.$$

So we have
\begin{align*}
\sum_{t=m+1}^T \bE\left[ \bI_t(\bT)^2 \right] - \sum_{t=m+1}^T \bE\left[ \bI_t(\tbT)^2 \right] 
= & \sum_{t=m+1}^{t_K-1} \left( \bE\left[ \bI_t(\bT)^2 \right] - \bE\left[ \bI_t(\tbT)^2 \right] \right) + \sum_{t=t_K}^{T} \left( \bE\left[ \bI_t(\bT)^2 \right] - \bE\left[ \bI_t(\tbT)^2 \right] \right) \\
\geq & \sum_{t=t_K}^{T} (4B^2) + 0 \\
= & 4(T-t_K+1)B^2 \\
> & 0
\end{align*}

Next we focus on the cross-product terms. For any $m+1 \leq t < t' \leq T$ such that $t \leq t_K-1$, $O_{\bT}(t,t') = O_{\tbT}(t,t')$ is totally unchanged.
$$\bE[\bI_t(\bT) \bI_{t'}(\bT)] - \bE[\bI_t(\tbT) \bI_{t'}(\tbT)] = 0.$$
For any $t_K \leq t < t' \leq T$, since $t' - m \leq T-m \leq t_K-1$, so $f_{\tbT}(t'-m) < t_{K}$ and $\left| O_{\tbT}(t,t') \right| \geq 1$ must contain an element.
Moreover, $O_{\tbT}(t,t') \subset O_{\bT}(t,t')$.
So
$$\bE[\bI_t(\bT) \bI_{t'}(\bT)] - \bE[\bI_t(\tbT) \bI_{t'}(\tbT)] \geq (2^{2+1} - 2^{1+1}) B^2 \geq 4B^2 > 0.$$

So we have
\begin{align*}
& \sum_{m+1 \leq t < t' \leq T} \bE\left[ \bI_t(\bT)\bI_{t'}(\bT) \right] - \sum_{m+1 \leq t < t' \leq T} \bE\left[ \bI_t(\tbT)\bI_{t'}(\tbT) \right] \\
= & \sum_{\substack{m+1 \leq t < t' \leq T \\ t \leq t_K-1}} \left( \bE\left[ \bI_t(\bT)\bI_{t'}(\bT) \right] - \bE\left[ \bI_t(\tbT)\bI_{t'}(\tbT) \right] \right) + \sum_{t_K \leq t < t' \leq T} \left( \bE\left[ \bI_t(\bT)\bI_{t'}(\bT) \right] - \bE\left[ \bI_t(\tbT)\bI_{t'}(\tbT) \right] \right) \\
\geq & 0
\end{align*}

Combine both square terms and cross-product terms we know that $$r(\bT, \bY) > r(\tbT, \bY).$$
\Halmos \endproof

\subsubsection{Proof of Lemma~\ref{lem:MiddleStructure}.}
\label{sec:proof:lem:MiddleStructure}

\proof{Proof of Lemma~\ref{lem:MiddleStructure}.}

Recall that we denote $t_0 = 1$ and $t_{K+1} = T+1$.
First, from Lemma~\ref{lem:StartEndStructure}, $t_1 \geq m+2, t_K \leq T-m$.
So $k=1$ and $k=K$ cases both hold.
Next, when $2 \leq k \leq K-1$, we prove by contradiction.

Suppose there exists some optimal design $\bT$, such that $\exists 2 \leq k \leq K-1, s.t. \ t_{k+1} - t_{k-1} \leq m-1.$
Denote $$\bK = \{k \in\{2:K-1\} \left| t_{k+1} - t_{k-1} \leq m-1 \right.\}.$$
Since $\bK \ne \emptyset$, pick $j = \max\bK$ to be the largest element in $\bK$.
Apparently $j \leq K-1$ since $j \in \{2:K-1\}$.
We also know that $t_{j+2} \geq t_j + m,$ because otherwise $j+1 \in \bK$, which contradicts the maximality of $j$.

We now construct another design $\tbT$ such that $\left| \tbT \right| = K = \left| \bT \right| -1$, and the $K$ elements are $\tbT = \{\ttt_0 = 1, \ttt_1 = t_1, ..., \ttt_{j-1}= t_{j-1}, \ttt_{j} = t_{j+1}, ..., \ttt_{K-1} = t_K\}$.

\begin{table}[!htb]
\TABLE{An example of two regular switchback experiments $\bT$ and $\tbT$ when $m=4$ and $t_j = t_{j+1} - 1 = t_{j-1} + 2$. \label{tbl:MiddleStructure:TwoTs}}
{
\begin{tabular}{p{1.3cm} | p{1.3cm} | p{1.3cm} | p{1.3cm} | p{1.3cm} | p{1.3cm} | p{1.3cm} | p{1.3cm} | p{1.3cm} | p{1.3cm} }
\hline
 & ... & $t_{j-1}$ & $t_{j-1}+1$ & $t_j$ & $t_{j+1}$ & $t_{j+1}+1$ & $t_{j+1} + 2$ & $t_{j+2}$ & ... \\
\hline
$\bT$ & ... & $\checkmark$ & $-$ & $\checkmark$ & $\checkmark$ & $-$ & $-$ & $\checkmark$ & ...\\
$\tbT$ & ... & $\checkmark$ & $-$ & $-$ & $\checkmark$ & $-$ & $-$ & $\checkmark$ & ...\\
\hline
\end{tabular}
}
{Each checkmark beneath a number indicates that this number is within that set; and each dash beneath a number indicates that this number is not within that set. For example, the checkmark $\checkmark$ beneath number $t_j$ indicates that $t_j \in \bT$; and the dash $-$ beneath number $t_j$ indicates that $t_j \ne \tbT$.}
\end{table}

Next we argue that when $\bY = \bY^{+}$ or $\bY = \bY^{-}$,
$$r(\bT, \bY) > r(\tbT, \bY),$$
which suggests that $\bT$ is not the optimal design.

First focus on the squared terms.
When $t \leq t_j - 1$, $f^m_{\bT}(t) = f^m_{\tbT}(t)$ is totally unchanged.
$$\bE[\bI_t(\bT)^2] - \bE[\bI_t(\tbT)^2] = 0.$$
When $t_j \leq t \leq t_j+m-1$, this suggests that $t-m \leq t_J-1$ so that $f_{\tbT} \leq t_j-1$. So $t_j \notin f^m_{\tbT}(t), t_j \in f^m_{\bT}(t)$.
And all the other determining randomization points are unchanged.
So $f^m_{\tbT}(t) \subset f^m_{\bT}(t)$ and $f^m_{\bT}(t) - \{t_j\} = f^m_{\tbT}(t)$ and $\left| f^m_{\tbT}(t) \right| \geq 1$.
$$\bE[\bI_t(\bT)^2] - \bE[\bI_t(\tbT)^2] \geq (2^{2+1} - 2^{1+1}) B^2 = 4B^2.$$
When $t_j +m \leq t \leq T$, either (i) $f_{\bT}(t-m) = t_j$, in which case $f_{\tbT}(t-m) = t_{j-1}$. This is the only difference between $f^m_{\bT}(t)$ and $f^m_{\tbT}(t)$, i.e., $f^m_{\bT}(t) - \{t_j\} = f^m_{\tbT}(t) - \{t_{j-1}\}$.
So $\left| f^m_{\bT}(t) \right| = \left| f^m_{\tbT}(t) \right|$.
The second case is (ii) $f_{\bT}(t-m) \geq t_{j+1}$, in which case $f^m_{\bT}(t) = f^m_{\tbT}(t)$. Both cases suggest that
$$\bE[\bI_t(\bT)^2] - \bE[\bI_t(\tbT)^2] = 0.$$

So we have
\begin{align*}
& \sum_{t=m+1}^T \bE\left[ \bI_t(\bT)^2 \right] - \sum_{t=m+1}^T \bE\left[ \bI_t(\tbT)^2 \right] \\
= & \sum_{t=m+1}^{t_j-1} \left( \bE\left[ \bI_t(\bT)^2 \right] - \bE\left[ \bI_t(\tbT)^2 \right] \right) + \sum_{t=t_j}^{t_j+m-1} \left( \bE\left[ \bI_t(\bT)^2 \right] - \bE\left[ \bI_t(\tbT)^2 \right] \right) + \sum_{t=t_j+m}^{T} \left( \bE\left[ \bI_t(\bT)^2 \right] - \bE\left[ \bI_t(\tbT)^2 \right] \right) \\
\geq & 0 + \sum_{t=t_j}^{t_j+m-1} (4B^2) + 0\\
= & 4(m-1)B^2 \\
> & 0
\end{align*}

Next we focus on the cross-product terms.
Let $m+1 \leq t < t' \leq T$.
There are many cases which we summarize in Table~\ref{tbl:SummaryCrossTerms}
\begin{table}[!htb]
\TABLE{Summary of the differences between cross-product terms under two regular switchback experiments $\bT$ and $\tbT$. \label{tbl:SummaryCrossTerms}}
{
\begin{tabular}{|l|c|c|}
\hline
 & $\bT$         & $\tbT$         \\ \hline
$m+1 \leq t \leq t_{j-1}, t < t' \leq T$ & \multicolumn{2}{c|}{unchanged} \\ \hline
$t_{j-1} \leq t \leq t_j-1, t < t' \leq t_j+m-1$ & \multicolumn{2}{c|}{unchanged} \\ \hline
$t_{j-1} \leq t \leq t_j-1, t_j +m \leq t' \leq t_{j+1}+m-1$ & 0             & $4B^2$         \\ \hline
$t_{j-1} \leq t \leq t_{j}-1, t_{j+1}+m \leq t' \leq T$ & \multicolumn{2}{c|}{unchanged}         \\ \hline
$t_j \leq t < t' \leq t_j+m-1$ & $2^{\left| O_{\bT}(t,t') \right|+1}B^2$            & $2^{\left| O_{\tbT}(t,t') \right|+1}B^2$              \\ \hline
$t_j \leq t \leq t_j+m-1, t_j+m \leq t' \leq T$ & \multicolumn{2}{c|}{unchanged} \\ \hline
$t_j + m \leq t < t' \leq T$ & \multicolumn{2}{c|}{unchanged} \\ \hline
\end{tabular}
}
{}
\end{table}

We explain Table~\ref{tbl:SummaryCrossTerms}.

When $m+1 \leq t \leq t_{j-1}, t < t' \leq T$, all the overlapping randomization points are earlier than $t_{j-1}-1$, i.e., $\forall a \in O_{\bT}(t,t'), a \leq t_{j-1}-1; \forall a \in O_{\tbT}(t,t'), a \leq t_{j-1}-1$.
So $t_j \notin O_{\bT}(t,t')$, and the overlapping randomization points are unchanged, i.e., $O_{\bT}(t,t') = O_{\tbT}(t,t')$.

When $t_{j-1} \leq t \leq t_j-1, t < t' \leq t_j+m-1$, all the overlapping randomization points are earlier than $t_{j-1}$, i.e., $\forall a \in O_{\bT}(t,t'), a \leq t_{j-1}; \forall a \in O_{\tbT}(t,t'), a \leq t_{j-1}$.
So $t_j \notin O_{\bT}(t,t')$, and the overlapping randomization points are unchanged, i.e., $O_{\bT}(t,t') = O_{\tbT}(t,t')$.

When $t_{j-1} \leq t \leq t_j-1, t_j +m \leq t' \leq t_{j+1}+m-1$, changing from $\bT$ to $\tbT$ increases the expected values.
This is because $t'-m \geq t_j > t$. So first, $O_{\bT}(t,t') = \emptyset$.
But $f_{\tbT}(t'-m) = t_{j-1}$ and $t_{j-1} \in f^m_{\tbT}(t)$, which suggests that $t_{j-1} \in O_{\tbT}(t,t').$
Also, $\forall a \in f^m_{\tbT}(t'), a \geq t_{j-1}; \forall a \in f^m_{\bT}(t), a \leq t_{j-1}$, which suggests that $t_{j-1}$ is the only overlapping element.
So, $O_{\tbT}(t,t') = \{t_{j-1}\}$.
In this case,
$$\bE[\bI_t(\bT) \bI_{t'}(\bT)] - \bE[\bI_t(\tbT) \bI_{t'}(\tbT)] = (0 - 2^{1+1}) B^2 = -4B^2.$$

When $t_{j-1} \leq t \leq t_{j}-1, t_{j+1}+m \leq t' \leq T$, since $t'-m \geq t_{j+1} > t_j > t$, $O_{\bT}(t, t') = O_{\tbT}(t, t') = \emptyset.$

When $t_j \leq t < t' \leq t_j+m-1$, $t_j \in O_{\bT}(t, t')$ and $t_j \notin O_{\tbT}(t, t')$.
And all the other overlapping randomization points are unchanged, so $O_{\bT}(t, t') - \{t_j\} = O_{\tbT}(t, t')$ and $\left| O_{\tbT}(t, t') \right| \geq 1.$
In this case,
$$\bE[\bI_t(\bT) \bI_{t'}(\bT)] - \bE[\bI_t(\tbT) \bI_{t'}(\tbT)] \geq (2^{2+1} - 2^{1+1}) B^2 = 4B^2.$$

When $t_j \leq t \leq t_j+m-1, t_j+m \leq t' \leq T$, either (i) $f^m_{\bT}(t'-m) = t_j$, in which case $f_{\tbT}(t'-m) = t_{j-1}$.
This is the only difference between $O_{\bT}(t, t')$ and $O_{\tbT}(t, t')$, i.e., $O_{\bT}(t, t') - \{t_j\} = O_{\tbT}(t, t') - \{t_{j-1}\}$.
$\left| O_{\bT}(t, t') \right| = \left| O_{\tbT}(t, t') \right|$.
The second case is (ii) $f_{\bT}(t'-m) \geq t_{j+1}$, in which case $O_{\bT}(t, t') = O_{\tbT}(t, t')$ is unchanged.
Both cases suggest that $\bE[\bI_t(\bT) \bI_{t'}(\bT)] - \bE[\bI_t(\tbT) \bI_{t'}(\tbT)]=0$.

When $t_j + m \leq t < t' \leq T$, either (i) $f^m_{\bT}(t'-m) = t_j$, in which case $f_{\tbT}(t'-m) = t_{j-1}$.
This is the only difference between $O_{\bT}(t, t')$ and $O_{\tbT}(t, t')$, i.e., $O_{\bT}(t, t') - \{t_j\} = O_{\tbT}(t, t') - \{t_{j-1}\}$.
$\left| O_{\bT}(t, t') \right| = \left| O_{\tbT}(t, t') \right|$.
The second case is (ii) $f_{\bT}(t'-m) \geq t_{j+1}$, in which case $O_{\bT}(t, t') = O_{\tbT}(t, t')$ is unchanged.
Both cases suggest that $\bE[\bI_t(\bT) \bI_{t'}(\bT)] - \bE[\bI_t(\tbT) \bI_{t'}(\tbT)]=0$.

So we have
\begin{align*}
& \sum_{m+1 \leq t < t' \leq T} \bE\left[ \bI_t(\bT)\bI_{t'}(\bT) \right] - \sum_{m+1 \leq t < t' \leq T} \bE\left[ \bI_t(\tbT)\bI_{t'}(\tbT) \right] \\
= & \sum_{\substack{t_{j-1} \leq t \leq t_j-1 \\ t_j +m \leq t' \leq t_{j+1}+m-1}} \left( \bE\left[ \bI_t(\bT)\bI_{t'}(\bT) \right] - \bE\left[ \bI_t(\tbT)\bI_{t'}(\tbT) \right] \right) + \sum_{t_j \leq t < t' \leq t_j+m-1} \left( \bE\left[ \bI_t(\bT)\bI_{t'}(\bT) \right] - \bE\left[ \bI_t(\tbT)\bI_{t'}(\tbT) \right] \right) \\
\geq & \sum_{\substack{t_{j-1} \leq t \leq t_j-1 \\ t_j +m \leq t' \leq t_{j+1}+m-1}} \left( -4B^2 \right) + \sum_{t_j \leq t < t' \leq t_j+m-1} \left( 4B^2 \right) \\
= & - (t_j - t_{j-1})(t_{j+1} - t_j) 4B^2 + \frac{m(m-1)}{2} 4B^2 \\
\geq & 0
\end{align*}
where the last inequality is because $j \in \bK$, $t_{j+1} - t_{j-1} \leq m-1$, so $(t_j - t_{j-1})(t_{j+1} - t_j) \leq \frac{(m-1)^2}{4} \leq \frac{m(m-1)}{2}$.

Combine both square terms and cross-product terms we know that $$r(\bT, \bY) > r(\tbT, \bY).$$
\Halmos \endproof

\subsubsection{Proof of Lemma~\ref{lem:RiskFunctionExplicit}.}
\label{sec:Proof:lem:RiskFunctionExplicit}

\proof{Proof of Lemma~\ref{lem:RiskFunctionExplicit}.}

Think of $\bE[\bI_t^2]$ as $\bE[\bI_t \bI_t]$, so that
$r(\eta_{\bT, \bQ}, \bY) = \sum_{t=m+1}^T \sum_{t'=m+1}^T \bE[\bI_t \bI_{t'}]$.
Then we can decompose the risk function to be
\begin{align}
(T-m)^2 \cdot r(\eta_{\bT, \bQ}, \bY) & = \sum_{\substack{m+1 \leq t, t' \leq T \\ \min\{t,t'\} \leq t_1-1}} \bE[\bI_t \bI_{t'}] + \sum_{k=1}^{K-1} \left( \sum_{\substack{t_k \leq t, t' \leq T \\ \min\{t,t'\} \leq t_{k+1}-1}} \bE[\bI_t \bI_{t'}] \right) + \sum_{t_K \leq t, t' \leq T} \bE[\bI_t \bI_{t'}] \label{eqn:RiskFunctionThreeBlocks}
\end{align}

The core of this proof is to carefully count how many values can each $\bE[\bI_t \bI_{t'}], \forall t, t' \in \{m+1:T\}$ take.
See Table~\ref{tbl:IllustratorCounting} for an illustration.

\begin{table}[!htb]
\TABLE{Illustrator of the different values of $\bE[\bI_t \bI_t]$, when $T=17, m=4, \bT=\{1,6,8,13\}$. \label{tbl:IllustratorCounting}}
{
\begin{tabular}{p{0.4cm}p{0.4cm}p{0.4cm}p{0.4cm}p{0.4cm}p{0.4cm}p{0.4cm}p{0.4cm}p{0.4cm}p{0.4cm}p{0.4cm}p{0.4cm}p{0.4cm}p{0.4cm}p{0.4cm}p{0.4cm}p{0.4cm}p{1.6cm}}
($1$ & $2$ & $3$ & $4$) & $5$ & $6$ & $7$ & $8$ & $9$ & $10$ & $11$ & $12$ & $13$ & $14$ & $15$ & $16$ & $17$ \\
($\checkmark$ & $-$ & $-$ & \multicolumn{1}{c}{$-$)} & $-$                    & $\checkmark$                    & $-$                    & $\checkmark$                     & $-$                    & $-$                    & $-$                    & $-$                    & $\checkmark$                    & $-$                    & $-$                    & $-$                    & $-$                    \\ \cline{5-17} 
      &       &       & \multicolumn{1}{l|}{$-$}  & \multicolumn{1}{l|}{$4$} & \multicolumn{1}{l|}{$4$} & \multicolumn{1}{l|}{$4$} & \multicolumn{1}{l|}{$4$} & \multicolumn{1}{l|}{$4$}  & \multicolumn{1}{l|}{}  & \multicolumn{1}{l|}{}  & \multicolumn{1}{l|}{}  & \multicolumn{1}{l|}{}  & \multicolumn{1}{l|}{}  & \multicolumn{1}{l|}{}  & \multicolumn{1}{l|}{}  & \multicolumn{1}{l|}{} & \\ [5pt] \cline{5-17} 
      &       &       & \multicolumn{1}{l|}{$\checkmark$}  & \multicolumn{1}{l|}{$4$} & \multicolumn{1}{l|}{$\bm{8}$} & \multicolumn{1}{l|}{$\bm{8}$} & \multicolumn{1}{l|}{$\bm{8}$} & \multicolumn{1}{l|}{$\bm{8}$}  & \multicolumn{1}{l|}{$4$} & \multicolumn{1}{l|}{$4$} & \multicolumn{1}{l|}{} & \multicolumn{1}{l|}{}  & \multicolumn{1}{l|}{}  & \multicolumn{1}{l|}{}  & \multicolumn{1}{l|}{}  & \multicolumn{1}{l|}{}  & \multicolumn{1}{l}{}  \\ [5pt] \cline{5-17} 
      &       &       & \multicolumn{1}{l|}{$-$}  & \multicolumn{1}{l|}{$4$} & \multicolumn{1}{l|}{$\bm{8}$} & \multicolumn{1}{l|}{$\bm{8}$} & \multicolumn{1}{l|}{$\bm{8}$} & \multicolumn{1}{l|}{$\bm{8}$}  & \multicolumn{1}{l|}{$4$} & \multicolumn{1}{l|}{$4$} & \multicolumn{1}{l|}{} & \multicolumn{1}{l|}{}  & \multicolumn{1}{l|}{}  & \multicolumn{1}{l|}{}  & \multicolumn{1}{l|}{}  & \multicolumn{1}{l|}{}  & \multicolumn{1}{l}{}  \\ [5pt] \cline{5-17} 
      &       &       & \multicolumn{1}{l|}{$\checkmark$}  & \multicolumn{1}{l|}{$4$} & \multicolumn{1}{l|}{$\bm{8}$} & \multicolumn{1}{l|}{$\bm{8}$} & \multicolumn{1}{l|}{$\bm{16}$} & \multicolumn{1}{l|}{$\bm{16}$} & \multicolumn{1}{l|}{$\bm{8}$} & \multicolumn{1}{l|}{$\bm{8}$} & \multicolumn{1}{l|}{$4$} & \multicolumn{1}{l|}{$4$} & \multicolumn{1}{l|}{$4$} & \multicolumn{1}{l|}{$4$} & \multicolumn{1}{l|}{$4$} & \multicolumn{1}{l|}{} & \multicolumn{1}{l}{}  \\ [5pt] \cline{5-17} 
      &       &       & \multicolumn{1}{l|}{$-$}  & \multicolumn{1}{l|}{$4$}  & \multicolumn{1}{l|}{$\bm{8}$} & \multicolumn{1}{l|}{$\bm{8}$} & \multicolumn{1}{l|}{$\bm{16}$} & \multicolumn{1}{l|}{$\bm{16}$}  & \multicolumn{1}{l|}{$\bm{8}$} & \multicolumn{1}{l|}{$\bm{8}$} & \multicolumn{1}{l|}{$4$} & \multicolumn{1}{l|}{$4$} & \multicolumn{1}{l|}{$4$} & \multicolumn{1}{l|}{$4$} & \multicolumn{1}{l|}{$4$} & \multicolumn{1}{l|}{} & \multicolumn{1}{l}{}  \\ [5pt] \cline{5-17} 
      &       &       & \multicolumn{1}{l|}{$-$}  & \multicolumn{1}{l|}{}  & \multicolumn{1}{l|}{$4$} & \multicolumn{1}{l|}{$4$} & \multicolumn{1}{l|}{$\bm{8}$}  & \multicolumn{1}{l|}{$\bm{8}$} & \multicolumn{1}{l|}{$\bm{8}$} & \multicolumn{1}{l|}{$\bm{8}$} & \multicolumn{1}{l|}{$4$} & \multicolumn{1}{l|}{$4$} & \multicolumn{1}{l|}{$4$} & \multicolumn{1}{l|}{$4$} & \multicolumn{1}{l|}{$4$} & \multicolumn{1}{l|}{} & \multicolumn{1}{l}{} \\ [5pt] \cline{5-17} 
      &       &       & \multicolumn{1}{l|}{$-$}  & \multicolumn{1}{l|}{}  & \multicolumn{1}{l|}{$4$} & \multicolumn{1}{l|}{$4$} & \multicolumn{1}{l|}{$\bm{8}$}  & \multicolumn{1}{l|}{$\bm{8}$} & \multicolumn{1}{l|}{$\bm{8}$} & \multicolumn{1}{l|}{$\bm{8}$} & \multicolumn{1}{l|}{$4$} & \multicolumn{1}{l|}{$4$} & \multicolumn{1}{l|}{$4$} & \multicolumn{1}{l|}{$4$} & \multicolumn{1}{l|}{$4$} & \multicolumn{1}{l|}{} & \multicolumn{1}{l}{} \\ [5pt] \cline{5-17} 
      &       &       & \multicolumn{1}{l|}{$-$}  & \multicolumn{1}{l|}{}  & \multicolumn{1}{l|}{}  & \multicolumn{1}{l|}{}  & \multicolumn{1}{l|}{$4$}  & \multicolumn{1}{l|}{$4$} & \multicolumn{1}{l|}{$4$} & \multicolumn{1}{l|}{$4$} & \multicolumn{1}{l|}{$4$} & \multicolumn{1}{l|}{$4$} & \multicolumn{1}{l|}{$4$} & \multicolumn{1}{l|}{$4$} & \multicolumn{1}{l|}{$4$} & \multicolumn{1}{l|}{}  \\ [5pt] \cline{5-17} 
      &       &       & \multicolumn{1}{l|}{$\checkmark$}  & \multicolumn{1}{l|}{}  & \multicolumn{1}{l|}{}  & \multicolumn{1}{l|}{}  & \multicolumn{1}{l|}{$4$}  & \multicolumn{1}{l|}{$4$} & \multicolumn{1}{l|}{$4$} & \multicolumn{1}{l|}{$4$} & \multicolumn{1}{l|}{$4$} & \multicolumn{1}{l|}{$\bm{8}$} & \multicolumn{1}{l|}{$\bm{8}$} & \multicolumn{1}{l|}{$\bm{8}$} & \multicolumn{1}{l|}{$\bm{8}$} & \multicolumn{1}{l|}{$4$} & \multicolumn{1}{l}{} \\ [5pt] \cline{5-17} 
      &       &       & \multicolumn{1}{l|}{$-$}  & \multicolumn{1}{l|}{}  & \multicolumn{1}{l|}{}  & \multicolumn{1}{l|}{}  & \multicolumn{1}{l|}{$4$}  & \multicolumn{1}{l|}{$4$} & \multicolumn{1}{l|}{$4$} & \multicolumn{1}{l|}{$4$} & \multicolumn{1}{l|}{$4$} & \multicolumn{1}{l|}{$\bm{8}$} & \multicolumn{1}{l|}{$\bm{8}$} & \multicolumn{1}{l|}{$\bm{8}$} & \multicolumn{1}{l|}{$\bm{8}$} & \multicolumn{1}{l|}{$4$} & \multicolumn{1}{l}{} \\ [5pt] \cline{5-17} 
      &       &       & \multicolumn{1}{l|}{$-$}  & \multicolumn{1}{l|}{}  & \multicolumn{1}{l|}{}  & \multicolumn{1}{l|}{}  & \multicolumn{1}{l|}{$4$}  & \multicolumn{1}{l|}{$4$} & \multicolumn{1}{l|}{$4$} & \multicolumn{1}{l|}{$4$} & \multicolumn{1}{l|}{$4$} & \multicolumn{1}{l|}{$\bm{8}$} & \multicolumn{1}{l|}{$\bm{8}$} & \multicolumn{1}{l|}{$\bm{8}$} & \multicolumn{1}{l|}{$\bm{8}$} & \multicolumn{1}{l|}{$4$} & \multicolumn{1}{l}{}  \\ [5pt] \cline{5-17} 
      &       &       & \multicolumn{1}{l|}{$-$}  & \multicolumn{1}{l|}{}  & \multicolumn{1}{l|}{}  & \multicolumn{1}{l|}{}  & \multicolumn{1}{l|}{$4$}  & \multicolumn{1}{l|}{$4$} & \multicolumn{1}{l|}{$4$} & \multicolumn{1}{l|}{$4$} & \multicolumn{1}{l|}{$4$} & \multicolumn{1}{l|}{$\bm{8}$} & \multicolumn{1}{l|}{$\bm{8}$} & \multicolumn{1}{l|}{$\bm{8}$} & \multicolumn{1}{l|}{$\bm{8}$} & \multicolumn{1}{l|}{$4$} & \multicolumn{1}{l}{} \\ [5pt] \cline{5-17} 
      &       &       & \multicolumn{1}{l|}{$-$}  & \multicolumn{1}{l|}{}  & \multicolumn{1}{l|}{}  & \multicolumn{1}{l|}{}  & \multicolumn{1}{l|}{}   & \multicolumn{1}{l|}{}  & \multicolumn{1}{l|}{}  & \multicolumn{1}{l|}{}  & \multicolumn{1}{l|}{}  & \multicolumn{1}{l|}{$4$} & \multicolumn{1}{l|}{$4$} & \multicolumn{1}{l|}{$4$} & \multicolumn{1}{l|}{$4$} & \multicolumn{1}{l|}{$4$} & \multicolumn{1}{l}{} \\ [5pt] \cline{5-17} 
\end{tabular}
}
{In the second line, each checkmark beneath number $t$ indicates that period $t \in \bT$, i.e. there is a randomization point at period $t$. This table illustrates different values of $\bE[\bI_t \bI_{t'}]$ when $t, t' \in \{m+1, T\}$, where the zero values are omitted. The $B^2$ magnitudes are also omitted.}
\end{table}

First we calculate the first block from equation \eqref{eqn:RiskFunctionThreeBlocks}.
Because $t_1 \geq m+2$, for any $t, t'$ such that $m+1 \leq \min\{t,t'\} \leq t_1 - 1$, $m+1 \leq \max\{t,t'\} \leq t_1+m-1$, we know that the only overlapping randomization point is $t_0$. So $\bE[\bI_t \bI_{t'}] = 4B^2.$
For any $t, t'$ such that $m+1 \leq \min\{t,t'\} \leq t_1 - 1$, $t_1+m \leq \max\{t,t'\} \leq T$, there is no overlapping randomization point so $\bE[\bI_t \bI_{t'}] = 0.$
\begin{align*}
\sum_{\substack{m+1 \leq t, t' \leq T \\ \min\{t,t'\} \leq t_1-1}} \bE[\bI_t \bI_{t'}] = B^2 \left( 4 \cdot ((t_{1} - 1)^2 - m^2) \right)
\end{align*}

Then we calculate the second block from equation \eqref{eqn:RiskFunctionThreeBlocks}.
For any $k \in [K-1]$, consider $t_{k} - t_{k-1}$ and $t_{k+1} - t_{k}$, which jointly determine the values of $\bE[\bI_t \bI_{t'}]$ for any $t, t'$, such that $t_k \leq \min\{t,t'\} \leq t_{k+1}-1$ and $t_k \leq \max\{t,t'\} \leq T$.
We will go over each of the four cases below.

\textbf{(1)} When $t_{k} - t_{k-1} \geq m, t_{k+1} - t_{k} \geq m$.
Due to Lemma~\ref{lem:HandyExpression}, for all $t, t' \in \{t_k : t_k+m-1\}$, $\bE[\bI_t \bI_{t'}] = 8B^2$, because both $t_{k-1} \leq t-m \leq t_k-1$ and $t_{k-1} \leq t'-m \leq t_k-1$, and both $t_{k-1}$ and $t_k$ are overlapping randomization points.
For all $t, t'$ such that $t_k \leq \min\{t,t'\} \leq t_{k+1}-1$ and $t_k+m \leq \max\{t,t'\} \leq t_{k+1}+m-1$, $\bE[\bI_t \bI_{t'}] = 4B^2$, because $t_k \leq \min\{t,t'\} \leq t_{k+1}-1$ and $t_k \leq \max\{t,t'\}-m \leq t_{k+1}-1$ so only $t_k$ is the overlapping randomization point.
For all $t, t'$ such that $t_k \leq \min\{t,t'\} \leq t_{k+1}-1$ and $t_{k+1}+m \leq \max\{t,t'\} \leq T$, $\bE[\bI_t \bI_{t'}] = 0$.

In this case,
\begin{align*}
\sum_{\substack{t_k \leq t, t' \leq T \\ \min\{t,t'\} \leq t_{k+1}-1}} \bE[\bI_t \bI_{t'}] = B^2 \left( 8 \cdot m^2 + 4 \cdot ((m+t_{k+1} - t_k)^2 - 2m^2) \right)
\end{align*}

\textbf{(2)} When $t_{k} - t_{k-1} \geq m, t_{k+1} - t_{k} < m$.
Due to Lemma~\ref{lem:HandyExpression}, for all t, t' such that $t_k \leq \min\{t,t'\} \leq t_{k+1}-1$, $t_k \leq \max\{t,t'\} \leq t_k+m-1$, $\bE[\bI_t \bI_{t'}] = 8B^2$, because both $t,t' \leq t_k+m-1$, so $t_{k-1} \leq t-m \leq t_k-1$ and $t_{k-1} \leq t'-m \leq t_k-1$, and both $t_{k-1}$ and $t_k$ are overlapping randomization points.
For all $t, t'$ such that $t_k \leq \min\{t,t'\} \leq t_{k+1}-1$ and $t_k+m \leq \max\{t,t'\} \leq t_{k+1}+m-1$, $\bE[\bI_t \bI_{t'}] = 4B^2$, because $t_k \leq \min\{t,t'\} \leq t_{k+1}-1$ and $t_k \leq \max\{t,t'\}-m \leq t_{k+1}-1$ so only $t_k$ is the overlapping randomization point.
For all $t, t'$ such that $t_k \leq \min\{t,t'\} \leq t_{k+1}-1$ and $t_{k+1}+m \leq \max\{t,t'\} \leq T$, $\bE[\bI_t \bI_{t'}] = 0$.

In this case,
\begin{align*}
\sum_{\substack{t_k \leq t, t' \leq T \\ \min\{t,t'\} \leq t_{k+1}-1}} \bE[\bI_t \bI_{t'}] = B^2 \left( 8 \cdot (m^2 - (m-t_{k+1}+t_k)^2) + 4 \cdot ((m+t_{k+1} - t_k)^2 - 2m^2 + (m-t_{k+1}-t_k)^2) \right)
\end{align*}

\textbf{(3)} When $t_{k} - t_{k-1} < m, t_{k+1} - t_{k} \geq m$.
Due to Lemma~\ref{lem:HandyExpression}, for all $t, t' \in \{t_k:t_{k-1}+m-1\}$, $\bE[\bI_t \bI_{t'}] = 16B^2$, because $t-m \leq t_{k-1}-1 \leq t_k \leq t$ and $t'-m \leq t_{k-1}-1 \leq t_k \leq t'$ so $t_{k-2}, t_{k-1}, t_k$ are three determining randomization points.
Also $t_k - t_{k-2} \geq m$ so $t_{k-2} \leq \min\{t,t'\}-m$ and $t_{k-3}$ is not a determining randomization point.
For all $t,t'$ such that $t_k \leq \min\{t,t'\} \leq t_k+m-1, t_{k-1}+m \leq \max\{t,t'\} \leq t_k+m-1$, $\bE[\bI_t \bI_{t'}] = 8B^2$, because $\min\{t,t'\} -m \leq t_k-1$ and $t_{k-1} \leq \max\{t,t'\}-m \leq t_k-1$ so $t_{k-1}$ and $t_k$ are two determining randomization point.
For all $t,t'$ such that $t_k \leq \min\{t,t'\} \leq t_{k+1}-1, t_k+m \leq \max\{t,t'\} \leq t_{k+1}+m-1$, $\bE[\bI_t \bI_{t'}] = 4B^2$, because $t_k \leq \max\{t,t'\}-m$ so $t_k$ is the only determining randomization point.
For all $t,t'$ such that $t_k \leq \min\{t,t'\} \leq t_{k+1}-1, t_{k+1}+m \leq \max\{t,t'\} \leq T$, $\bE[\bI_t \bI_{t'}] = 0.$

In this case,
\begin{align*}
\sum_{\substack{t_k \leq t, t' \leq T \\ \min\{t,t'\} \leq t_{k+1}-1}} \bE[\bI_t \bI_{t'}] = & B^2 \left( 16 \cdot (m-t_k+t_{k-1})^2 + 8 \cdot (m^2 - (m-t_{k}+t_{k-1})^2) \right. \\
& \qquad \left. + 4 \cdot ((m+t_{k+1} - t_k)^2 - 2m^2) \right)
\end{align*}

\textbf{(4)} When $t_{k} - t_{k-1} < m, t_{k+1} - t_{k} < m$.
Due to Lemma~\ref{lem:HandyExpression}, for all $t, t' \in \{t_k:t_{k-1}+m-1\}$, $\bE[\bI_t \bI_{t'}] = 16B^2$, because $t-m \leq t_{k-1}-1 \leq t_k \leq t$ and $t'-m \leq t_{k-1}-1 \leq t_k \leq t'$ so $t_{k-2}, t_{k-1}, t_k$ are three determining randomization points.
Also $t_k - t_{k-2} \geq m$ so $t_{k-2} \leq \min\{t,t'\}-m$ and $t_{k-3}$ is not a determining randomization point.
For all $t,t'$ such that $t_k \leq \min\{t,t'\} \leq t_{k+1}-1, t_{k-1}+m \leq \max\{t,t'\} \leq t_k+m-1$, $\bE[\bI_t \bI_{t'}] = 8B^2$, because $\min\{t,t'\} -m < t_k-1$ and $t_{k-1} \leq \max\{t,t'\}-m \leq t_k-1$ so $t_{k-1}$ and $t_k$ are two determining randomization points.
For all $t,t'$ such that $t_k \leq \min\{t,t'\} \leq t_{k+1}-1, t_k+m \leq \max\{t,t'\} \leq t_{k+1}+m-1$, $\bE[\bI_t \bI_{t'}] = 4B^2$, because $t_k \leq \max\{t,t'\}-m$ so $t_k$ is the only determining randomization point.
For all $t,t'$ such that $t_k \leq \min\{t,t'\} \leq t_{k+1}-1, t_{k+1}+m \leq \max\{t,t'\} \leq T$, $\bE[\bI_t \bI_{t'}] = 0.$

In this case,
\begin{align*}
\sum_{\substack{t_k \leq t, t' \leq T \\ \min\{t,t'\} \leq t_{k+1}-1}} \bE[\bI_t \bI_{t'}] = & B^2 \left( 16 \cdot (m-t_k+t_{k-1})^2 + 8 \cdot (m^2 - (m-t_{k}+t_{k-1})^2 - (m-t_{k+1}+t_k)^2) \right. \\
& \qquad \left. + 4 \cdot ((m+t_{k+1} - t_k)^2 - 2m^2 + (m-t_{k+1}+t_k)^2) \right)
\end{align*}

Finally we calculate the third block from equation \eqref{eqn:RiskFunctionThreeBlocks}.
Observe that $T - t_{K} \geq m$.
\textbf{(1)} When $t_{K} - t_{K-1} \geq m$.
Due to Lemma~\ref{lem:HandyExpression}, for all $t, t' \in \{t_K : t_K+m-1\}$, $\bE[\bI_t \bI_{t'}] = 8B^2$, because both $t_{K-1} \leq t-m \leq t_K-1$ and $t_{K-1} \leq t'-m \leq t_K-1$, and both $t_{K-1}$ and $t_K$ are overlapping randomization points.
For all $t,t'$ such that $t_K \leq \min\{t,t'\} \leq T, t_K+m \leq \max\{t,t'\} \leq T$, $\bE[\bI_t \bI_{t'}] = 4B^2$, because $t_K \leq \max\{t,t'\}-m$ so $t_K$ is the only determining randomization point.

In this case,
\begin{align*}
\sum_{t_K \leq t, t' \leq T} \bE[\bI_t \bI_{t'}] = B^2 \left( 8 \cdot m^2 + 4 \cdot ((T+1 - t_K)^2 - m^2) \right)
\end{align*}

\textbf{(2)} When $t_{K} - t_{K-1} < m$.
Due to Lemma~\ref{lem:HandyExpression}, for all $t, t' \in \{t_K:t_{K-1}+m-1\}$, $\bE[\bI_t \bI_{t'}] = 16B^2$, because $t-m \leq t_{K-1}-1 \leq t_K \leq t$ and $t'-m \leq t_{K-1}-1 \leq t_K \leq t'$ so $t_{K-2}, t_{K-1}, t_K$ are three determining randomization points.
Also $t_K - t_{K-2} \geq m$ so $t_{K-2} \leq \min\{t,t'\}-m$ and $t_{K-3}$ is not a determining randomization point.
For all $t,t'$ such that $t_K \leq \min\{t,t'\} \leq t_K+m-1, t_{K-1}+m \leq \max\{t,t'\} \leq t_K+m-1$, $\bE[\bI_t \bI_{t'}] = 8B^2$, because $\min\{t,t'\} -m \leq t_K-1$ and $t_{K-1} \leq \max\{t,t'\}-m \leq t_K-1$ so $t_{K-1}$ and $t_K$ are two determining randomization points.
For all $t,t'$ such that $t_K \leq \min\{t,t'\} \leq T, t_K+m \leq \max\{t,t'\} \leq T$, $\bE[\bI_t \bI_{t'}] = 4B^2$, because $t_K \leq \max\{t,t'\}-m$ so $t_K$ is the only determining randomization point.

In this case,
\begin{align*}
\sum_{t_K \leq t, t' \leq T} \bE[\bI_t \bI_{t'}] = & B^2 \left( 16 \cdot (m-t_K+t_{K-1})^2 + 8 \cdot (m^2 - (m-t_{K}+t_{K-1})^2) + 4 \cdot ((T+1 - t_K)^2 - m^2) \right)
\end{align*}

Now we combine all above together.

Note that whenever there exists $k \in \{2:K\}$ such that $(t_{k} - t_{k-1}) < m$, this suggests that in $\sum_{\substack{t_k \leq t, t' \leq T \\ \min\{t,t'\} \leq t_{k+1}-1}} \bE[\bI_t \bI_{t'}]$ there is a $16 (m-t_k+t_{k-1})^2$; but in $\sum_{\substack{t_{k-1} \leq t, t' \leq T \\ \min\{t,t'\} \leq t_{k}-1}} \bE[\bI_t \bI_{t'}]$ there is a $8 (-(m-t_k+t_{k-1})^2)$.
So when we sum them up, we break $16 (m-t_k+t_{k-1})^2$ into two $8 (m-t_k+t_{k-1})^2$, which cancels in two sumations.
By telescoping,
\begin{align*}
(T-m)^2 \cdot r(\eta_{\bT, \bQ}, \bY) & = \sum_{\substack{m+1 \leq t, t' \leq T \\ \min\{t,t'\} \leq t_1-1}} \bE[\bI_t \bI_{t'}] + \sum_{k=1}^{K-1} \left( \sum_{\substack{t_k \leq t, t' \leq T \\ \min\{t,t'\} \leq t_{k+1}-1}} \bE[\bI_t \bI_{t'}] \right) + \sum_{t_K \leq t, t' \leq T} \bE[\bI_t \bI_{t'}] \\
& = 4B^2 \cdot \left( (t_1-1)^2-m^2 \right) + \sum_{k=1}^{K-1} B^2 \cdot \left( 8 m^2 + 4\left( (m+t_{k+1}-t_k)^2 - 2m^2 + \left( (m-t_{k+1}+t_k)^+ \right)^2 \right) \right) \\
& \qquad + B^2 \cdot \left( 8 m^2 + 4\left( (T+1-t_K)^2 - m^2 \right) \right) \\
& = B^2 \cdot \left\{ 4 \sum_{k=0}^{K} (t_{k+1} - t_{k})^2 + 8 m (t_K - t_1) + 4 m^2 K - 4 m^2 + 4 \sum_{k=1}^{K-1} [(m-t_{k+1}+t_{k})^+]^2\right\}
\end{align*}
which finishes the proof.
\Halmos \endproof

\subsection{Optimal Solutions to the Subset Selection Problem in Theorem~\ref{thm:OptimalDesign}}
\label{sec:OptimalSolutionDiscussion}

\subsubsection{Proof of Theorem~\ref{thm:OptimalDesign}.}
\label{sec:Proof:thm:OptimalDesign}

\proof{Proof of Theorem~\ref{thm:OptimalDesign}.}

Consider the problem as we have introduced in \eqref{eqn:TheProblem}.
Due to Lemma~\ref{lem:AdversaryStrategy}, $\bY^{+} = \left\{Y_t(\bm{1}_{m+1}) = Y_t(\bm{0}_{m+1}) = B \right\}_{t \in \{m+1:T\}}$
and
$\bY^{-} = \left\{Y_t(\bm{1}_{m+1}) = Y_t(\bm{0}_{m+1}) = -B \right\}_{t \in \{m+1:T\}}$
are the only two dominating strategies for the adversarial selection of potential outcomes.

Then due to Lemma~\ref{lem:StartEndStructure} and Lemma~\ref{lem:MiddleStructure}, the optimal design of switchback experiment must satisfy the following three conditions.
\begin{align*}
t_1 \geq m+2, & & t_K \leq T-m & & t_{k+1} - t_{k-1} \geq m, \ \forall k \in [K].
\end{align*}

Due to Lemma~\ref{lem:RiskFunctionExplicit}, the risk function of the optimal design of experiment is given by
\begin{align*}
r(\eta_{\bT, \bQ}, \bY) = \frac{1}{(T-m)^2}\left\{ 4 \sum_{k=1}^{K+1} (t_{k} - t_{k-1})^2 + 8 m (t_K - t_1) + 4 m^2 K - 4 m^2 + 4 \sum_{k=2}^{K} [(m-t_k+t_{k-1})^+]^2\right\} B^2
\end{align*}

So if we further take minimum over $\bT \subset [T]$ in the above risk function, we find the optimal solution to the original problem introduced in \eqref{eqn:TheProblem}.
Note that $B^2$ is a constant and irrelevant to our decisions, and that $T$ and $m$ are inputs.
So we solve, for any given $T$ and $m$, the following subset selection problem:
\begin{align*}
\min_{\bT \subset [T]} \left\{ 4 \sum_{k=0}^{K} (t_{k+1} - t_{k})^2 + 8 m (t_K - t_1) + 4 m^2 K - 4 m^2 + 4 \sum_{k=1}^{K-1} [(m-t_{k+1}+t_{k})^+]^2\right\},
\end{align*}
as stated in \eqref{eqn:SubsetSelection}.

In particular, if there exists some constant $n \in \bN, n \geq 4$, such that $T = n m$, we can explicitly find the optimal design of experiment.
Take the continuous relaxation of this problem, such that for any $K$, $\{1<t_1<t_2<...<t_K<T+1\} \in [1,T+1]^K$.
\begin{align*}
\min_{\substack{K \in \bN, \\ \{1<t_1<t_2<...<t_K<T+1\} \in [1,T+1]^K}} \left\{ 4 \sum_{k=0}^{K} (t_{k+1} - t_{k})^2 + 8 m (t_K - t_1) + 4 m^2 K - 4 m^2 + 4 \sum_{k=1}^{K-1} [(m-t_{k+1}+t_{k})^+]^2\right\}
\end{align*}
The relaxed problem provides a lower bound to the original subset selection problem as stated in \eqref{eqn:SubsetSelection}.
We will argue later that it is a lucky coincidence that the optimal solution to this relaxed problem is also an integer solution.

First we argue that $t_1-t_0=t_{K+1}-t_K$.
This is because otherwise if $t_1-t_0 \ne t_{K+1}-t_K$ then denote $a = \frac{t_1-t_0+t_{K+1}-t_K}{2}$.
We could always pick for any $k \in \{1:K\}$, $\ttt_k = t_k + a - t_1 + 1$, such that $t_{k+1} - t_k$ is unchanged for any $k \in \{1:K-1\}$.
The only change in the objective value comes from
$$\left(2a^2\right) - \left((t_1 - t_0)^2 + (t_{K+1}-t_K)^2 \right) < 0,$$
which suggests that $t_1-t_0 \ne t_{K+1}-t_K$ is not optimal.

Second, similarly, we argue that for any $k' < k'' \in [K-1]$, $t_{k'+1} - t_{k'} = t_{k''+1} - t_{k''}$
This is because otherwise if $t_{k'+1}-t_{k'} \ne t_{k''+1}-t_{k''}$ then denote $b = \frac{t_{k'+1}-t_{k'}+t_{k''+1}-t_{k''}}{2}$.
We could always pick for any $k \in \{k'+1:k''\}$, $\ttt_k = t_k + b - (t_{k'+1} - t_{k'})$, such that $t_{k+1} - t_k$ is unchanged for any $k \in \{k'+1:k''-1\}$.
The only change in the objective value comes from
$$\left(2b^2 + 2((m-b)^+)^2\right) - \left((t_{k'+1}-t_{k'})^2 + (t_{k''+1}-t_{k''})^2 + ((m-t_{k'+1}+t_{k'})^+)^2 + ((m-t_{k''+1}+t_{k''})^+)^2 \right) < 0,$$
where $x^2+((m-x)^+)^2$ is convex and the inequality holds due to Jensen's Inequality.
This inequality suggests that $t_{k'+1}-t_{k'} \ne t_{k''+1}-t_{k''}$ is not optimal.

With the above two structural results, we can assume that there exists $a, b > 0$, such that $t_1-t_0 = t_{K+1}-t_K = a$, and $t_{k+1}-t_{k} = b, \forall k \in [K-1]$
Also, it must be satisfied that $2a+(K-1)b = T$.
Next we replace $K-1=\frac{T-2a}{b}$ into the relaxed problem, to have
\begin{align*}
\min_{a,b>0} & \left\{ 4 (2 a^2 + (K-1)b^2) + 8 m (K-1) b + 4 m^2 (K-1) + 4 (K-1) ((m-b)^+)^2\right\} \\
= \min_{a,b>0} & \left\{ 8 a^2 + 4 (T-2a)b + 8 m (T-2a) + 4 m^2 \frac{T-2a}{b} + 4 \frac{T-2a}{b} ((m-b)^+)^2\right\}
\end{align*}

Either when $b \geq m$, the above is to minimize
\begin{align*}
\min_{a,b>0} & \left\{ 8 a^2 + 4 (T-2a)b + 8 m (T-2a) + 4 m^2 \frac{T-2a}{b} \right\}
\end{align*}
Note that
\begin{align*}
8 a^2 + 4 (T-2a)b + 8 m (T-2a) + 4 m^2 \frac{T-2a}{b} = & 8 a^2 + 8 m (T-2a) + 4 (T-2a) \left(b+ \frac{m^2}{b}\right) \\
\geq & 8 a^2 + 16 m (T-2a) \\
= & 8 (a-2m)^2 +16mT -32m^2 \\
\geq & 16mT -32m^2
\end{align*}
where the first inequality takes equality if and only if $b = \frac{m^2}{b}$, which suggests $b=m$;
the second inequality takes equality if and only if $a=2m$.

Or when $b \leq m$, the above is to minimize
\begin{align*}
\min_{a,b>0} & \left\{ 8 a^2 + 4 (T-2a)b + 8 m (T-2a) + 4 m^2 \frac{T-2a}{b} + 4 \frac{T-2a}{b} (m-b)^2 \right\}
\end{align*}

Note that
\begin{align*}
8 a^2 + 4 (T-2a)b + 8 m (T-2a) + 4 m^2 \frac{T-2a}{b} + 4 \frac{T-2a}{b} (m-b)^2 = & 8 a^2 + 8 (T-2a) \left( b + \frac{m^2}{b} \right) \\
\geq & 8 a^2 + 16 m (T-2a) \\
= & 8 (a-2m)^2 +16mT -32m^2 \\
\geq & 16mT -32m^2
\end{align*}
where the first inequality takes equality if and only if $b = \frac{m^2}{b}$, which suggests $b=m$;
the second inequality takes equality if and only if $a=2m$.

Combining both cases, the optimal solution is when $a=2m$ and $b=m$, which happens to be an integer solution, thus optimal for the subset selection problem.
Translating into $t_1, ..., t_K$ this suggests that $t_1 = 2m+1, t_2 = 3m+1, ..., t_K = (n-2)m+1$.
\Halmos \endproof

\subsubsection{Solutions in the Imperfect Cases.}

It is always worth noting that we are taking a design of experiments perspective.
So when practically we have control of $T$, we can pick $T$ to be some multiples of $m$, which fits our Theorem~\ref{thm:OptimalDesign} perfectly.
If we do not have control of $T$, we can always pick a smaller $T'$ such that $T' = \lfloor T/m \rfloor \cdot m$ is some multiples of $m$.

Nonetheless, from an optimization perspective, we establish the following optimal structure for the subset selection problem as in \eqref{eqn:SubsetSelection}. Recall that $t_{K+1}=T+1$.

\begin{lemma}
\label{lem:StructureSubsetSelection}
Under Assumptions~\ref{asp:nonanticipating}--\ref{asp:BoundedPO}, the optimal design of regular switchback experiment must satisfy the following two conditions,
\begin{align*}
\left| (t_1-t_0) - (t_{K+1}-t_K) \right| \leq 1, & & \left| (t_{j+1}-t_j) - (t_{j'+1}-t_{j'}) \right| \leq 1, \forall 1 \leq j, j' \leq K-1.
\end{align*}
\end{lemma}

\proof{Proof of Lemma~\ref{lem:StructureSubsetSelection}.}
Prove by contradiction.

\noindent \textbf{Case 1.} Suppose there exists some optimal design $\bT$, such that $(t_1-t_0) - (t_{K+1}-t_K) \geq 2$.
We now construct another design $\tbT$, such that $\left| \tbT \right| = K = \left| \bT \right|$, and the $K$ elements are $\tbT = \{\ttt_0 = 1, \ttt_1 = t_1-1, \ttt_2 = t_2-1, ..., \ttt_K = t_K-1\}$.
Now check the expression as in \eqref{eqn:SubsetSelection}.
Note that $\ttt_{k+1} - \ttt_k = t_{k+1} - t_k$ is unchanged for any $k\in[K-1]$; $\ttt_K - \ttt_1 = t_K - t_1$ is unchanged; and $m-\ttt_{k+1} - \ttt_k = m- t_{k+1} - t_k$ in unchanged for any $k\in[K-1]$.
But $(\ttt_1 - \ttt_0)^2 + (\ttt_{K+1} - \ttt_K)^2 = (t_1 - t_0 - 1)^2 + (t_{K+1} - t_K + 1)^2 \leq (t_1 - t_0)^2 + (t_{K+1} - t_K)^2$, because $(t_1 - t_0) - (t_{K+1} - t_K) \geq 2$ and due to convexity.

Similarly, if there exists some optimal design $\bT$, such that $(t_{K+1}-t_K) - (t_1-t_0) \geq 2$, then construct another design $\tbT = \{\ttt_0 = 1, \ttt_1 = t_1+1, \ttt_2 = t_2+1, ..., \ttt_K = t_K+1\}$.

\noindent \textbf{Case 2.} Suppose there exists some optimal design $\bT$, and there exists $1 \leq j < j' \leq K-1$ such that $(t_{j+1}-t_j) - (t_{j'+1}-t_{j'}) \geq 2$.
We now construct another design $\tbT$, such that $\left| \tbT \right| = K = \left| \bT \right|$, and the $K$ elements are $\tbT = \{\ttt_0 = 1, \ttt_1 = t_1, ..., \ttt_j = t_j, \ttt_{j+1} = t_{j+1}-1, ..., \ttt_{j'} = t_{j'}-1, \ttt_{j'+1} = t_{j'+1},..., \ttt_K = t_K\}$.
Now check the expression as in \eqref{eqn:SubsetSelection}.
Note that $\ttt_{k+1} - \ttt_k = t_{k+1} - t_k$ is unchanged for any $k\in\{0:K\}$ except $j$ and $j'$; $\ttt_K - \ttt_1 = t_K - t_1$ is unchanged; and $m-\ttt_{k+1} - \ttt_k = m- t_{k+1} - t_k$ in unchanged for any $k\in[K-1]$ except $j$ and $j'$.
Now focus on $j$ and $j'$.
\begin{align*}
& (\ttt_{j+1} - \ttt_j)^2 + (\ttt_{j'+1} - \ttt_{j'})^2 + [(m - \ttt_{j+1} + \ttt_{j})^+]^2 + [(m - \ttt_{j'+1} + \ttt_{j'})^+]^2 \\
= & (t_{j+1} - t_j - 1)^2 + (t_{j'+1} - t_{j'} + 1)^2 + [(m - t_{j+1} + t_{j} + 1)^+]^2 + [(m - t_{j'+1} + t_{j'} - 1)^+]^2 \\
\leq & (t_{j+1} - t_j)^2 + (t_{j'+1} - t_{j'})^2 + [(m - t_{j+1} + t_{j})^+]^2 + [(m - t_{j'+1} + t_{j'})^+]^2
\end{align*}
To see why this inequality holds, define $g(x)=x^2+[(m-x)^+]^2$ and note that $g(x)$ is a univariate convex function. The inequality holds due to $(t_{j+1}-t_j) - (t_{j'+1}-t_{j'}) \geq 2$ and convexity.

Similarly, if there exists some optimal design $\bT$, and there exists $1 \leq j < j' \leq K-1$ such that $(t_{j'+1}-t_{j'}) - (t_{j+1}-t_j) \geq 2$.
Then construct another design $\tbT = \{\ttt_0 = 1, \ttt_1 = t_1, ..., \ttt_j = t_j, \ttt_{j+1} = t_{j+1}+1, ..., \ttt_{j'} = t_{j'}+1, \ttt_{j'+1} = t_{j'+1},..., \ttt_K = t_K\}$.

Combine both cases we finish the proof.
\Halmos \endproof

\section{Proofs and Discussions from Section~\ref{sec:InferenceAndTesting}}
In the first two sub-Sections of Section~\ref{sec:InferenceAndTesting} we focus on the case when $p=m$.
In Sections~\ref{sec:InferenceAppendix:ExtraNotations}--\ref{sec:proof:thm:AsymptoticNormality} in the appendix, we also focus on the case when $p=m$, and use only $m$ instead of $p$.
In Sections~\ref{sec:FurtherMisspecified}--\ref{sec:proof:coro:AsymptoticNormality}, we will use both $p$ and $m$. Recall that $m$ is the order of the carryover effect, and $p$ is the experimenter's knowledge of $m$.

\subsection{Extra Notations Used in the Proofs from Section~\ref{sec:InferenceAndTesting}}
\label{sec:InferenceAppendix:ExtraNotations}

For any $t \in \{m+1:T\}$, we use the notations of $\bI_t$ as defined in \eqref{eqn:defn:bI}.
Denote
\begin{align*}
\bbI_0 = & \sum_{t=m+1}^{2m} \bI_t & & \\
\bbI_k = & \sum_{t=(k+1)m+1}^{(k+2)m} \bI_t, & & \forall k \in [K] \\
\bbI_{K+1} = & \sum_{t=(K+2)m+1}^{(K+3)m} \bI_t & &
\end{align*}
It is worth noting that under the optimal design as suggested by Theorem~\ref{thm:OptimalDesign}, when $T/m = n \in \bN$ is an integer, we have $K = n-3$.
So $(K+3)m = T$.
See Example~\ref{exa:bbINotation} below.

\example
[An Optimal Design and Its $\bbI_k$ Notations]
\label{exa:bbINotation}
When $T=12$, $p=m=2$, the optimal design of regular switchback experiment is $\bT^*=\{1,5,7,9\}$, and $K=3$.
The $\bbI_k$ notations are defined below.
Each $\bbI_k$ spans $m=2$ periods.
See Table~\ref{tbl:bbINotation}.
\begin{table}[!thb]
\TABLE{An example of the optimal design $\bT^*$ and its $\bbI_k$ notations when $T=12$ and $p=m=2$. \label{tbl:bbINotation}}
{
\begin{tabular}{| p{1.78cm} | p{0.88cm} | p{0.88cm} | p{0.88cm} | p{0.88cm} | p{0.88cm} | p{0.88cm} | p{0.88cm} | p{0.88cm} | p{0.88cm} | p{0.88cm} | p{0.88cm} | p{0.88cm} |}
\hline
  & 1            & 2   & 3          & 4         & 5            & 6       & 7            & 8        & 9            & 10       & 11           & 12       \\ \hline
$\bT^*$ & $\checkmark$ & $-$ & $-$        & $-$       & $\checkmark$ & $-$     & $\checkmark$ & $-$      & $\checkmark$ & $-$      & $-$ & $-$      \\ \hline
  & \multicolumn{2}{l|}{$-$}     & \multicolumn{2}{l|}{} & \multicolumn{2}{l|}{} & \multicolumn{2}{l|}{} & \multicolumn{2}{l|}{} & \multicolumn{2}{l|}{} \\ [-1em]
$\{\bbI_k\}_{k=0}^{K+1}$ & \multicolumn{2}{l|}{}     & \multicolumn{2}{l|}{$\bbI_0$} & \multicolumn{2}{l|}{$\bbI_1$} & \multicolumn{2}{l|}{$\bbI_2$} & \multicolumn{2}{l|}{$\bbI_3$} & \multicolumn{2}{l|}{$\bbI_4$} \\ \hline
\end{tabular}
}
{}
\end{table}
\Halmos \endexample

Using the above notation, we could write
\begin{align*}
\widehat{\tau}_m - \tau_m = \frac{1}{T-m} \sum_{k=0}^{K+1} \bbI_k,
\end{align*}
and so
\begin{align*}
\var(\widehat{\tau}_m) = \frac{1}{(T-m)^2} \var \left( \sum_{k=0}^{K+1} \bbI_k \right).
\end{align*}

\subsection{Proof of Lemma~\ref{lem:HTvariance}}
\label{sec:proof:thm:Variance}
The proof of Lemma~\ref{lem:HTvariance} resembles the proof of Lemmas~\ref{lem:square-terms} and~\ref{lem:cross-products}.
The trick here is to observe that for any $k \in [K]$, the values of all the variables $\bI_t$, where $(k+1)m+1 \leq t \leq (k+2)m$, are all determined by the randomization at time $km+1$ and $(k+1)m+1$.
Since they are all correlated, we can use $\bbI_k$ to stand for $\sum_{t=(k+1)m+1}^{(k+2)m} \bI_t$ for short.

\proof{Proof of Lemma~\ref{lem:HTvariance}.}
First observe that $\bbI_k$ has zero mean for each $k \in \{0:K+1\}$.
So we can decompose the variance into squared terms and cross-product terms,
\begin{align*}
(T-m)^2 \var(\widehat{\tau}_m) = \var \left( \sum_{k=0}^{K+1} \bbI_k \right) = \sum_{k=0}^{K+1} \bE\left[\bbI_k^2\right] + \sum_{0 \leq k < k' \leq K+1} 2 \bE\left[\bbI_k \bbI_{k'}\right].
\end{align*}
We focus on the variance of the squared terms first,
\begin{align*}
\bE\left[\bbI_k^2\right] = \left\{
\begin{aligned}
& \bar{Y}_0(\bm{1}_{m+1})^2 + \bar{Y}_0(\bm{0}_{m+1})^2 + 2 \bar{Y}_0(\bm{1}_{m+1}) \bar{Y}_0(\bm{0}_{m+1}), & & \text{if \ } k=0 \\
& 3 \bar{Y}_k(\bm{1}_{m+1})^2 + 3 \bar{Y}_k(\bm{0}_{m+1})^2 + 2 \bar{Y}_k(\bm{1}_{m+1}) \bar{Y}_k(\bm{0}_{m+1}), & & \text{if \ } 1 \leq k \leq K \\
& \bar{Y}_{K+1}(\bm{1}_{m+1})^2 + \bar{Y}_{K+1}(\bm{0}_{m+1})^2 + 2 \bar{Y}_{K+1}(\bm{1}_{m+1}) \bar{Y}_{K+1}(\bm{0}_{m+1}), & & \text{if \ } k=K+1 \\
\end{aligned}
\right.
\end{align*}
This is because when $k=0$ or $k=K+1$, then with probability $1/2$, $\bbI_k = \bar{Y}_0(\bm{1}_{m+1}) + \bar{Y}_0(\bm{0}_{m+1})$; with probability $1/2$, $\bbI_k = - \bar{Y}_0(\bm{1}_{m+1}) - \bar{Y}_0(\bm{0}_{m+1})$.
When $k\in[K]$, with probability $1/4$, $\bbI_k = 3 \bar{Y}_0(\bm{1}_{m+1}) + \bar{Y}_0(\bm{0}_{m+1})$; with probability $1/2$, $\bbI_k = - \bar{Y}_0(\bm{1}_{m+1}) + \bar{Y}_0(\bm{0}_{m+1})$; with probability $1/4$, $\bbI_k = - \bar{Y}_0(\bm{1}_{m+1}) -3 \bar{Y}_0(\bm{0}_{m+1})$.

Then for the cross-product terms, if $k' - k \geq 2$, then $\bbI_k$ and $\bbI_{k'}$ are independent, i.e., $\bE\left[\bbI_k \bbI_{k'}\right] = 0$.
If $k' - k = 1$, then
\begin{align*}
\bE\left[\bbI_k \bbI_{k+1}\right] = (\bar{Y}_{k}(\bm{1}_{m+1}) + \bar{Y}_{k}(\bm{0}_{m+1})) \cdot (\bar{Y}_{k+1}(\bm{1}_{m+1}) + \bar{Y}_{k+1}(\bm{0}_{m+1}))
\end{align*}
This is because the values of $\bbI_k$ and $\bbI_{k+1}$ are determined by the realization at $3$ randomization points, $W_{km+1}, W_{(k+1)m+1}, W_{(k+2)m+1}$.
With probability $1/8$, $\bbI_k \bbI_{k+1} = (3 \bar{Y}_{k}(\bm{1}_{m+1}) + \bar{Y}_{k}(\bm{0}_{m+1})) \cdot (3 \bar{Y}_{k+1}(\bm{1}_{m+1}) + \bar{Y}_{k+1}(\bm{0}_{m+1}))$;
with probability $1/8$, $\bbI_k \bbI_{k+1} = (3 \bar{Y}_{k}(\bm{1}_{m+1}) + \bar{Y}_{k}(\bm{0}_{m+1})) \cdot (- \bar{Y}_{k+1}(\bm{1}_{m+1}) + \bar{Y}_{k+1}(\bm{0}_{m+1}))$;
with probability $1/8$, $\bbI_k \bbI_{k+1} = (- \bar{Y}_{k}(\bm{1}_{m+1}) + \bar{Y}_{k}(\bm{0}_{m+1})) \cdot (3 \bar{Y}_{k+1}(\bm{1}_{m+1}) + \bar{Y}_{k+1}(\bm{0}_{m+1}))$;
with probability $1/8$, $\bbI_k \bbI_{k+1} = (- \bar{Y}_{k}(\bm{1}_{m+1}) + \bar{Y}_{k}(\bm{0}_{m+1})) \cdot (- \bar{Y}_{k+1}(\bm{1}_{m+1}) + \bar{Y}_{k+1}(\bm{0}_{m+1}))$;
with probability $1/8$, $\bbI_k \bbI_{k+1} = (- \bar{Y}_{k}(\bm{1}_{m+1}) + \bar{Y}_{k}(\bm{0}_{m+1})) \cdot (- \bar{Y}_{k+1}(\bm{1}_{m+1}) + \bar{Y}_{k+1}(\bm{0}_{m+1}))$;
with probability $1/8$, $\bbI_k \bbI_{k+1} = (- \bar{Y}_{k}(\bm{1}_{m+1}) + \bar{Y}_{k}(\bm{0}_{m+1})) \cdot (- \bar{Y}_{k+1}(\bm{1}_{m+1}) -3 \bar{Y}_{k+1}(\bm{0}_{m+1}))$;
with probability $1/8$, $\bbI_k \bbI_{k+1} = (- \bar{Y}_{k}(\bm{1}_{m+1}) -3 \bar{Y}_{k}(\bm{0}_{m+1})) \cdot (- \bar{Y}_{k+1}(\bm{1}_{m+1}) + \bar{Y}_{k+1}(\bm{0}_{m+1}))$;
with probability $1/8$, $\bbI_k \bbI_{k+1} = (- \bar{Y}_{k}(\bm{1}_{m+1}) -3 \bar{Y}_{k}(\bm{0}_{m+1})) \cdot (- \bar{Y}_{k+1}(\bm{1}_{m+1}) -3 \bar{Y}_{k+1}(\bm{0}_{m+1}))$.

Combining the squared terms and the cross-product terms we finish the proof.
\Halmos \endproof

\subsection{Discssions and proof of Corollary~\ref{coro:VarianceUpperBound}}
\label{sec:proof:coro:VarianceUpperBound}

We first provide the details of the two variance upper bounds here.
\begin{multline*}
\var^\mathsf{U1}(\widehat{\tau}_m) = \frac{1}{(T-m)^2} \left\{ 3 \left[ \bar{Y}_0(\bm{1}_{m+1})^2 + \bar{Y}_0(\bm{0}_{m+1})^2 \right] + \sum_{k=1}^{n-3} 6 \left[ \bar{Y}_k(\bm{1}_{m+1})^2 + \bar{Y}_k(\bm{0}_{m+1})^2 \right] \right.\\
+ \left. 4 \left[ \bar{Y}_{n-2}(\bm{1}_{m+1})^2 + \bar{Y}_{n-2}(\bm{0}_{m+1})^2 \right] + \sum_{k=0}^{n-3} 2 \left[ \bar{Y}_k(\bm{1}_{m+1}) \cdot \bar{Y}_{k+1}(\bm{1}_{m+1}) + \bar{Y}_k(\bm{0}_{m+1}) \cdot  \bar{Y}_{k+1}(\bm{0}_{m+1}) \right] \right\},
\end{multline*}
and
\begin{multline*}
\var^\mathsf{U2}(\widehat{\tau}_m) = \frac{1}{(T-m)^2} \left\{ 4 \left[ \bar{Y}_0(\bm{1}_{m+1})^2 + \bar{Y}_0(\bm{0}_{m+1})^2 \right] + \sum_{k=1}^{n-3} 8 \left[ \bar{Y}_k(\bm{1}_{m+1})^2 + \bar{Y}_k(\bm{0}_{m+1})^2 \right] \right. \\
+ \left. 4 \left[ \bar{Y}_{n-2}(\bm{1}_{m+1})^2 + \bar{Y}_{n-2}(\bm{0}_{m+1})^2 \right] \vphantom{\sum_{k=1}^{n-3}} \right\}.
\end{multline*}

We prove Corollary~\ref{coro:VarianceUpperBound} using the basic inequality that $2 x y \leq x^2 + y^2$.
Such an inequality is commonly used to find a conservative upper bound of the variance.

\proof{Proof of Corollary~\ref{coro:VarianceUpperBound}.}
From Lemma~\ref{lem:HTvariance}, the variance of the estimator is given by
\begin{align*}
& (T-m)^2 \var(\widehat{\tau}_m) \\
\leq & 2 \left\{ \bar{Y}_0(\bm{1}_{m+1})^2 + \bar{Y}_0(\bm{0}_{m+1})^2 \right\} + \sum_{k=1}^{n-3} 4 \left\{ \bar{Y}_k(\bm{1}_{m+1})^2 + \bar{Y}_k(\bm{0}_{m+1})^2 \right\} + 2 \left\{ \bar{Y}_{n-2}(\bm{1}_{m+1})^2 + \bar{Y}_{n-2}(\bm{0}_{m+1})^2 \right\} \\
& + \sum_{k=0}^{n-3} 2 \left[ \bar{Y}_k(\bm{1}_{m+1}) + \bar{Y}_k(\bm{0}_{m+1}) \right] \cdot \left[ \bar{Y}_{k+1}(\bm{1}_{m+1}) + \bar{Y}_{k+1}(\bm{0}_{m+1})\right] \\
\leq & 2 \left\{ \bar{Y}_0(\bm{1}_{m+1})^2 + \bar{Y}_0(\bm{0}_{m+1})^2 \right\} + \sum_{k=1}^{n-3} 4 \left\{ \bar{Y}_k(\bm{1}_{m+1})^2 + \bar{Y}_k(\bm{0}_{m+1})^2 \right\} + 2 \left\{ \bar{Y}_{n-2}(\bm{1}_{m+1})^2 + \bar{Y}_{n-2}(\bm{0}_{m+1})^2 \right\} \\
& + \sum_{k=0}^{n-3} \left\{ 2 \bar{Y}_k(\bm{1}_{m+1})\bar{Y}_{k+1}(\bm{1}_{m+1}) + 2 \bar{Y}_k(\bm{0}_{m+1})\bar{Y}_{k+1}(\bm{0}_{m+1}) + \bar{Y}_k(\bm{1}_{m+1})^2 + \bar{Y}_k(\bm{0}_{m+1})^2 + \bar{Y}_{k+1}(\bm{1}_{m+1})^2 + \bar{Y}_{k+1}(\bm{0}_{m+1})^2 \right\} \\
\leq & 3 \left\{ \bar{Y}_0(\bm{1}_{m+1})^2 + \bar{Y}_0(\bm{0}_{m+1})^2 \right\} + \sum_{k=1}^{n-3} 6 \left\{ \bar{Y}_k(\bm{1}_{m+1})^2 + \bar{Y}_k(\bm{0}_{m+1})^2 \right\} + 3 \left\{ \bar{Y}_{n-2}(\bm{1}_{m+1})^2 + \bar{Y}_{n-2}(\bm{0}_{m+1})^2 \right\} \\
& + \sum_{k=0}^{n-3} \left\{ \bar{Y}_k(\bm{1}_{m+1})^2 + \bar{Y}_k(\bm{0}_{m+1})^2 + \bar{Y}_{k+1}(\bm{1}_{m+1})^2 + \bar{Y}_{k+1}(\bm{0}_{m+1})^2 \right\} \\
= & 4 \left\{ \bar{Y}_0(\bm{1}_{m+1})^2 + \bar{Y}_0(\bm{0}_{m+1})^2 \right\} + \sum_{k=1}^{n-3} 8 \left\{ \bar{Y}_k(\bm{1}_{m+1})^2 + \bar{Y}_k(\bm{0}_{m+1})^2 \right\} + 4 \left\{ \bar{Y}_{n-2}(\bm{1}_{m+1})^2 + \bar{Y}_{n-2}(\bm{0}_{m+1})^2 \right\}
\end{align*}
where the first inequality suggests $\var(\widehat{\tau}_m) \leq \var^{\mathsf{U1}}(\widehat{\tau}_m)$, and the last inequality suggests $\var^{\mathsf{U1}}(\widehat{\tau}_m) \leq \var^{\mathsf{U2}}(\widehat{\tau}_m)$.

The unbiasedness part is due to the estimator of the variances being Horvitz-Thompson type estimators.
\Halmos \endproof

\subsection{Proof of Theorem~\ref{thm:AsymptoticNormality}}
\label{sec:proof:thm:AsymptoticNormality}

We prove Theorem~\ref{thm:AsymptoticNormality} by using Lemma~\ref{lem:RomanoWolfCLT}.
In particular, we derive $B_{n,k,a}^2$, and then construct some proper $\Delta_n, K_n$, and $L_n$.

\proof{Proof of Theorem~\ref{thm:AsymptoticNormality}.}
In the $n$-replica experiment, $\widehat{\tau}_m - \tau_m = \frac{1}{(n-1)m} \sum_{k=0}^{n-2} \bbI_k$, and $\var(\widehat{\tau}_m) = \frac{1}{(n-1)^2m^2} \var\left( \sum_{k=0}^{n-2} \bbI_k \right)$.
To use the language from Lemma~\ref{lem:RomanoWolfCLT}, denote $d=n-1$.
Denote for any $i \in [n-1]$, $X_{n,i} = \frac{1}{(n-1)m} \bbI_{i-1}$ so we know that $\phi = 1$, i.e., $\{X_{n,1}, X_{n,2}, ...\}$ is a sequence of $1$-dependent random variables.

First note that $B_{n}^2 = \var(\widehat{\tau}_m)$, and we calculate $B_{n,k,a}^2$ as follows.
\begin{align*}
B^2_{n,k,a} & = \frac{1}{(n-1)^2 m^2} \var\left( \sum_{i=a}^{a+k-1} \bbI_{i-1} \right) \\
& \leq \frac{1}{(n-1)^2 m^2} \left\{ \sum_{i=a}^{a+k-1} \left[ 3 \bar{Y}_{i-1}(\bm{1}_{m+1})^2 + 3 \bar{Y}_{i-1}(\bm{0}_{m+1})^2 + 2 \bar{Y}_{i-1}(\bm{1}_{m+1}) \bar{Y}_{i-1}(\bm{0}_{m+1}) \right]\right. \\
& \qquad \qquad \qquad \quad + \left. \sum_{i=a}^{a+k-2} 2 [\bar{Y}_{i-1}(\bm{1}_{m+1}) + \bar{Y}_{i-1}(\bm{0}_{m+1})] \cdot [\bar{Y}_{i}(\bm{1}_{m+1}) + \bar{Y}_{i}(\bm{0}_{m+1})] \right\} \\
& \leq \frac{8 k m^2 B^2 + 8 (k-1) m^2 B^2}{(n-1)^2 m^2} \\
& \leq \frac{16 k B^2}{(n-1)^2}
\end{align*}

Pick $\gamma = 0, \delta = 1$, then $\Delta_n = B^3 / (n-1)^3$, $K_n = 16B^2 / (n-1)^2$, and $L_n = \var(\widehat{\tau}_m) / (n-1)$.

We check that all the five conditions from Lemma~\ref{lem:RomanoWolfCLT} are satisfied.
\begin{enumerate}
\item $\bE\left| X_{n,i} \right|^3 \leq \Delta_n = B^3 / (n-1)^3$, because all the potential outcomes are bounded by $B$, so that $X_{n,i} \leq B/(n-1)$.
\item $B^2_{n,k,a}/ k \leq K_n = 16B^2 / (n-1)^2$.
\item $B_n^2/(n-1) \geq L_n = \var(\widehat{\tau}_m) / (n-1)$.
\item $K_n / L_n = 16B^2 / (n-1) \var(\widehat{\tau}_m) = O(1)$, where the last equality is due to Assumption~\ref{asp:NonNegligibleVariance}.
\item $\Delta_n / L_n^{3/2} = B^3 / (n-1)^{3/2} \var(\widehat{\tau}_m)^{3/2} = O(1)$, where the last equality is due to Assumption~\ref{asp:NonNegligibleVariance}.
\end{enumerate}

Due to Lemma~\ref{lem:RomanoWolfCLT},
\begin{align*}
\frac{\widehat{\tau}_m - \tau_m}{\sqrt{\var(\widehat{\tau}_m)}} \xrightarrow[]{D} \mathcal{N}(0,1).
\end{align*}
\Halmos \endproof

\subsection{Interpretation for the Horvitz-Thompson Estimator under Misspecified $m$ Case}
\label{sec:FurtherMisspecified}

For the remainder of this section, we discuss the cases when $m$ is misspecified.
Throughout this section in the appendix, we use both $p$ and $m$.
Recall that $m$ is the order of the carryover effect, and $p$ is the experimenter's knowledge of $m$.

As we have discussed in Section~\ref{sec:Misspecifiedm}, all our estimation and inference methods will hold when $p \geq m$. When $p<m$, the Horvitz-Thompson estimator as we defined in \eqref{eqn:estimator} will no longer be unbiased in estimating the lag-$p$ causal estimand as we defined in \eqref{eqn:estimand}. However, we can still interpret the Horvitz-Thompson estimator as we defined in \eqref{eqn:estimator}.

When $p<m$, the lag-$p$ effect in \eqref{eqn:estimand} is not well defined. Instead, we define the $m$-misspecified lag-$p$ causal effect that pads the $p+1$ assignments with the earlier observed treatments. 
\begin{multline}
\label{eqn:estimandgeneral}
\tau_p^{(m)}(\bY) = \frac{1}{T-p} \left\{\sum_{t=p+1}^m \left[ Y_t(\bm{w}^\obs_{1:t-p-1}, \bm{1}_{p+1}) - Y_t(\bm{w}^\obs_{1:t-p-1}, \bm{0}_{p+1}) \right] + \right. \\
\left. \sum_{t=m+1}^{T} \left[ Y_t(\bm{w}_{t-m:t-p-1}^{\obs},\bm{1}_{p+1}) - Y_t(\bm{w}_{t-m:t-p-1}^{\obs},\bm{0}_{p+1}) \right] \right\}.
\end{multline}
This is a special case of the weighted lag-$p$ causal effect introduced in \cite{bojinov2019time}. Similarly to the average lag-$p$ causal effect, $\tau_p^{(m)}(\bY)$ captures how administering $p+1$ consecutive treatments as opposed to $p+1$ consecutive controls impact the outcomes at time $t$, \emph{conditional on} the observed assignment path up to time $t-p-1$.\footnote{See \cite[Section~3]{bojinov2019time} for an extended discussion.}
See Section~\ref{sec:simu:misspecifiedm} for numerical results.

When $p>m$, Proposition~\ref{thm:HTUnbiased} still holds, i.e., $\bE[\widehat{\tau}_p] = \tau_p(\bY) = \tau_m(\bY)$.
When $p<m$, sometimes we have to slightly augment the results and study the conditional expectation.

Define $f_\bT: [T] \to \bT$ to be the ``determining randomization point of period $t$,''
$$f_\bT(t) = \max \left\{ j \left| j \in \bT, j \leq t \right. \right\}$$
such that, it is the realization at time $f_\bT(t)$ that uniquely determines the assignment at time $t$, \emph{i.e.} $W_t = W_{f_\bT(t)}, \forall t \in [T]$.
See Example~\ref{exa:misspecifiedm} for an illustration of $f_\bT(\cdot)$.
When $\bT$ is clear from the context we drop the subscript and use $f(\cdot) = f_{\bT}(\cdot)$.
Depending on if $f(t-p) \leq t-m$, we establish an analogy of Proposition~\ref{thm:HTUnbiased} for the $p<m$ case.

\begin{proposition}
[Conditional Unbiasedness of the Estimator when $m$ is Misspecified]
\label{prop:HTConditionallyUnbiased}
Under Assumptions~\ref{asp:nonanticipating} and~\ref{asp:nomcarryover}, for $p<m$, at each time $t \geq m+1$, the Horvitz-Thompson estimator is either unbiased for the lag-$m$ causal effect when $f(t-p) \leq t-m$, or conditionally unbiased for the $m$-misspecified lag-$p$ causal effect when $f(t-p) > t-m$.
When $p+1 \leq t \leq m$, the Horvitz-Thompson estimator is either unbiased for the lag-$t$ causal effect when $f(t-p) = 1$, or conditionally unbiased for the $m$-misspecified lag-$t$ causal effect when $f(t-p) > 1$.
\end{proposition}
To remove the conditional expectation, we can further take an outer loop of expectation averaged over the past assignment paths.
Although this is somewhat different from the average lag-$p$ effect introduced earlier in \eqref{eqn:estimand}, it does capture the impact of a sequence of treatment relative to a sequence of controls.

All the mathematical expressions of Proposition~\ref{prop:HTConditionallyUnbiased}, as well its proof, are stated in Section~\ref{sec:proof:prop:HTConditionallyUnbiased} in the Appendix.
See Example~\ref{exa:misspecifiedm} below for a specific illustration of Proposition~\ref{prop:HTConditionallyUnbiased}.
For a numerical illustration of the estimand and estimator in more general setups, see Section~\ref{sec:simu:misspecifiedm}.

\example
[Misspecified $m$]
\label{exa:misspecifiedm}
Suppose $T=4, m=2, p=1, \bT=\{1,3\}$. Then the determining randomization points are $f_\bT(1) = 1, f_\bT(2) = 1, f_\bT(3) = 3, f_\bT(4) = 3$, and
\begin{align*}
\bE\left[ Y_2^{\obs} \frac{\bI{\{\bm{W}_{1:2} = (1,1)\}}}{\Pr(\bm{W}_{1:2} = (1,1))} - Y_2^{\obs} \frac{\bI{\{\bm{W}_{1:2} = (0,0)\}}}{{\Pr(\bm{W}_{1:2} = (0,0))}}  \right] & = Y_2(1,1) - Y_2(0,0) \\
\bE\left[ Y_3^{\obs} \frac{\bI{\{\bm{W}_{2:3} = (1,1)\}}}{\Pr(\bm{W}_{2:3} = (1,1))} - Y_3^{\obs} \frac{\bI{\{\bm{W}_{2:3} = (0,0)\}}}{{\Pr(\bm{W}_{2:3} = (0,0))}}  \right] & = Y_3(1,1,1) - Y_3(0,0,0) \\
\bE\left[ Y_4^{\obs} \frac{\bI{\{\bm{W}_{3:4} = (1,1)\}}}{\Pr(\bm{W}_{3:4} = (1,1))} - Y_4^{\obs} \frac{\bI{\{\bm{W}_{3:4} = (0,0)\}}}{{\Pr(\bm{W}_{3:4} = (0,0))}}  \right] & = \frac{1}{2}\left[ Y_4(1,1,1) + Y_4(0,1,1) - Y_4(0,0,0) - Y_4(1,0,0) \right]
\end{align*}
Note that this is the $2$-misspecified lag-$1$ causal effect.
\Halmos \endexample

\subsection{Unbiasedness of the Horvitz-Thompson Estimator when $m$ is Misspecified}
\label{sec:proof:prop:HTConditionallyUnbiased}

We state here the omitted mathematics in Proposition~\ref{prop:HTConditionallyUnbiased}.

Under Assumptions~\ref{asp:nonanticipating} and~\ref{asp:nomcarryover}, for $p<m$, at each time $t \geq m+1$, the Horvitz-Thompson estimator is either unbiased for the lag-$m$ causal effect when $f(t-p) \leq t-m$, i.e.,
\begin{align*}
\bE_{\bm{W}_{1:T} \sim \eta_{\bT, \bQ}}\left[ Y_t^{\obs} \frac{\bI{\{\bm{W}_{t-p:t} = \bm{1}_{p+1}\}}}{\Pr(\bm{W}_{t-p:t} = \bm{1}_{p+1})} - Y_t^{\obs} \frac{\bI{\{\bm{W}_{t-p:t} = \bm{0}_{p+1}\}}}{{\Pr(\bm{W}_{t-p:t} = \bm{0}_{p+1})}}  \right] = Y_t(\bm{1}_{m+1}) - Y_t(\bm{0}_{m+1}),
\end{align*}
or conditionally unbiased for the $m$-misspecified lag-$p$ causal effect when $f(t-p) > t-m$, i.e.,
\begin{multline*}
\bE_{\bm{W}_{1:T} \sim \eta_{\bT, \bQ}}\left[ \left\{Y_t^{\obs} \frac{\bI{\{\bm{W}_{t-p:t} = \bm{1}_{p+1}\}}}{\Pr(\bm{W}_{t-p:t} = \bm{1}_{p+1})} - Y_t^{\obs} \frac{\bI{\{\bm{W}_{t-p:t} = \bm{0}_{p+1}\}}}{{\Pr(\bm{W}_{t-p:t} = \bm{0}_{p+1})}}  \right\} - \right. \\
\left. \left\{ Y_t(\bm{w}_{t-m:f(t-p)-1}^{\obs},\bm{1}_{t-f(t-p)+1}) - Y_t(\bm{w}_{t-m:f(t-p)-1}^{\obs},\bm{0}_{t-f(t-p)+1}) \right\} \left| \bm{W}_{t-m:f(t-p)-1} = \bm{w}^{\obs}_{t-m:f(t-p)-1} \vphantom{\frac{\bI{\{\bm{W}_{t-p:t} = \bm{1}_{p+1}\}}}{\Pr(\bm{W}_{t-p:t} = \bm{1}_{p+1})}} \right. \right] = 0. 
\end{multline*}
When $p+1 \leq t \leq m$, the Horvitz-Thompson estimator is either unbiased for the lag-$t$ causal effect when $f(t-p) = 1$, i.e.,
\begin{align*}
\bE_{\bm{W}_{1:T} \sim \eta_{\bT, \bQ}}\left[ Y_t^{\obs} \frac{\bI{\{\bm{W}_{t-p:t} = \bm{1}_{p+1}\}}}{\Pr(\bm{W}_{t-p:t} = \bm{1}_{p+1})} - Y_t^{\obs} \frac{\bI{\{\bm{W}_{t-p:t} = \bm{0}_{p+1}\}}}{{\Pr(\bm{W}_{t-p:t} = \bm{0}_{p+1})}}  \right] = Y_t(\bm{1}_{t}) - Y_t(\bm{0}_{t}),
\end{align*}
or conditionally unbiased for the $m$-misspecified lag-$t$ causal effect when $f(t-p) > 1$, i.e.,
\begin{multline*}
\bE_{\bm{W}_{1:T} \sim \eta_{\bT, \bQ}}\left[ \left\{Y_t^{\obs} \frac{\bI{\{\bm{W}_{t-p:t} = \bm{1}_{p+1}\}}}{\Pr(\bm{W}_{t-p:t} = \bm{1}_{p+1})} - Y_t^{\obs} \frac{\bI{\{\bm{W}_{t-p:t} = \bm{0}_{p+1}\}}}{{\Pr(\bm{W}_{t-p:t} = \bm{0}_{p+1})}}  \right\} - \right. \\
\left. \left\{ Y_t(\bm{w}_{1:f(t-p)-1}^{\obs},\bm{1}_{t-f(t-p)+1}) - Y_t(\bm{w}_{1:f(t-p)-1}^{\obs},\bm{0}_{t-f(t-p)+1}) \right\} \left| \bm{W}_{1:f(t-p)-1} = \bm{w}^{\obs}_{1:f(t-p)-1} \vphantom{\frac{\bI{\{\bm{W}_{t-p:t} = \bm{1}_{p+1}\}}}{\Pr(\bm{W}_{t-p:t} = \bm{1}_{p+1})}} \right. \right] = 0. 
\end{multline*}

To remove the conditional expectation, we can further take an outer loop of expectation averaged over the past assignment paths.
So the estimator is estimating a weighted average of lag-$p$ effects.
When $t \geq m+1$,
\begin{align*}
\sum_{\bm{w}_{t-m:f(t-p)-1}} \Pr(\bm{W}_{t-m:f(t-p)-1} = \bm{w}_{t-m:f(t-p)-1}) (Y_t(\bm{w}_{t-m:f(t-p)-1},\bm{1}_{t-f(t-p)+1}) - Y_t(\bm{w}_{t-m:f(t-p)-1},\bm{0}_{t-f(t-p)+1})), \end{align*}
and when $p+1 \leq t \leq m$,
\begin{align*}
\sum_{\bm{w}_{1:f(t-p)-1}} \Pr(\bm{W}_{1:f(t-p)-1} = \bm{w}_{1:f(t-p)-1}) (Y_t(\bm{w}_{1:f(t-p)-1},\bm{1}_{t-f(t-p)+1}) - Y_t(\bm{w}_{1:f(t-p)-1},\bm{0}_{t-f(t-p)+1})). 
\end{align*}

We prove Proposition~\ref{prop:HTConditionallyUnbiased} as follows.

\proof{Proof of Proposition~\ref{prop:HTConditionallyUnbiased}.}
Focus on any specific $t \in \{m+1:T\}$.

When $f(t-p) \leq t-m$, both $0 < \Pr(\bm{W}_{t-p:t} = \bm{1}_{p+1}), \Pr(\bm{W}_{t-p:t} = \bm{0}_{p+1}) < 1$.
With probability $\Pr(\bm{W}_{t-p:t} = \bm{1}_{p+1}) \ne 0$, $\bI{\{\bm{W}_{t-p:t} = \bm{1}_{p+1}\}} = 1$, and $Y_t^{\obs} = Y_t(\bm{1}_{m+1})$.
So $\bE\left[ Y_t^{\obs} \frac{\bI{\{\bm{W}_{t-p:t} = \bm{1}_{p+1}\}}}{\Pr(\bm{W}_{t-p:t} = \bm{1}_{p+1})} \right] = Y_t(\bm{1}_{m+1})$.
Similarly $\bE\left[ Y_t^{\obs} \frac{\bI{\{\bm{W}_{t-p:t} = \bm{0}_{p+1}\}}}{\Pr(\bm{W}_{t-p:t} = \bm{0}_{p+1})} \right] = Y_t(\bm{0}_{m+1})$.
So
\begin{align*}
\bE_{\bm{W}_{1:T} \sim \eta_{\bT, \bQ}}\left[ \left\{Y_t^{\obs} \frac{\bI{\{\bm{W}_{t-p:t} = \bm{1}_{p+1}\}}}{\Pr(\bm{W}_{t-p:t} = \bm{1}_{p+1})} - Y_t^{\obs} \frac{\bI{\{\bm{W}_{t-p:t} = \bm{0}_{p+1}\}}}{{\Pr(\bm{W}_{t-p:t} = \bm{0}_{p+1})}}  \right\} \right] = Y_t(\bm{1}_{m+1}) - Y_t(\bm{0}_{m+1}).
\end{align*}

When $f(t-p) > t-m$, both $0 < \Pr\left(\bm{W}_{t-p:t} = \bm{1}_{p+1} \left| \bm{W}_{t-m:f(t-p)-1} = \bm{w}^{\obs}_{t-m:f(t-p)-1} \right. \right) < 1$ and $0 < \Pr\left(\bm{W}_{t-p:t} = \bm{0}_{p+1} \left| \bm{W}_{t-m:f(t-p)-1} = \bm{w}^{\obs}_{t-m:f(t-p)-1} \right.\right) < 1$.
Conditional on $\bm{W}_{t-m:f(t-p)-1} = \bm{w}^{\obs}_{t-m:f(t-p)-1}$,
we know that with probability $\Pr\left(\bm{W}_{t-p:t} = \bm{1}_{p+1} \left| \bm{W}_{t-m:f(t-p)-1} = \bm{w}^{\obs}_{t-m:f(t-p)-1} \right. \right) \ne 0$, $\bI{\{\bm{W}_{t-p:t} = \bm{1}_{p+1}\}} = 1$, and $Y_t^{\obs} = Y_t(\bm{w}^{\obs}_{t-m:f(t-p)-1},\bm{1}_{t-f(t-p)+1})$.
So $$\bE_{\bm{W}_{1:T} \sim \eta_{\bT, \bQ}} \left[ Y_t^{\obs} \frac{\bI{\{\bm{W}_{t-p:t} = \bm{1}_{p+1}\}}}{\Pr(\bm{W}_{t-p:t} = \bm{1}_{p+1})} - Y_t(\bm{w}_{t-m:f(t-p)-1}^{\obs},\bm{1}_{t-f(t-p)+1}) \left| \bm{W}_{t-m:f(t-p)-1} = \bm{w}^{\obs}_{t-m:f(t-p)-1} \right. \right] = 0.$$
Similarly, we have
$$\bE_{\bm{W}_{1:T} \sim \eta_{\bT, \bQ}} \left[ Y_t^{\obs} \frac{\bI{\{\bm{W}_{t-p:t} = \bm{0}_{p+1}\}}}{\Pr(\bm{W}_{t-p:t} = \bm{0}_{p+1})} - Y_t(\bm{w}_{t-m:f(t-p)-1}^{\obs},\bm{0}_{t-f(t-p)+1}) \left| \bm{W}_{t-m:f(t-p)-1} = \bm{w}^{\obs}_{t-m:f(t-p)-1} \right. \right] = 0,$$
which finishes the proof.
\Halmos \endproof

\subsection{Asymptotic Normality when $m$ is Misspecified}
\label{sec:proof:coro:AsymptoticNormality}

The proof of Corollary~\ref{coro:MisspecifiedmCLT} consists of two parts: $p>m$ and $p<m$. 
When $p>m$ we consult Theorems~\ref{lem:HTvariance} and~\ref{thm:AsymptoticNormality}.
When $p<m$ we prove Corollary~\ref{coro:MisspecifiedmCLT} by using Lemma~\ref{lem:RomanoWolfCLT}.
In particular, we derive $B_{n,k,a}^2$, and then construct some proper $\Delta_n, K_n$, and $L_n$.

\proof{Proof of Corollary~\ref{coro:MisspecifiedmCLT}.}
The proof consists of two parts: $p>m$ and $p<m$. 
First, when $p>m$, we know that $\widehat{\tau}_p = \widehat{\tau}_m, \tau_p = \tau_m, \var(\widehat{\tau}_p) = \var(\widehat{\tau}_m)$.
Due to Theorems~\ref{lem:HTvariance} we prove part (i) the expression in \eqref{eqn:HTVariance}. Due to Theorem~\ref{thm:AsymptoticNormality} we know that
\begin{align*}
\frac{\widehat{\tau}_p - \tau_p}{\sqrt{\var(\widehat{\tau}_p)}} = \frac{\widehat{\tau}_m - \tau_m}{\sqrt{\var(\widehat{\tau}_m)}} \xrightarrow[]{D} \mathcal{N}(0,1).
\end{align*}

Second, when $p<m$, then we follow the same trick as in Theorem~\ref{thm:AsymptoticNormality}.
In the $n$-replica experiment, $\widehat{\tau}_p - \bE[\tau^{[m]}_p] = \frac{1}{(n-1)p} \sum_{k=0}^{n-2} \bbI_k$, and $\var(\widehat{\tau}_p) = \frac{1}{(n-1)^2p^2} \var\left( \sum_{k=0}^{n-2} \bbI_k \right)$.
To use the language from Lemma~\ref{lem:RomanoWolfCLT}, denote $d=n-1$.
Denote for any $i \in [n-1]$, $X_{n,i} = \frac{1}{(n-1)p} \bbI_{i-1}$.
We know that $\phi = \lceil \frac{m}{p} \rceil$, so that $\{X_{n,1}, X_{n,2}, ...\}$ is a sequence of $\phi$-dependent random variables. See Table~\ref{tbl:phiIllustration} for an illustration of $\phi$.

\begin{table}[!thb]
\TABLE{An illustration of $\phi$ when $m=5, p=3$. \label{tbl:phiIllustration}}
{
\begin{tabular}{p{0.8cm} p{0.8cm} p{0.8cm} p{0.8cm} p{0.8cm} p{0.8cm} p{0.8cm} p{0.8cm} p{0.8cm} p{0.8cm} p{0.8cm} p{0.8cm} p{0.8cm} p{0.8cm} p{0.8cm}}
&  &  &  &  &  & \multicolumn{6}{p{4.8cm}}{$\xrightarrow[\text{carryover \  effect}]{\hspace*{5.4cm}}$} &  &  &  \\ \hline
\multicolumn{1}{|p{0.8cm}|}{} & \multicolumn{1}{p{0.8cm}|}{$\hdots$} & \multicolumn{1}{p{0.8cm}|}{13}           & \multicolumn{1}{p{0.8cm}|}{14}  & \multicolumn{1}{p{0.8cm}|}{15}  & \multicolumn{1}{p{0.8cm}|}{16} & \multicolumn{1}{p{0.8cm}|}{17} & \multicolumn{1}{p{0.8cm}|}{18} & \multicolumn{1}{p{0.8cm}|}{19} & \multicolumn{1}{p{0.8cm}|}{20} & \multicolumn{1}{p{0.8cm}|}{21} & \multicolumn{1}{p{0.8cm}|}{22} & \multicolumn{1}{p{0.8cm}|}{23} & \multicolumn{1}{p{0.8cm}|}{24} & \multicolumn{1}{p{0.8cm}|}{$\hdots$} \\ \hline
\multicolumn{1}{|p{0.8cm}|}{$\bT^*$} & \multicolumn{1}{p{0.8cm}|}{}         & \multicolumn{1}{p{0.8cm}|}{$\checkmark$} & \multicolumn{1}{p{0.8cm}|}{$-$} & \multicolumn{1}{p{0.8cm}|}{$-$} & \multicolumn{1}{p{0.8cm}|}{$\checkmark$} & \multicolumn{1}{p{0.8cm}|}{$-$} & \multicolumn{1}{p{0.8cm}|}{$-$} & \multicolumn{1}{p{0.8cm}|}{$\checkmark$} & \multicolumn{1}{p{0.8cm}|}{$-$} & \multicolumn{1}{p{0.8cm}|}{$-$} & \multicolumn{1}{p{0.8cm}|}{$\checkmark$} & \multicolumn{1}{p{0.8cm}|}{$-$} & \multicolumn{1}{p{0.8cm}|}{$-$} & \multicolumn{1}{p{0.8cm}|}{}         \\ \hline
\multicolumn{1}{|l|}{$\{\bbI_k\}_{k=0}^{K+1}$} & \multicolumn{1}{l|}{}         & \multicolumn{3}{l|}{$\bbI_{3}$} & \multicolumn{3}{l|}{$\bbI_{4}$} & \multicolumn{3}{l|}{$\bbI_{5}$} & \multicolumn{3}{l|}{$\bbI_{6}$} & \multicolumn{1}{l|}{}         \\ \hline
\end{tabular}
}
{In this example $\phi = \lceil \frac{m}{p} \rceil = 2$. The arrow above numbers $17$ through $22$ means that the assignment on period $17$ affects the outcome on period $22$. So that $\bbI_4$ and $\bbI_6$ are correlated, but $\bbI_3$ and $\bbI_6$ are independent.}
\end{table}

First note that $B_{n}^2 = \var(\widehat{\tau}_p)$, and we calculate $B_{n,k,a}^2$ as follows. Note that $k \geq \phi+1$.
\begin{align*}
B^2_{n,k,a} & = \frac{1}{(n-1)^2 p^2} \var\left( \sum_{i=a}^{a+k-1} \bbI_{i-1} \right) \\
& \leq \frac{1}{(n-1)^2 p^2} \left( \sum_{i=a}^{a+k-1} \bE[\bbI_{i-1}^2] + \sum_{i=a}^{a+k-2} 2 \bE[\bbI_{i-1}\bbI_{i}] + ... + \sum_{i=a}^{a+k-1+\phi} 2 \bE[\bbI_{i-1}\bbI_{i-1+\phi}] \right)\\
& \leq \frac{C p^2 B^2}{(n-1)^2 p^2} \cdot \left( k+(k-1)+...+(k-\phi) \right)\\
& \leq \frac{(\phi+1) C k B^2}{(n-1)^2}
\end{align*}
where $C$ is some constant bounding the number of terms in each cross-product expectation $2 \bE[\bbI_{i-1}\bbI_{i}], ..., 2 \bE[\bbI_{i-1}\bbI_{i-1+\phi}]$; and $\phi+1$ is a constant as well.

Pick $\gamma = 0, \delta = 1$, then $\Delta_n = B^3 / (n-1)^3$, $K_n = (\phi+1) C B^2 / (n-1)^2$, and $L_n = \var(\widehat{\tau}_m) / (n-1)$.

We check that all the five conditions from Lemma~\ref{lem:RomanoWolfCLT} are satisfied.
\begin{enumerate}
\item $\bE\left| X_{n,i} \right|^3 \leq \Delta_n = B^3 / (n-1)^3$, because all the potential outcomes are bounded by $B$, so that $X_{n,i} \leq B/(n-1)$.
\item $B^2_{n,k,a}/ k \leq K_n = (\phi+1) C B^2 / (n-1)^2$.
\item $B_n^2/(n-1) \geq L_n = \var(\widehat{\tau}_m) / (n-1)$.
\item $K_n / L_n = (\phi+1) C B^2 / (n-1) \var(\widehat{\tau}_m) = O(1)$, where the last equality is due to Assumption~\ref{asp:NonNegligibleVariance}.
\item $\Delta_n / L_n^{3/2} = B^3 / (n-1)^{3/2} \var(\widehat{\tau}_m)^{3/2} = O(1)$, where the last equality is due to Assumption~\ref{asp:NonNegligibleVariance}.
\end{enumerate}

Due to Lemma~\ref{lem:RomanoWolfCLT},
\begin{align*}
\frac{\widehat{\tau}_p - \tau_p}{\sqrt{\var(\widehat{\tau}_p)}} \xrightarrow[]{D} \mathcal{N}(0,1).
\end{align*}

\Halmos \endproof


\section{Additional Simulation Results}

\subsection{Flexibility of the Outcome Models}
\label{sec:simu:Appendix:OutcomeModels}
As we will see below, it is easy to use the potential outcome framework to describe many complex relationships between assignments and outcomes.

We start with a simple model which originates from \citet{oman1988switch}:
\begin{align}
Y_t(\bm{w}_{1:t}) = \mu + \alpha_t + \delta w_t + \gamma w_{t-1} + \epsilon_t \label{eqn:simu:SimpleOutcomeModel}
\end{align}
where $\mu$ is a fixed effect; $\alpha_t$ is a fixed effect associated to period $t$; $\delta w_t$ is the contemporaneous effect, and $\gamma w_{t-1}$ is the carryover effect from period $t-1$; $\epsilon_t$ is the random noise in period $t$.
Such a model as well as a few very similar ones are widely used in the literature \citep{hedayat1978repeated, jones2014design}.

A more general variant from the above model is to consider carryover effects of any arbitrary order, which we have defined in \eqref{eqn:simu:MoreGeneral} in the main body of the paper.
\begin{align*}
Y_t(\bm{w}_{1:t}) = \mu + \alpha_t + \delta^{(1)} w_t + \delta^{(2)} w_{t-1} + ... + \delta^{(t)} w_1 + \epsilon_t
\end{align*}
where $\delta^{(1)}, \delta^{(2)}, ..., \delta^{(t)}$ are non-stochastic coefficients. The dotted terms are carryover effects of higher orders. And all the other parameters are as defined in \eqref{eqn:simu:SimpleOutcomeModel}.
We will run simulations based on this more general model, which enables us to test the performance of our proposed optimal design under a misspecified $m$.

The autoregressive model \citep{arellano2003panel} is even more general: $Y_1(w_{1}) = \delta_{1,1} w_1 + \epsilon_1$ and $\forall t > 1$
\begin{multline}
Y_t(\bm{w}_{1:t}) = \phi_{t,t-1} Y_{t-1}(\bm{w}_{1:t-1}) + \phi_{t,t-2} Y_{t-2}(\bm{w}_{1:t-2}) + ... + \phi_{t,1} Y_{1}(w_{1}) + \\
\delta_{t,t} w_t + \delta_{t,t-1} w_{t-1} + ... + \delta_{t,1} w_1 + \epsilon_t \label{eqn:simu:Autoregressive}
\end{multline}
where $\phi_{t,\ttt}$ and $\delta_{t,\ttt}$ are non-stochastic coefficients; the dotted terms are carryover effects of higher orders; $\epsilon_t$ is the random noise in period $t$.
We can iteratively replace $Y_t(w_{t})$ using a linear combination of $w_t, w_{t-1}, ..., w_1$.
So the autoregressive model in \eqref{eqn:simu:Autoregressive} can be written in a similar form of \eqref{eqn:simu:MoreGeneral}.
The only difference is that the coefficients are different and dependent on $t$.

\subsection{Comparison of Different Designs when Estimating Instantaneous Effects and Other Lag Effects}
\label{sec:simu:DifferentEstimands}

\subsubsection{Simulation setup.}
We consider a similar setup as in Section~\ref{sec:simu:RiskFunctions}.
We run simulations based on the outcome model as in \eqref{eqn:simu:MoreGeneral}.
We consider $T=120, p=m=2$ where $m$ is correctly identified.
For the outcome model, we consider $\mu = 0$, $\alpha_t = \log{(t)}$, and $\epsilon_t \sim N(0,1)$ are i.i.d. standard normal distributions.
For any $t >3$, let $\delta^{(t)} = 0$.
We will vary the values of $\delta^{(1)}, \delta^{(2)}, \delta^{(3)} \in \{1, 2\}$ and conduct experiments under $2^3=8$ different scenarios.

Different from Section~\ref{sec:simu:RiskFunctions}, instead of estimating the average lag-p causal effect as defined in \eqref{eqn:estimand}, we estimate the following family of causal effects.
For any non-negative integers $p,q$, define
\begin{align}
\label{eqn:OtherEstimands}
\tau_{p,q}(\bY) = \frac{1}{T-p} \sum_{t=p+1}^{T} [Y_t(\bm{0}_{p-q}, \bm{1}_{q+1}) - Y_t(\bm{0}_{p+1})].
\end{align}
Such a family of causal effects are already studied in the literature \citep{bojinov2017}.
When $q=0$, $\tau_{p,0}(\bY)$ is also known as the instantaneous treatment effect.
When $q=p$, $\tau_{p,p}(\bY)$ is the average lag-p causal effect as defined in \eqref{eqn:estimand}.
Since our belief is that the carryover effect is of order $p$, there is no reason we would like to estimate $\tau_{p,q}$ when $q>p$.

Similar to Section~\ref{sec:simu:RiskFunctions}, we compare the same three different designs of switchback experiments.
$\bT^*=\{1,5,7,...,117\}, \bT^\mathsf{H1}=\{1,2,3,...,120\}$, and $\bT^\mathsf{H2}=\{1,4,7,...,118\}$.
Although the primary purpose of running any of these three experiments is to estimate the average lag-p causal effect as defined in \eqref{eqn:estimand}, after running the experiment we can use the observed data to estimate the causal effects as defined in \eqref{eqn:OtherEstimands}.

In order to estimate such a family of causal effects, we again use the Horvitz-Thompson estimator.
However, when we use the optimal design as suggested by Theorem~\ref{thm:OptimalDesign}, on some periods we would have zero probability to observe $Y_t(\bm{1}_{p-q}, \bm{1}_{q+1})$.
This is because due to Theorem~\ref{thm:OptimalDesign}, we only randomize every $p$ periods.
Therefore, we define
\begin{align*}
\mathcal{T}_{p,q}(\eta_{\bT, \bQ}) = \left\{ t \left| \Pr(\bm{W}_{t-p:t} = (\bm{0}_{p-q},\bm{1}_{q+1})) \ne 0 \right.\right\}
\end{align*}
and then use the set $\mathcal{T}_{p,q}(\eta_{\bT, \bQ})$ to define the Horvitz-Thompson estimator,
\begin{multline}
\label{eqn:OtherEstimators}
\widehat{\tau}_{p,q} (\eta_{\bT, \bQ}, \bm{w}_{1:T}, \bY) = \\
\frac{1}{T-p} \sum_{\substack{{p+1\leq t \leq T} \\ t \in \mathcal{T}_{p,q}(\eta_{\bT, \bQ})}} \left\{ Y_t^{\obs} \frac{\bI{\{\bm{w}_{t-p:t} = (\bm{0}_{p-q},\bm{1}_{q+1})\}}}{\Pr(\bm{W}_{t-p:t} = (\bm{0}_{p-q},\bm{1}_{q+1}))} - Y_t^{\obs} \frac{\bI{\{\bm{w}_{t-p:t} = \bm{0}_{p+1}\}}}{{\Pr(\bm{W}_{t-p:t} = \bm{0}_{p+1})}} \right\}.
\end{multline}

We simulate one assignment path at a time, and conduct an experiment following this assignment path.
Since the outcome model is prescribed, we can calculate both the causal estimand and and the observed outcomes (along the simulated assignment path).
Then, we calculate the Horvitz-Thompson estimator based on the simulated assignment path and the simulated observed outcomes.
With both the estimand and estimator, we can calculate the loss function.
We repeat the above procedure enough ($100000$) times to obtain an accurate approximation of the risk function.

\begin{table}[!tb]
\TABLE{Simulation results for the risk function when estimating instantaneous effects and other lag effects. \label{tbl:simu:RIskFunctions:OtherEstimands}}
{
\begin{tabular}{| p{1cm} | p{1cm} | p{1cm} | p{1cm} | p{1.5cm} | p{1.5cm} | p{1.5cm} | p{1.5cm} | p{1.5cm} | p{1.5cm} |}
\multicolumn{10}{l}{$q=0$}\\
\hline
$\delta^{(1)}$ & $\delta^{(2)}$ & $\delta^{(3)}$ & $\tau_{2,0}$ & $\bE[\widehat{\tau}^*_{2,0}]$ & $\bE[\widehat{\tau}^\mathsf{H1}_{2,0}]$ & $\bE[\widehat{\tau}^\mathsf{H2}_{2,0}]$ & $r(\eta_{\bT^*})$ & $r(\eta_{\bT^\mathsf{H1}})$ & $r(\eta_{\bT^\mathsf{H2}})$ \\ \hline
1 & 1 & 1 & 1 & 0.998 & 1.002 & 1.001 & 2.66 & 2.35 & 3.71 \\ \hline
1 & 1 & 2 & 1 & 0.998 & 1.002 & 1.001 & 2.66 & 2.35 & 3.71 \\ \hline
1 & 2 & 1 & 1 & 0.998 & 1.002 & 1.001 & 2.66 & 2.35 & 3.71 \\ \hline
2 & 1 & 1 & 2 & 1.998 & 2.002 & 2.001 & 3.00 & 2.56 & 4.20 \\ \hline
1 & 2 & 2 & 1 & 0.998 & 1.002 & 1.001 & 2.66 & 2.35 & 3.71 \\ \hline
2 & 1 & 2 & 2 & 1.998 & 2.002 & 2.001 & 3.00 & 2.56 & 4.20 \\ \hline
2 & 2 & 1 & 2 & 1.998 & 2.002 & 2.001 & 3.00 & 2.56 & 4.20 \\ \hline
2 & 2 & 2 & 2 & 1.998 & 2.002 & 2.001 & 3.00 & 2.56 & 4.20 \\ \hline
\multicolumn{10}{l}{$q=1$}\\
\hline
$\delta^{(1)}$ & $\delta^{(2)}$ & $\delta^{(3)}$ & $\tau_{2,1}$ & $\bE[\widehat{\tau}^*_{2,1}]$ & $\bE[\widehat{\tau}^\mathsf{H1}_{2,1}]$ & $\bE[\widehat{\tau}^\mathsf{H2}_{2,1}]$ & $r(\eta_{\bT^*})$ & $r(\eta_{\bT^\mathsf{H1}})$ & $r(\eta_{\bT^\mathsf{H2}})$ \\ \hline
1 & 1 & 1 & 2 & 1.999 & 2.000 & 2.005 & 2.85 & 4.08 & 3.90 \\ \hline
1 & 1 & 2 & 2 & 1.999 & 2.000 & 2.005 & 2.85 & 4.08 & 3.90 \\ \hline
1 & 2 & 1 & 3 & 2.999 & 3.000 & 3.005 & 3.21 & 4.60 & 4.41 \\ \hline
2 & 1 & 1 & 3 & 2.999 & 3.000 & 3.005 & 3.21 & 4.60 & 4.41 \\ \hline
1 & 2 & 2 & 3 & 2.999 & 3.000 & 3.005 & 3.21 & 4.60 & 4.41 \\ \hline
2 & 1 & 2 & 3 & 2.999 & 3.000 & 3.005 & 3.21 & 4.60 & 4.41 \\ \hline
2 & 2 & 1 & 4 & 3.998 & 3.999 & 4.005 & 3.61 & 5.18 & 4.98 \\ \hline
2 & 2 & 2 & 4 & 3.998 & 3.999 & 4.005 & 3.61 & 5.18 & 4.98 \\ \hline
\multicolumn{10}{l}{$q=2$}\\
\hline
$\delta^{(1)}$ & $\delta^{(2)}$ & $\delta^{(3)}$ & $\tau_{2,2}$ & $\bE[\widehat{\tau}^*_{2,2}]$ & $\bE[\widehat{\tau}^\mathsf{H1}_{2,2}]$ & $\bE[\widehat{\tau}^\mathsf{H2}_{2,2}]$ & $r(\eta_{\bT^*})$ & $r(\eta_{\bT^\mathsf{H1}})$ & $r(\eta_{\bT^\mathsf{H2}})$ \\ \hline
1 & 1 & 1 & 3 & 3.016 & 3.012 & 3.002 & 7.96  & 10.22 & 8.11 \\ \hline
1 & 1 & 2 & 4 & 4.018 & 4.013 & 4.002 & 9.57  & 12.39 & 9.74 \\ \hline
1 & 2 & 1 & 4 & 4.018 & 4.013 & 4.002 & 9.57  & 12.39 & 9.74 \\ \hline
2 & 1 & 1 & 4 & 4.018 & 4.013 & 4.002 & 9.57  & 12.39 & 9.74  \\ \hline
1 & 2 & 2 & 5 & 5.020 & 5.015 & 5.003 & 11.34 & 14.81 & 11.52 \\ \hline
2 & 1 & 2 & 5 & 5.020 & 5.015 & 5.003 & 11.34 & 14.81 & 11.52 \\ \hline
2 & 2 & 1 & 5 & 5.020 & 5.015 & 5.003 & 11.34 & 14.81 & 11.52 \\ \hline
2 & 2 & 2 & 6 & 6.022 & 6.016 & 6.003 & 13.28 & 17.48 & 13.47 \\ \hline
\end{tabular}
}
{For each row, the random seed that generates the simulation setup is fixed. When $q=0$, the treatment effect is $\delta^{(1)}$; when $q=1$, the treatment effect is $\delta^{(1)} + \delta^{(2)}$; when $q=2$, the treatment effect is $\delta^{(1)} + \delta^{(2)} + \delta^{(3)}$.}
\end{table}

\subsubsection{Simulation results.}
We calculate the risk functions based on the outcome model in \eqref{eqn:simu:MoreGeneral}.
See Table~\ref{tbl:simu:RIskFunctions:OtherEstimands}.
For each $q$, as we vary the values of $\delta^{(1)}$, $\delta^{(2)}$ and $\delta^{(3)}$, the causal estimand is being changed.
For each $q$, all three estimators are able to reflect the change as the estimand changes.
This is because the Horvitz-Thompson estimator is unbiased.

Moving on to the risk function, we discuss $q=0,1,2$ separately.
When $q=0$, we see that the risk function associated with the first benchmark $\bT^\mathsf{H1}$ is smaller than the optimal design; and the second benchmark $\bT^\mathsf{H2}$ is much larger than the optimal design.
This is because the Horvitz-Thompson estimator as we defined in \eqref{eqn:OtherEstimators} does not use all the data, if there are periods that have zero probability to observe $Y_t(\bm{1}_{p-q}, \bm{1}_{q+1})$, such as when we are using $\bT^*$ or $\bT^\mathsf{H2}$.
When $q=1$, despite the fact that the Horvitz-Thompson estimator does not use all the data when we are using $\bT^*$ or $\bT^\mathsf{H2}$, we can see that the risk function associated with the first benchmark $\bT^\mathsf{H1}$ is larger than the optimal design; and the second benchmark $\bT^\mathsf{H2}$ is also larger.
Our proposed optimal design have the smallest risk.
When $q=2$, the table exactly recovers Table~\ref{tbl:simu:OneModelRisk}.
Again our proposed optimal design have the smallest risk

The simulation results across $q=0,1,2$ suggest that, whenever the primary interest is in estimating the instantaneous effects, we recommend to use a design of experiment that randomizes more frequently.
Whenever the primary interest is in estimating the lag effects, we recommend to use our optimal design as proposed in Theorem~\ref{thm:OptimalDesign}.
If the primary interest is undetermined or subject to future changes, our optimal design, combined with the proper Horvitz-Thompson estimator, still provides unbiased estimation whenever the causal estimand is specified.

\subsection{Additional Simulation Results for Section~\ref{sec:simu:AsymptoticNormality} Asymptotic Normality}
\label{sec:simu:additional:AsymptoticNormality}
In Section~\ref{sec:simu:AsymptoticNormality} we have only shown simulation results for the variance distribution, when $m$ is correctly specified and under $\delta=3$, see asymptotic normality.
In this section we provide additional simulation results under $\delta=1$ and $\delta=2$.

\begin{figure}[!htb]
\centering
\includegraphics[width=0.7\textwidth]{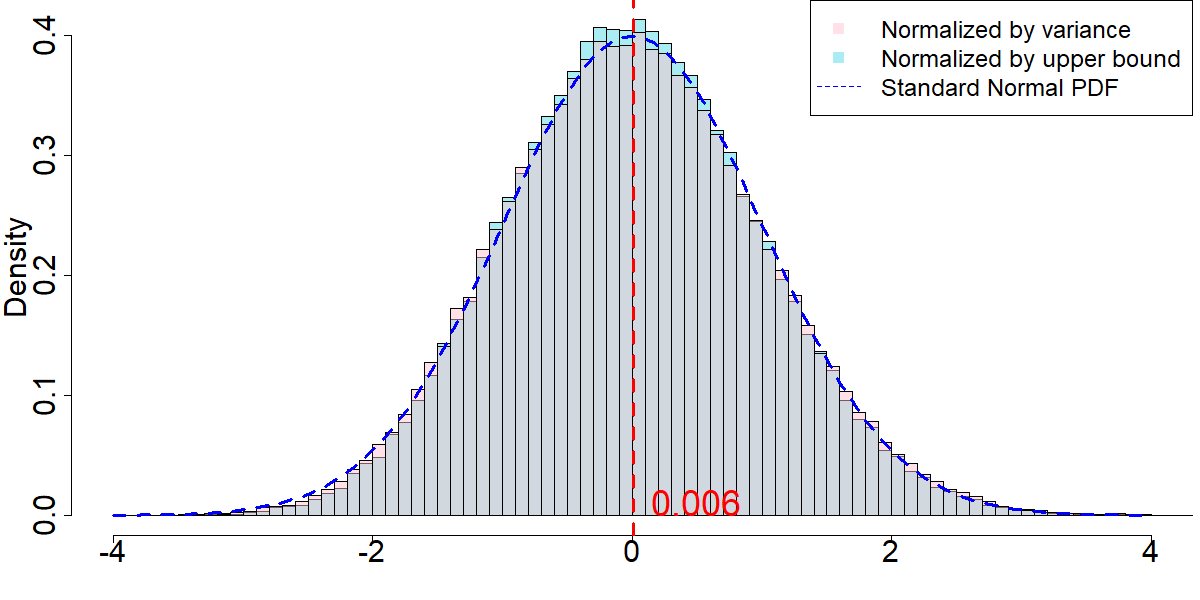}
\caption{Approximate normality of the randomization distribution when $m=2, p=2, \delta=1$.}
\label{fig:mCorrect:ApproximateNormality:delta=1}
\end{figure}

\begin{figure}[!htb]
\centering
\includegraphics[width=0.7\textwidth]{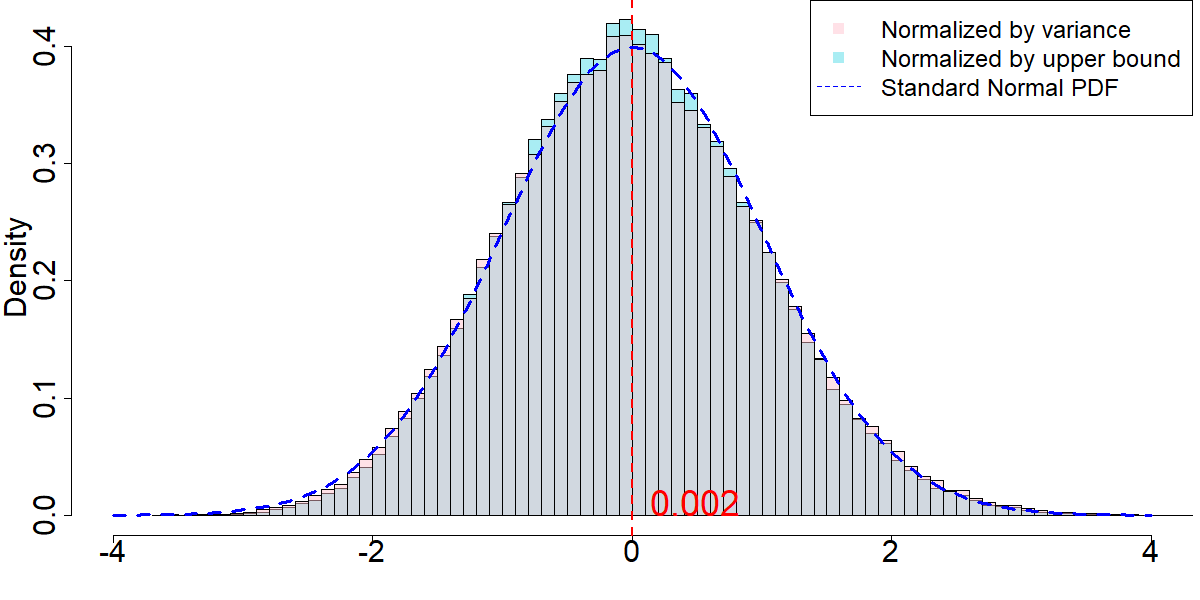}
\caption{Approximate normality of the randomization distribution when $m=2, p=3, \delta=1$.}
\label{fig:mOver:ApproximateNormality:delta=1}
\end{figure}

\begin{figure}[!htb]
\centering
\includegraphics[width=0.7\textwidth]{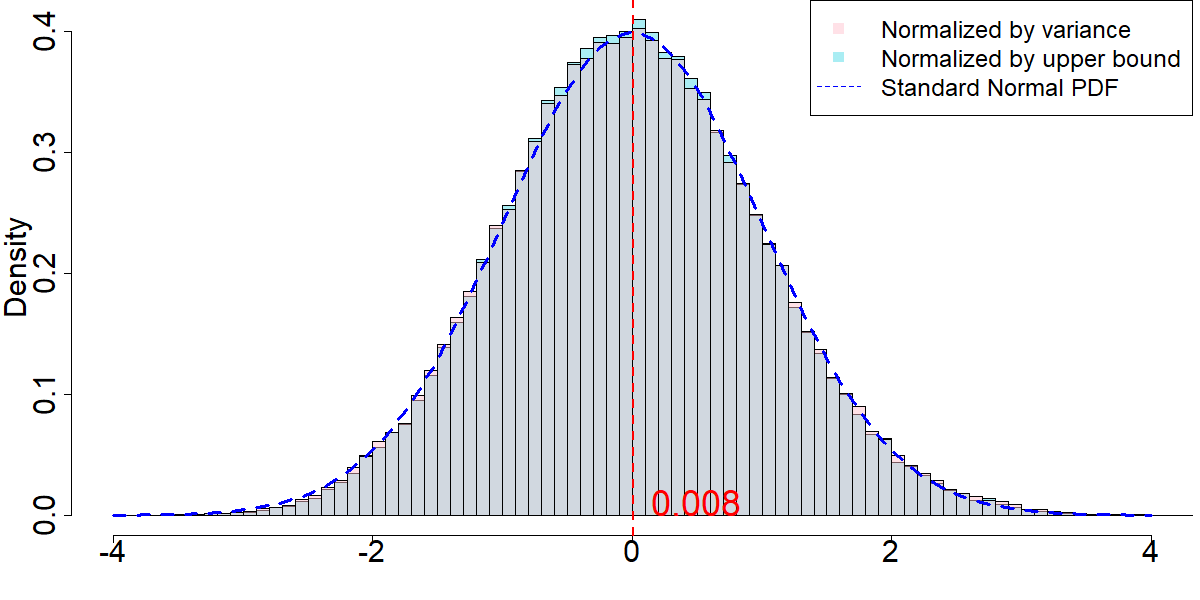}
\caption{Approximate normality of the randomization distribution when $m=2, p=1, \delta=1$.}
\label{fig:mUnder:ApproximateNormality:delta=1}
\end{figure}

\begin{figure}[!htb]
\centering
\includegraphics[width=0.7\textwidth]{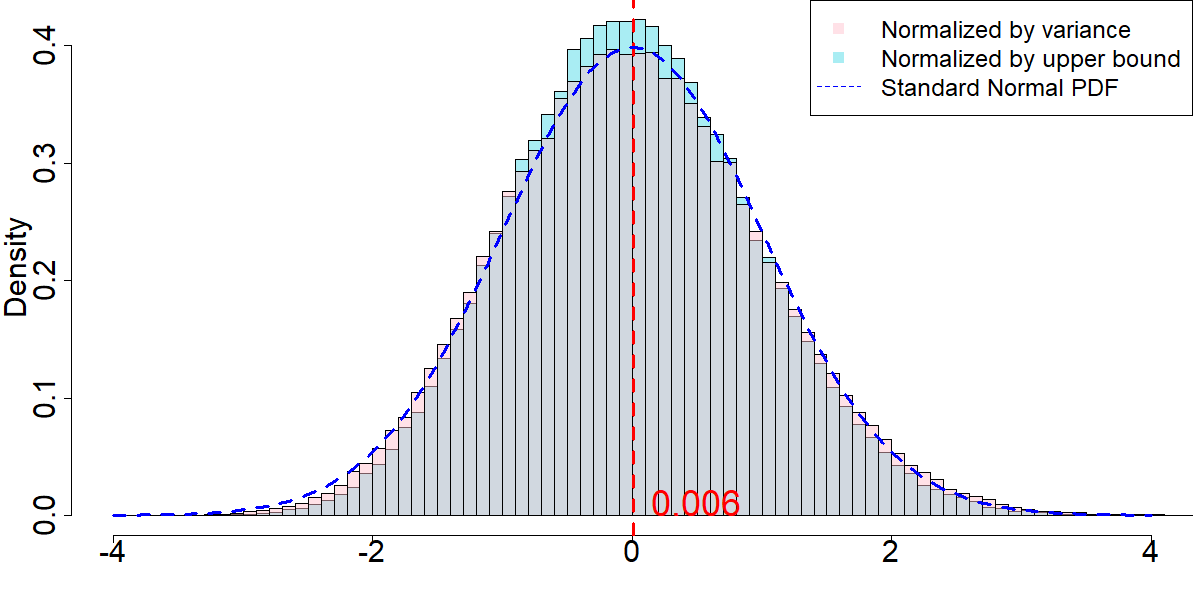}
\caption{Approximate normality of the randomization distribution when $m=2, p=2, \delta=2$.}
\label{fig:mCorrect:ApproximateNormality:delta=2}
\end{figure}

\begin{figure}[!htb]
\centering
\includegraphics[width=0.7\textwidth]{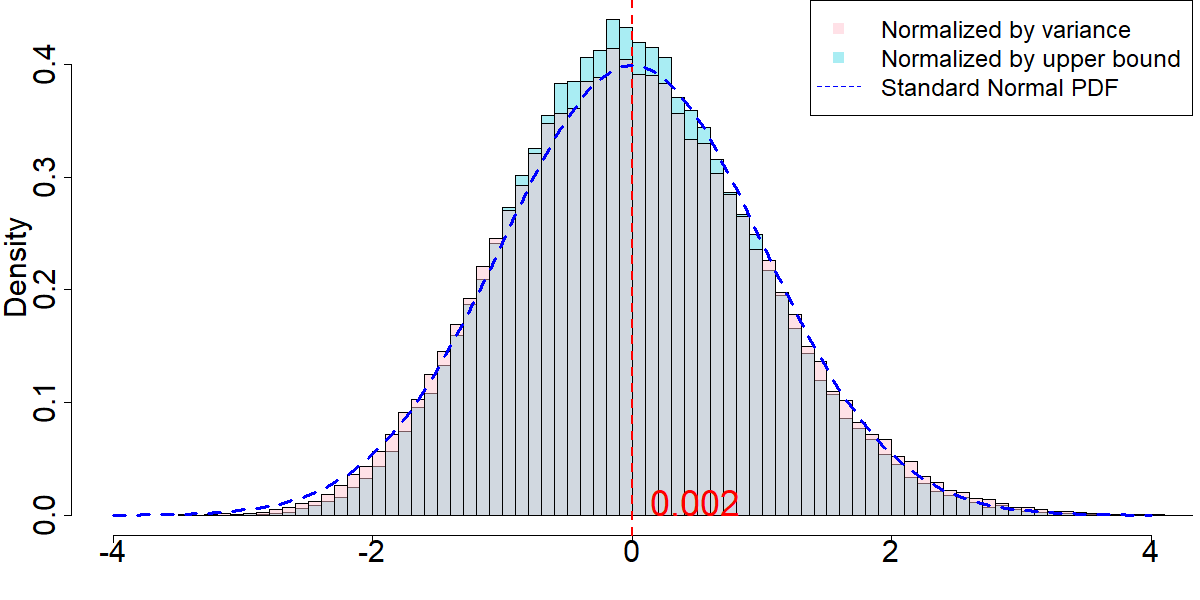}
\caption{Approximate normality of the randomization distribution when $m=2, p=3, \delta=2$.}
\label{fig:mOver:ApproximateNormality:delta=2}
\end{figure}

\begin{figure}[!htb]
\centering
\includegraphics[width=0.7\textwidth]{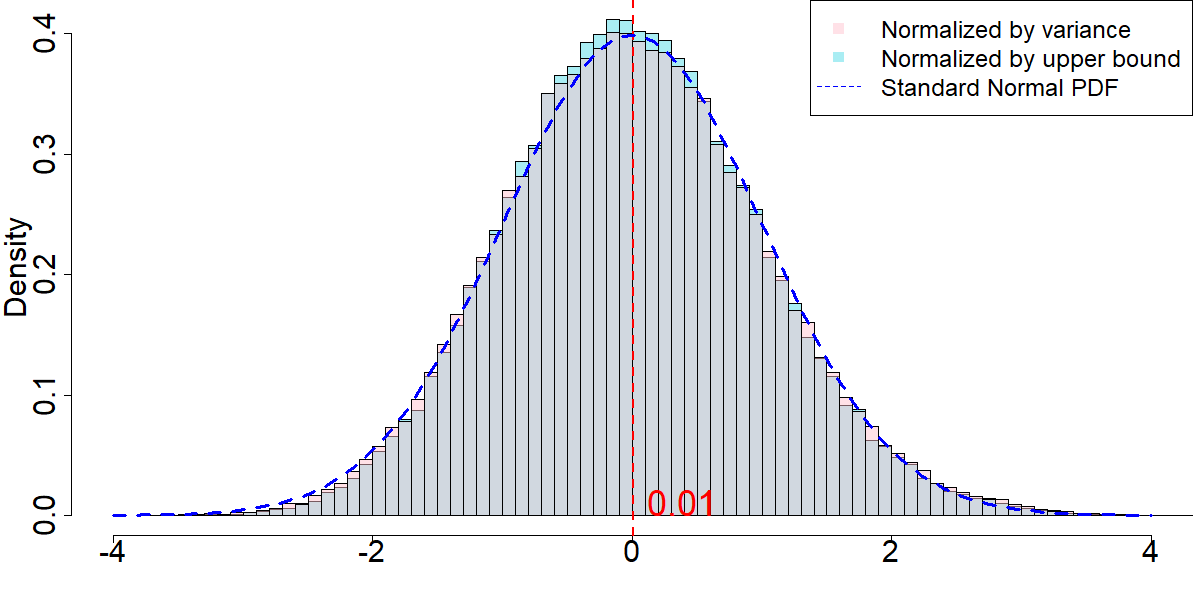}
\caption{Approximate normality of the randomization distribution when $m=2, p=1, \delta=2$.}
\label{fig:mUnder:ApproximateNormality:delta=2}
\end{figure}

\begin{figure}[!htb]
\centering
\includegraphics[width=0.7\textwidth]{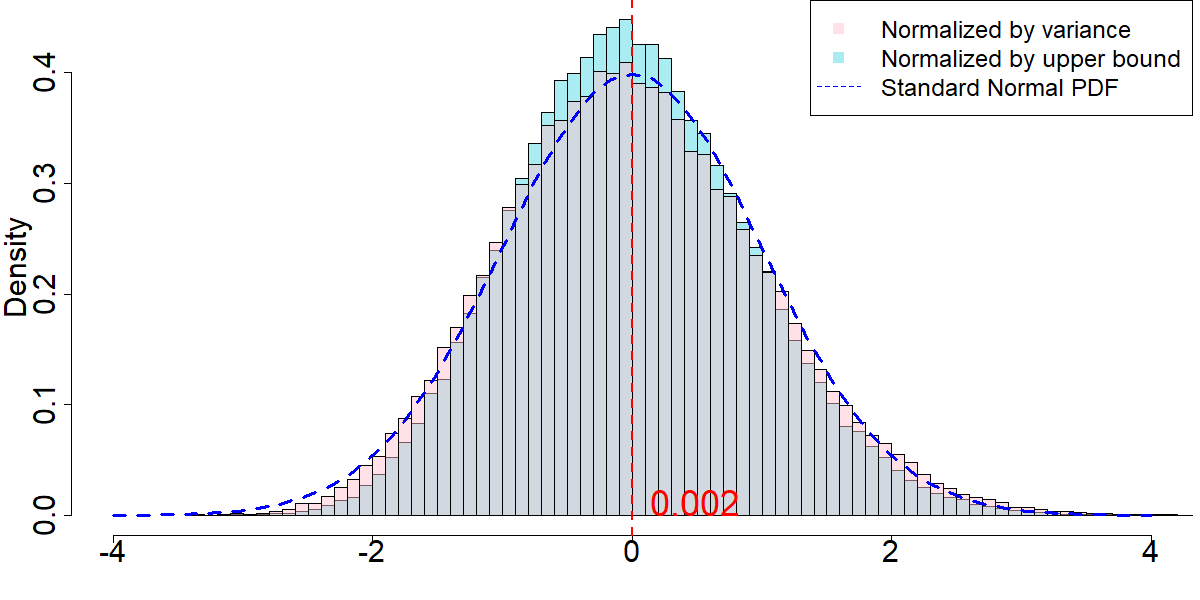}
\caption{Approximate normality of the randomization distribution when $m=2, p=3, \delta=3$.}
\label{fig:mOver:ApproximateNormality:delta=3}
\end{figure}

\begin{figure}[!htb]
\centering
\includegraphics[width=0.7\textwidth]{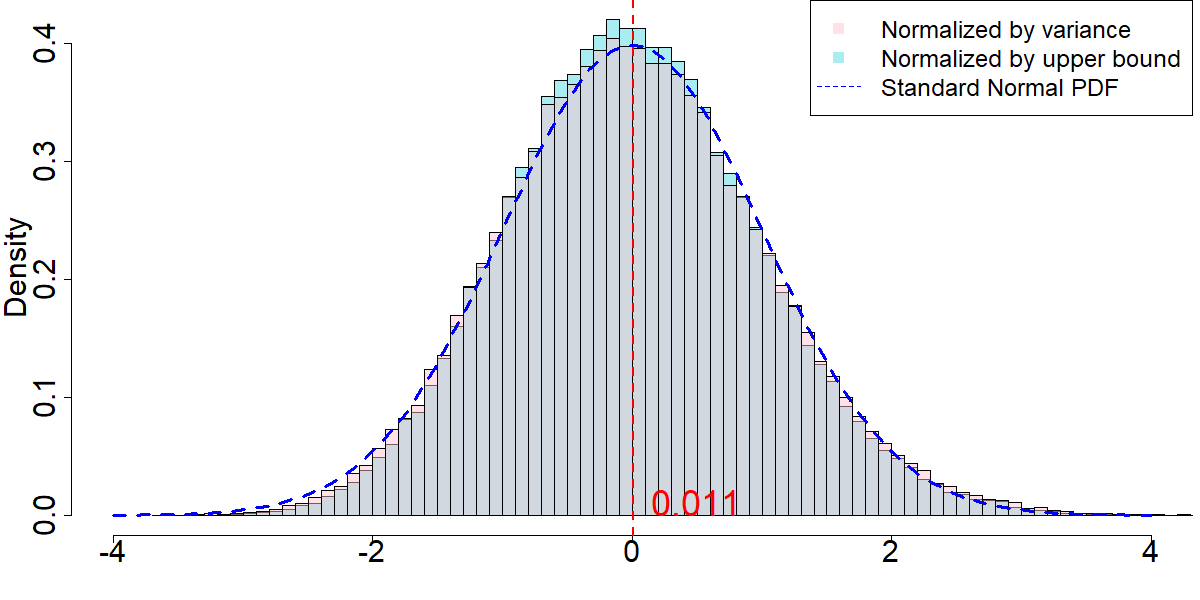}
\caption{Approximate normality of the randomization distribution when $m=2, p=1, \delta=3$.}
\label{fig:mUnder:ApproximateNormality:delta=3}
\end{figure}

See Figures~\ref{fig:mCorrect:ApproximateNormality:delta=1}--\ref{fig:mUnder:ApproximateNormality:delta=1} for simulation results under $\delta=1$;
See Figures~\ref{fig:mCorrect:ApproximateNormality:delta=2}--\ref{fig:mUnder:ApproximateNormality:delta=2} for simulation results under $\delta=2$.
See Figures~\ref{fig:mOver:ApproximateNormality:delta=3}--\ref{fig:mUnder:ApproximateNormality:delta=3} for simulation results under $\delta=3$.

By comparing all the results, we see that in all cases, the pink histograms approximately follow the standard normal distribution; whereas the light blue histograms, since the distributions are induced by normalizing the expectation of the conservative upper bound, are more concentrated around zero.
Furthermore, as $\delta$ increases, the light blue histograms become even more concentrated around zero, i.e., the distances between the light blue histograms and the pink histograms grow larger.




\subsubsection{Robustness Check.}
\begin{figure}[!htb]
\centering
\includegraphics[width=0.7\textwidth]{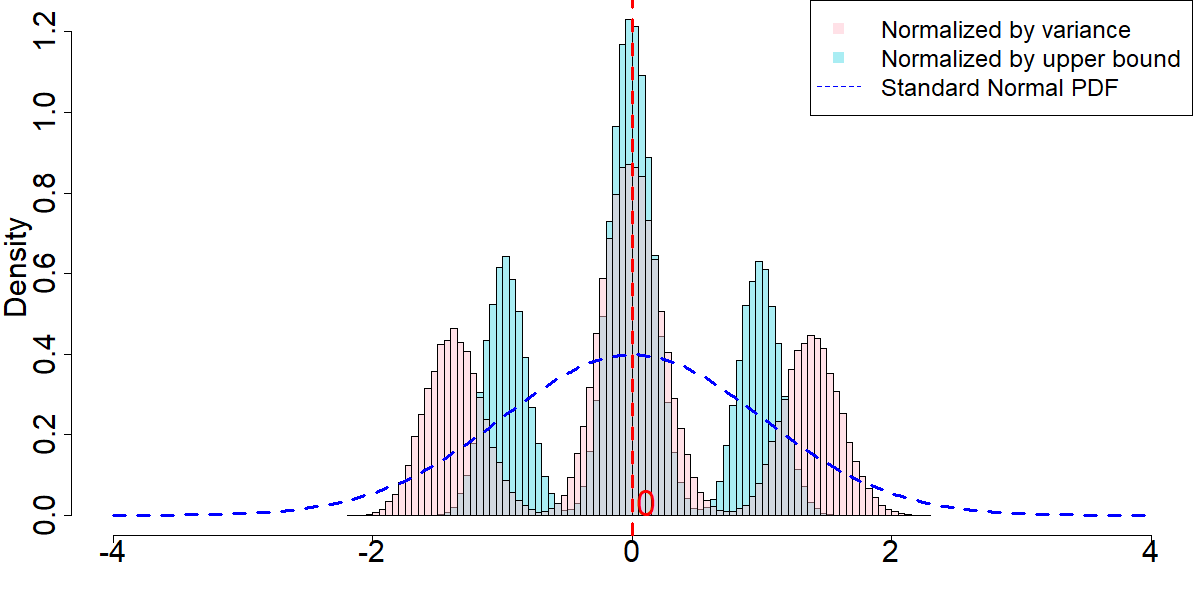}
\caption{Randomization distribution when random noises are Student's t-distributions, and when $m=2, p=2, \delta=2, T=120$.}
\label{fig:mCorrect:ApproximateNormality:delta=2;T=120}
\end{figure}

\begin{figure}[!htb]
\centering
\includegraphics[width=0.7\textwidth]{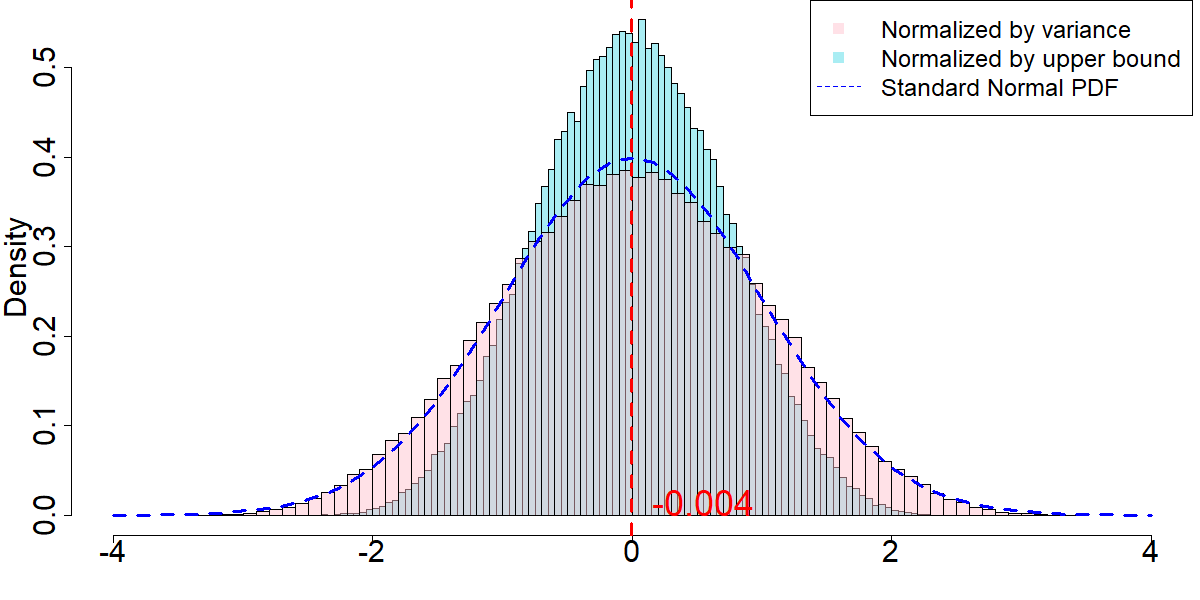}
\caption{Randomization distribution when random noises are Student's t-distributions, and when $m=2, p=2, \delta=2, T=1200$.}
\label{fig:mCorrect:ApproximateNormality:delta=2;T=1200}
\end{figure}

\begin{figure}[!htb]
\centering
\includegraphics[width=0.7\textwidth]{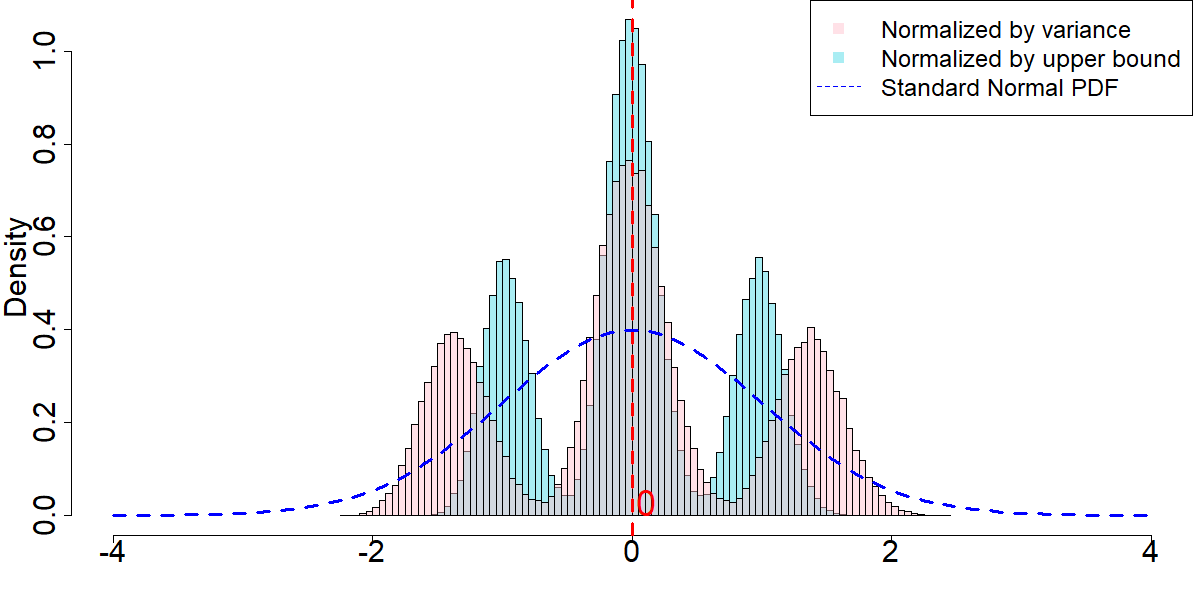}
\caption{Randomization distribution when random noises are Student's t-distributions, and when $m=2, p=2, \delta=3, T=120$.}
\label{fig:mCorrect:ApproximateNormality:delta=3;T=120}
\end{figure}

\begin{figure}[!htb]
\centering
\includegraphics[width=0.7\textwidth]{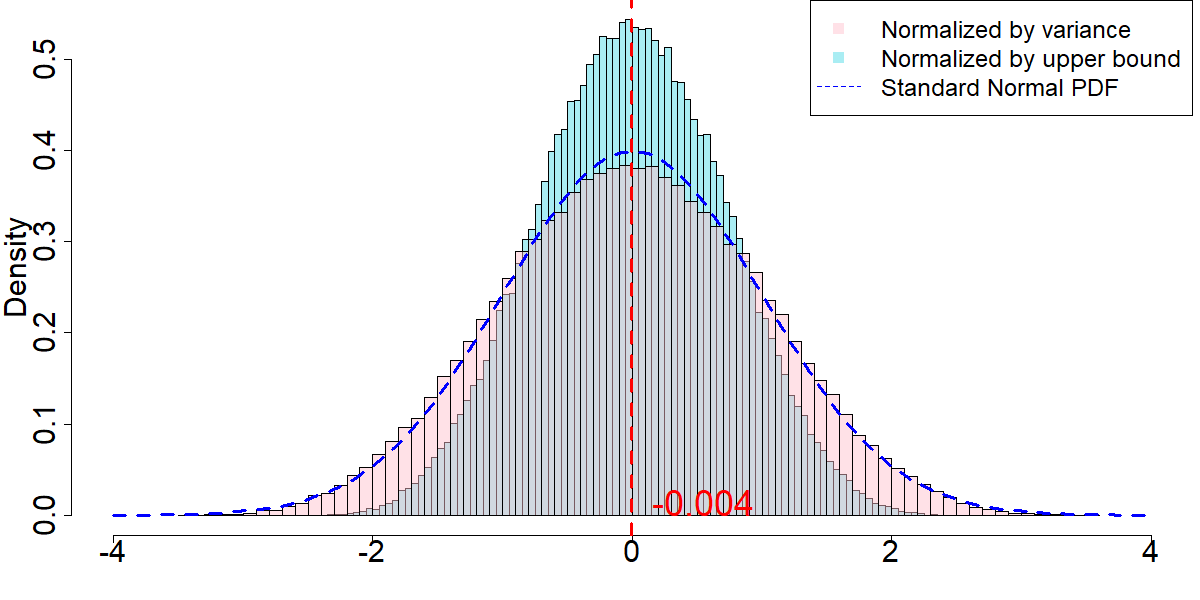}
\caption{Randomization distribution when random noises are Student's t-distributions, and when $m=2, p=2, \delta=3, T=1200$.}
\label{fig:mCorrect:ApproximateNormality:delta=3;T=1200}
\end{figure}

\begin{figure}[!htb]
\centering
\includegraphics[width=0.7\textwidth]{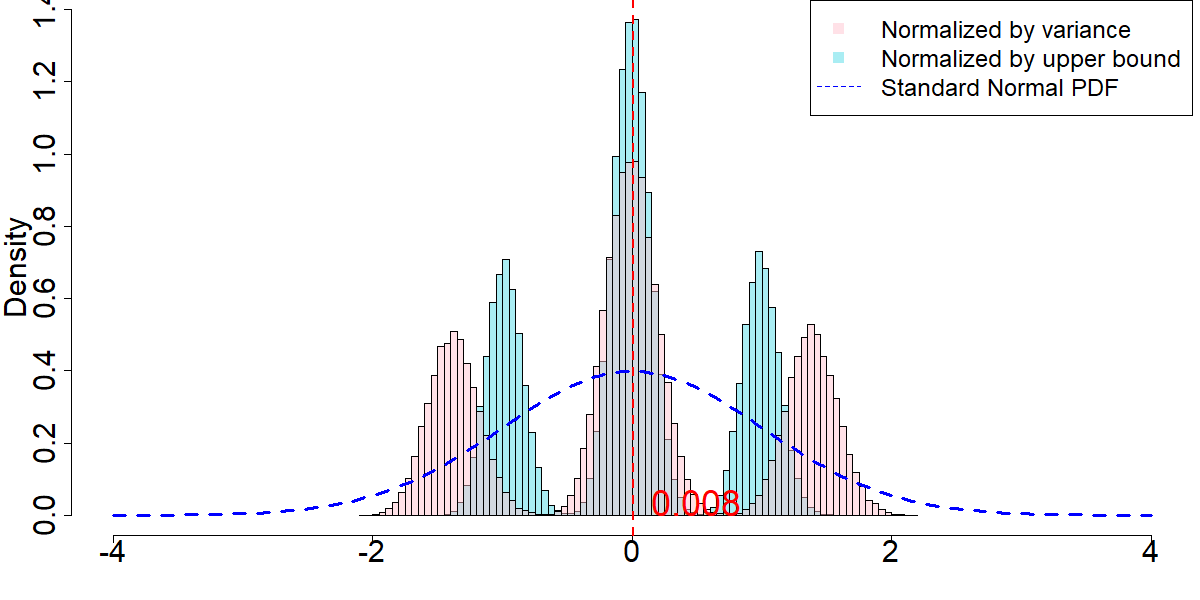}
\caption{Randomization distribution when random noises are Student's t-distributions, and when $m=2, p=3, \delta=1, T=120$.}
\label{fig:mOver:ApproximateNormality:delta=1;T=120}
\end{figure}

\begin{figure}[!htb]
\centering
\includegraphics[width=0.7\textwidth]{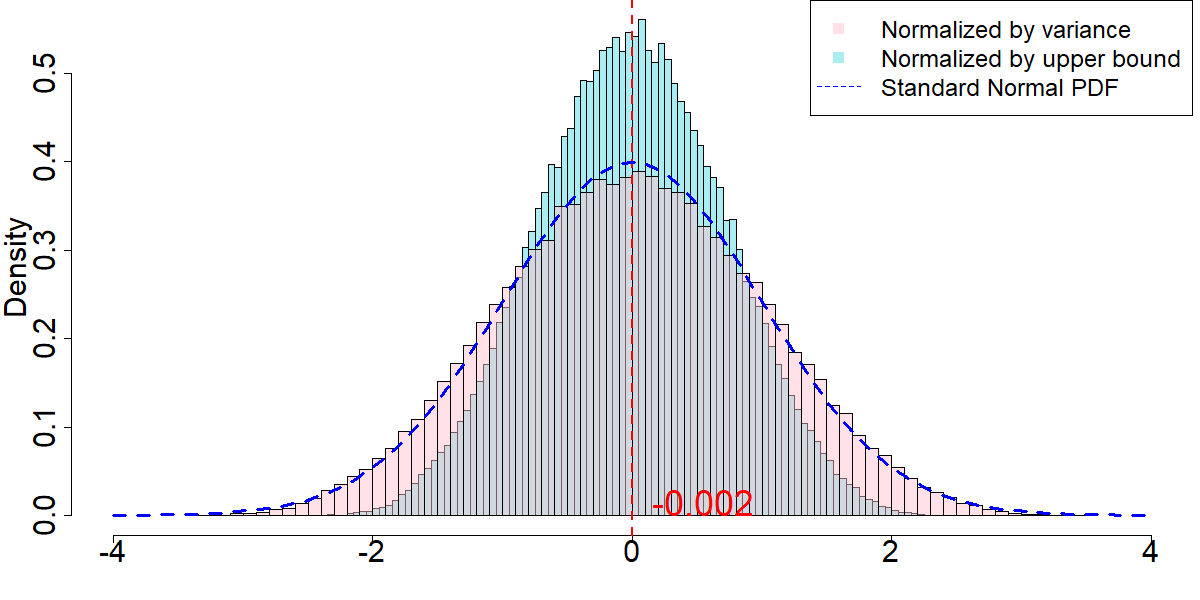}
\caption{Randomization distribution when random noises are Student's t-distributions, and when $m=2, p=3, \delta=1, T=1200$.}
\label{fig:mOver:ApproximateNormality:delta=1;T=1200}
\end{figure}

\begin{figure}[!htb]
\centering
\includegraphics[width=0.7\textwidth]{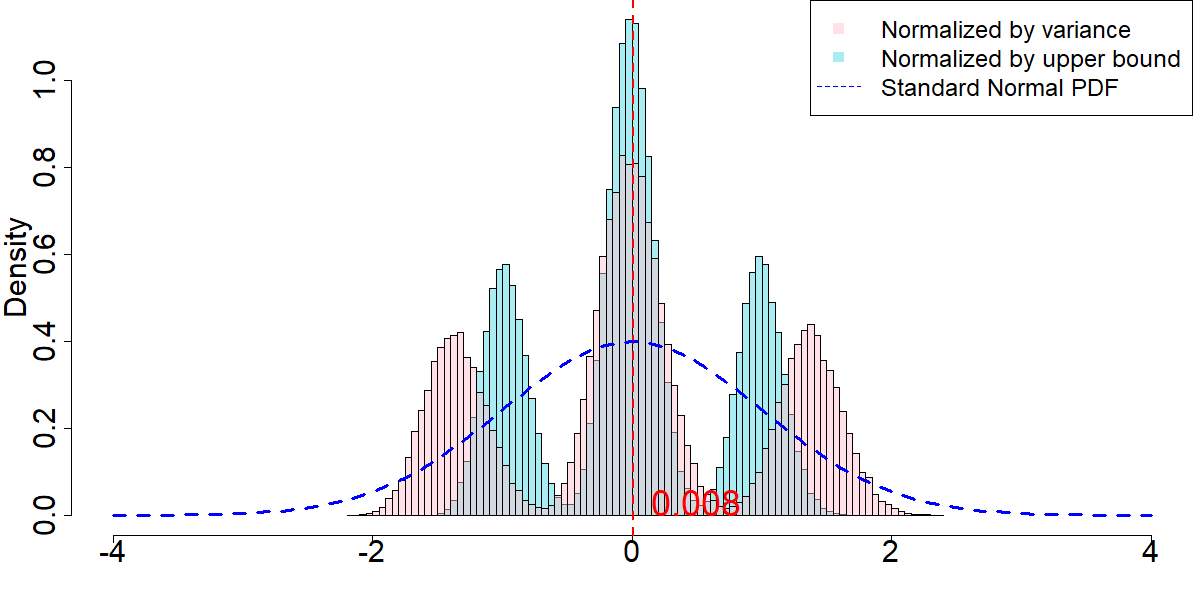}
\caption{Randomization distribution when random noises are Student's t-distributions, and when $m=2, p=3, \delta=2, T=120$.}
\label{fig:mOver:ApproximateNormality:delta=2;T=120}
\end{figure}

\begin{figure}[!htb]
\centering
\includegraphics[width=0.7\textwidth]{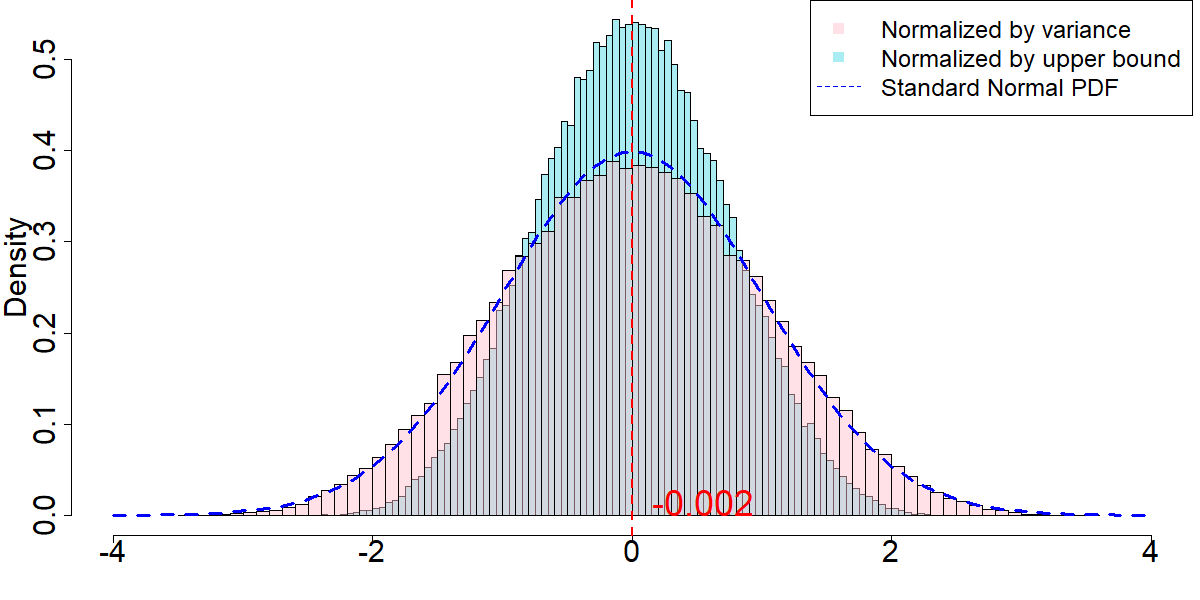}
\caption{Randomization distribution when random noises are Student's t-distributions, and when $m=2, p=3, \delta=2, T=1200$.}
\label{fig:mOver:ApproximateNormality:delta=2;T=1200}
\end{figure}

\begin{figure}[!htb]
\centering
\includegraphics[width=0.7\textwidth]{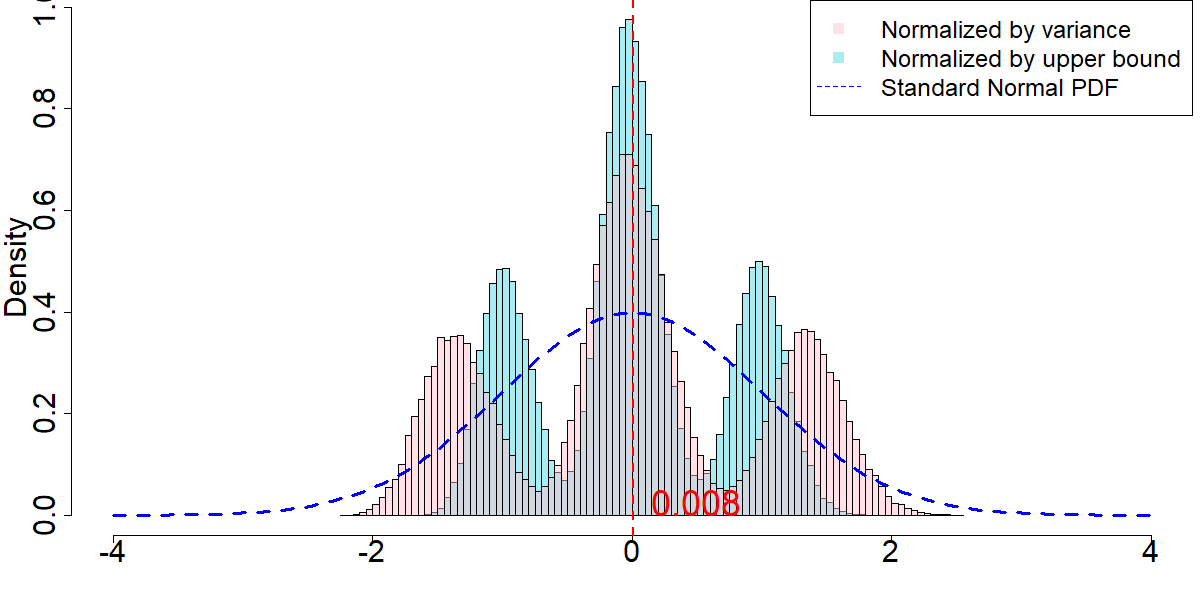}
\caption{Randomization distribution when random noises are Student's t-distributions, and when $m=2, p=3, \delta=3, T=120$.}
\label{fig:mOver:ApproximateNormality:delta=3;T=120}
\end{figure}

\begin{figure}[!htb]
\centering
\includegraphics[width=0.7\textwidth]{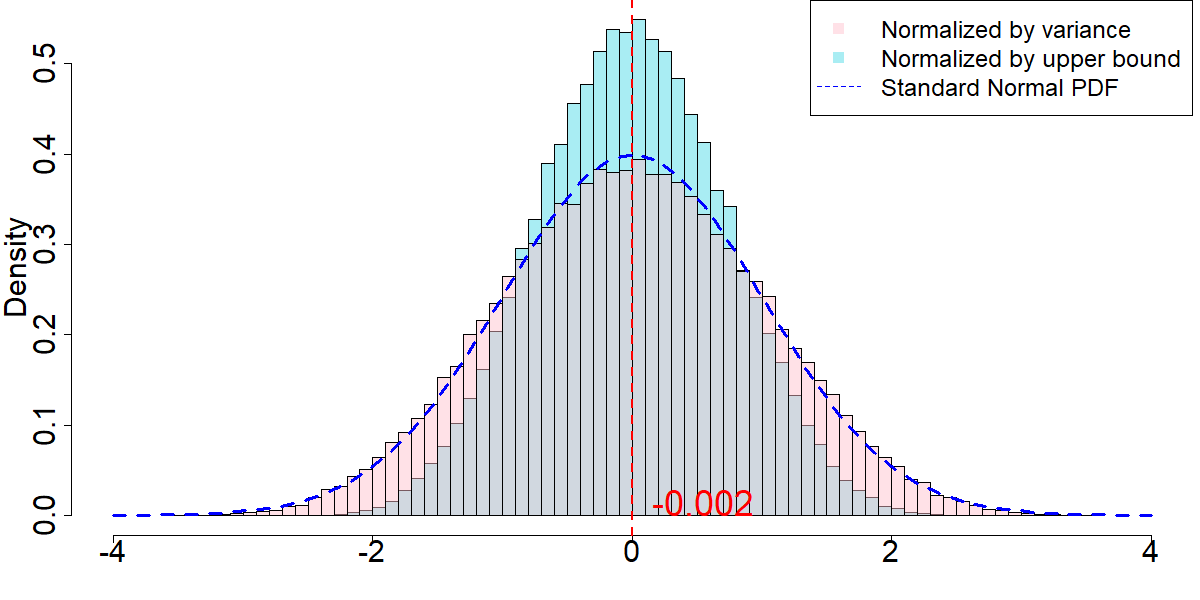}
\caption{Randomization distribution when random noises are Student's t-distributions, and when $m=2, p=3, \delta=3, T=1200$.}
\label{fig:mOver:ApproximateNormality:delta=3;T=1200}
\end{figure}

\begin{figure}[!htb]
\centering
\includegraphics[width=0.7\textwidth]{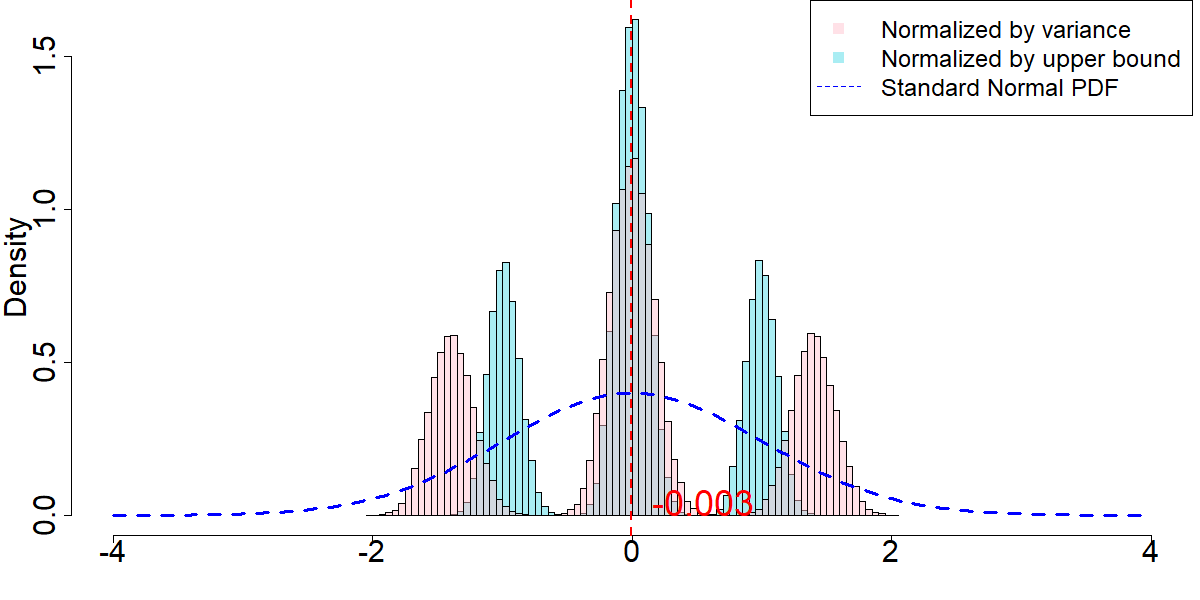}
\caption{Randomization distribution when random noises are Student's t-distributions, and when $m=2, p=1, \delta=1, T=120$.}
\label{fig:mUnder:ApproximateNormality:delta=1;T=120}
\end{figure}

\begin{figure}[!htb]
\centering
\includegraphics[width=0.7\textwidth]{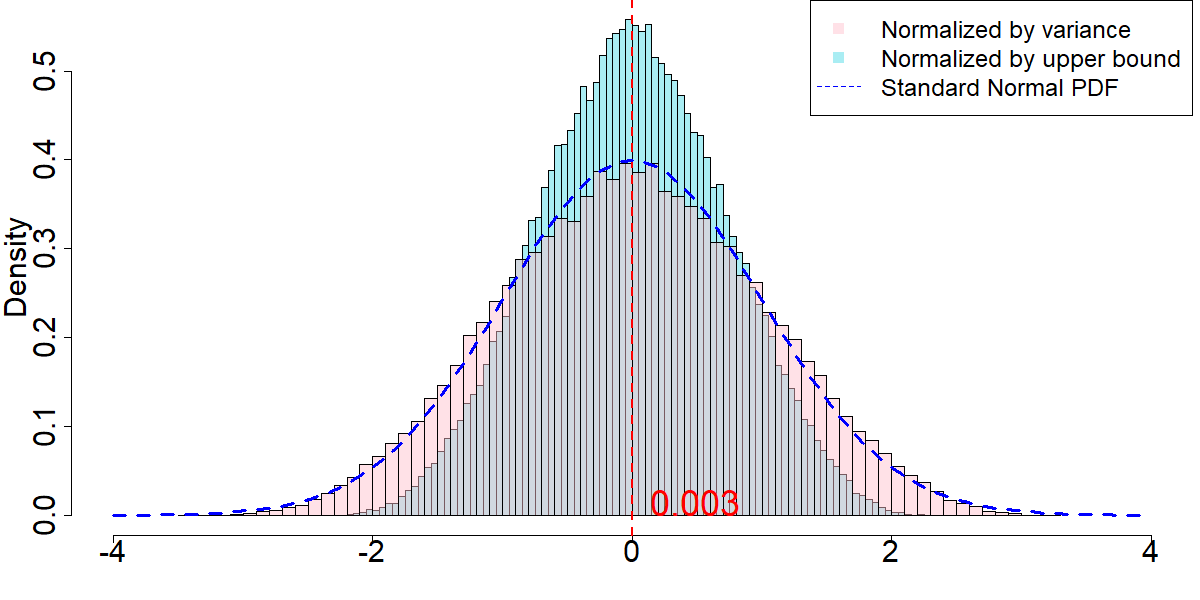}
\caption{Randomization distribution when random noises are Student's t-distributions, and when $m=2, p=1, \delta=1, T=1200$.}
\label{fig:mUnder:ApproximateNormality:delta=1;T=1200}
\end{figure}

\begin{figure}[!htb]
\centering
\includegraphics[width=0.7\textwidth]{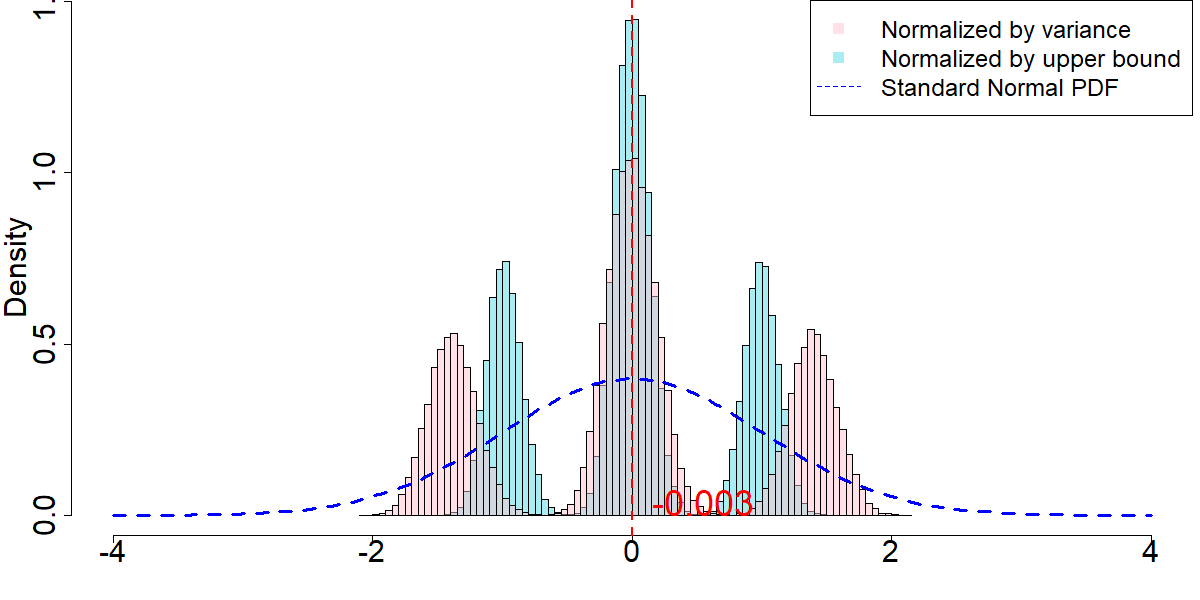}
\caption{Randomization distribution when random noises are Student's t-distributions, and when $m=2, p=1, \delta=2, T=120$.}
\label{fig:mUnder:ApproximateNormality:delta=2;T=120}
\end{figure}

\begin{figure}[!htb]
\centering
\includegraphics[width=0.7\textwidth]{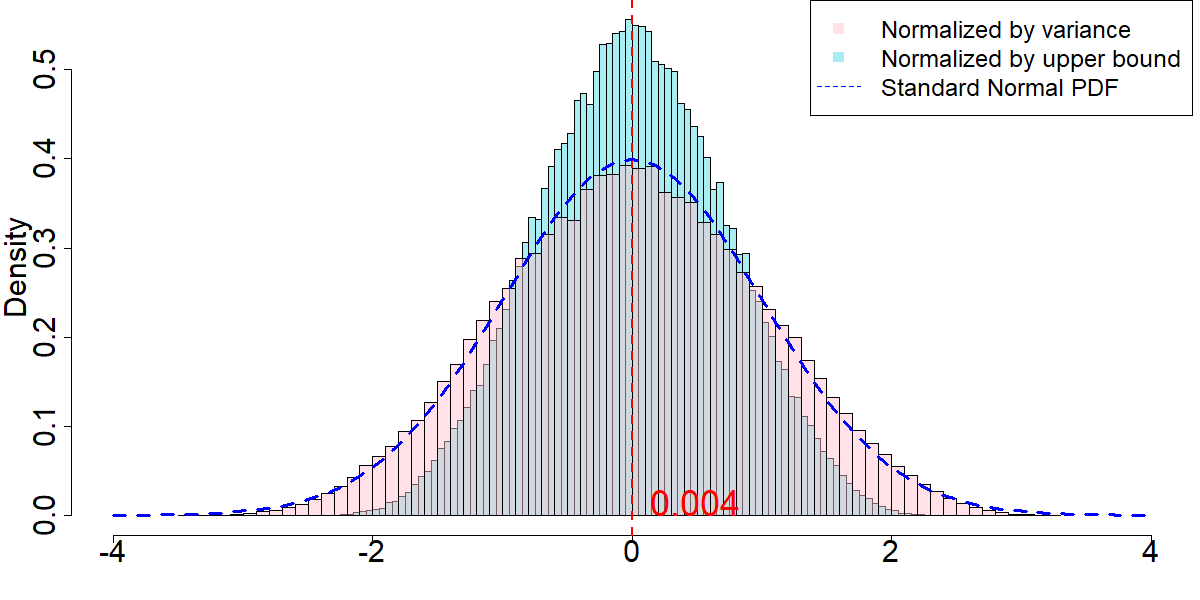}
\caption{Randomization distribution when random noises are Student's t-distributions, and when $m=2, p=1, \delta=2, T=1200$.}
\label{fig:mUnder:ApproximateNormality:delta=2;T=1200}
\end{figure}

\begin{figure}[!htb]
\centering
\includegraphics[width=0.7\textwidth]{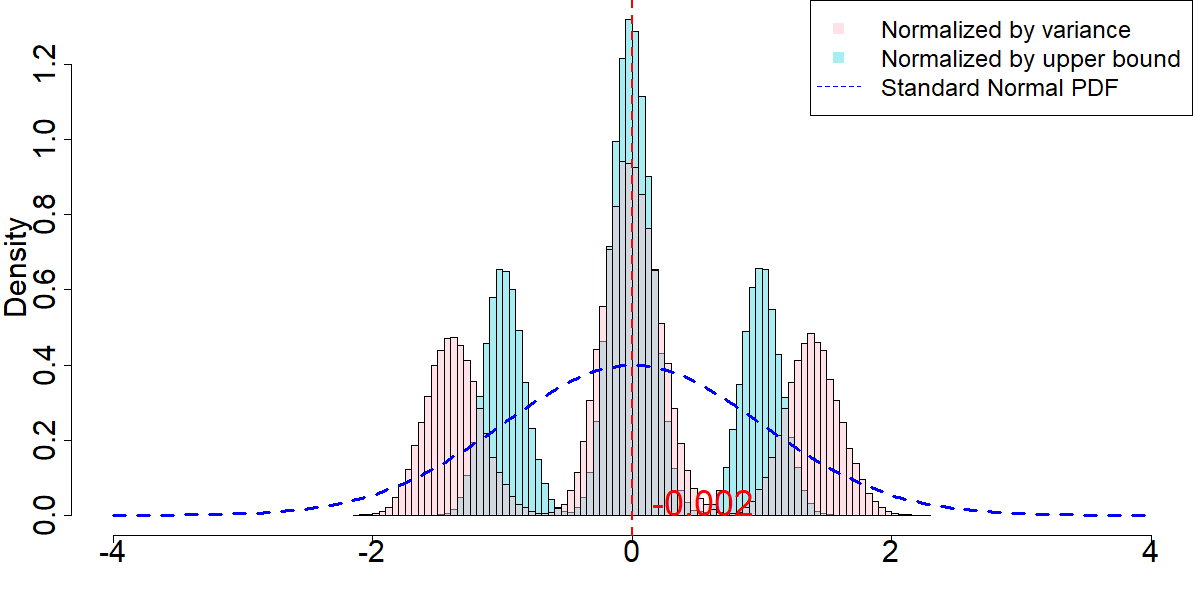}
\caption{Randomization distribution when random noises are Student's t-distributions, and when $m=2, p=1, \delta=3, T=120$.}
\label{fig:mUnder:ApproximateNormality:delta=3;T=120}
\end{figure}

\begin{figure}[!htb]
\centering
\includegraphics[width=0.7\textwidth]{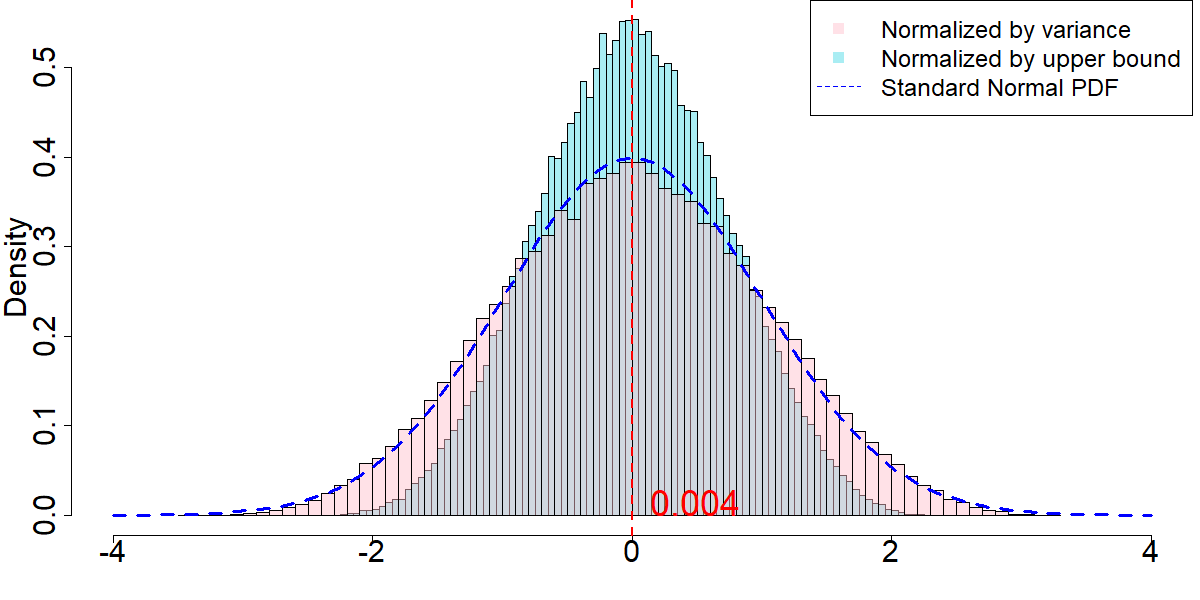}
\caption{Randomization distribution when random noises are Student's t-distributions, and when $m=2, p=1, \delta=3, T=1200$.}
\label{fig:mUnder:ApproximateNormality:delta=3;T=1200}
\end{figure}

In Section~\ref{sec:simu:Cauchy:AsymptoticNormality} we have shown results when $m=2,p=2,\delta=1$.
In this section we provide additional simulation results under other parameters.
When $T=120$, the empirical distributions as shown in the histograms are significantly different from normal distributions. See Figures~\ref{fig:mCorrect:ApproximateNormality:delta=1;T=120},~\ref{fig:mCorrect:ApproximateNormality:delta=2;T=120},~\ref{fig:mCorrect:ApproximateNormality:delta=3;T=120},~\ref{fig:mOver:ApproximateNormality:delta=1;T=120},~\ref{fig:mOver:ApproximateNormality:delta=2;T=120},~\ref{fig:mOver:ApproximateNormality:delta=3;T=120},~\ref{fig:mUnder:ApproximateNormality:delta=1;T=120},~\ref{fig:mUnder:ApproximateNormality:delta=2;T=120},~\ref{fig:mUnder:ApproximateNormality:delta=3;T=120}.
When $T=1200$, the empirical distributions as shown in the histograms are much closer to normal distributions. See Figures~\ref{fig:mCorrect:ApproximateNormality:delta=1;T=1200},~\ref{fig:mCorrect:ApproximateNormality:delta=2;T=1200},~\ref{fig:mCorrect:ApproximateNormality:delta=3;T=1200},~\ref{fig:mOver:ApproximateNormality:delta=1;T=1200},~\ref{fig:mOver:ApproximateNormality:delta=2;T=1200},~\ref{fig:mOver:ApproximateNormality:delta=3;T=1200},~\ref{fig:mUnder:ApproximateNormality:delta=1;T=1200},~\ref{fig:mUnder:ApproximateNormality:delta=2;T=1200},~\ref{fig:mUnder:ApproximateNormality:delta=3;T=1200}.
All the simulation results deliver the same message, that when $\epsilon_t$ noises are heavy tailed, the convergence to a standard normal distribution as we have shown in Theorem~\ref{thm:AsymptoticNormality} requires longer horizon.

Interestingly, if we make the comparison between the pink histogram and the light blue histogram, we can see how much gap it incurs when we replace the true variance with the conservative upper bound.
If we compare Figure~\ref{fig:mCorrect:ApproximateNormality:delta=1} and Figure~\ref{fig:mCorrect:ApproximateNormality:delta=1;T=1200}, then we find that the conservative upper bound is a better approximation of the true variance when the noises $\epsilon_t$ conform normal distributions, rather than heavy-tailed distributions.









\subsection{Additional Simulation Results for Section~\ref{sec:simu:RejectionRates} Rejection Rates}
\label{sec:simu:additional:RejectionRates}

\begin{figure}[!htb]
\begin{subfigure}{.33\textwidth}
\centering
\includegraphics[width=\textwidth]{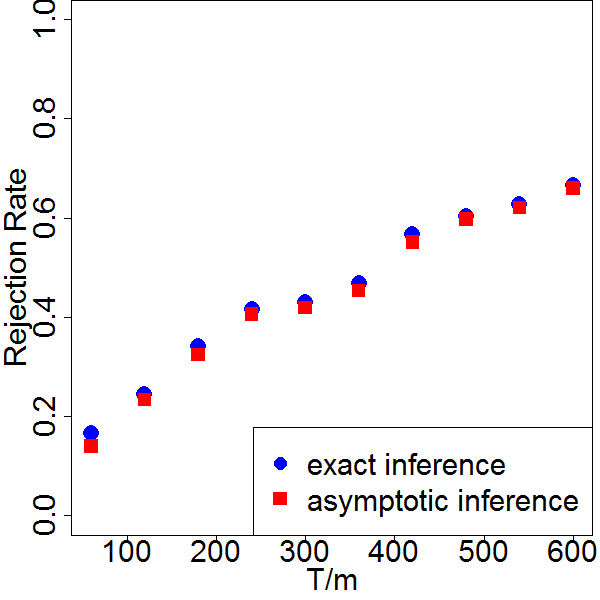}
\end{subfigure}\hfill
\begin{subfigure}{.33\textwidth}
\centering
\includegraphics[width=\textwidth]{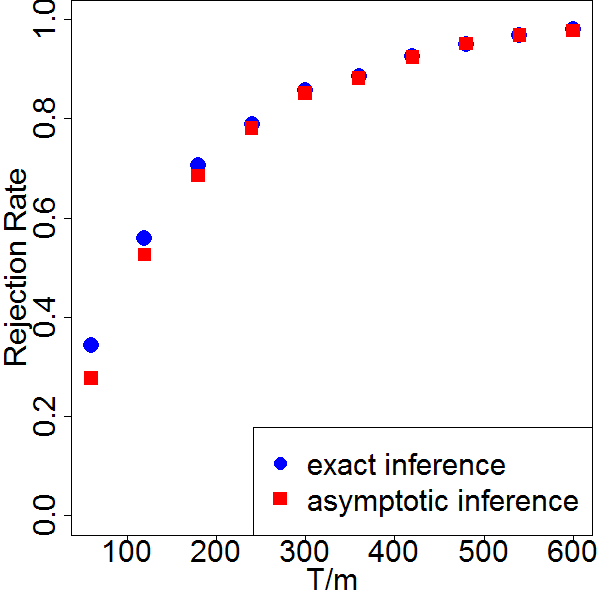}
\end{subfigure}\hfill
\begin{subfigure}{.33\textwidth}
\centering
\includegraphics[width=\textwidth]{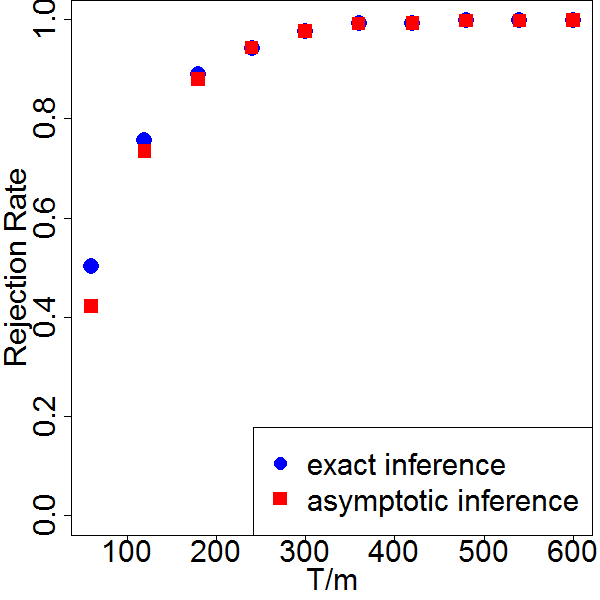}
\end{subfigure}
\caption{Rejection rates and their dependence on $T/m$, when the rejection threshold is $0.05$. Left: $\delta=1$; Middle: $\delta=2$; Right: $\delta=3$}
\label{fig:additional:RejectionRates005}
\end{figure}

\begin{figure}[!htb]
\begin{subfigure}{.33\textwidth}
\centering
\includegraphics[width=\textwidth]{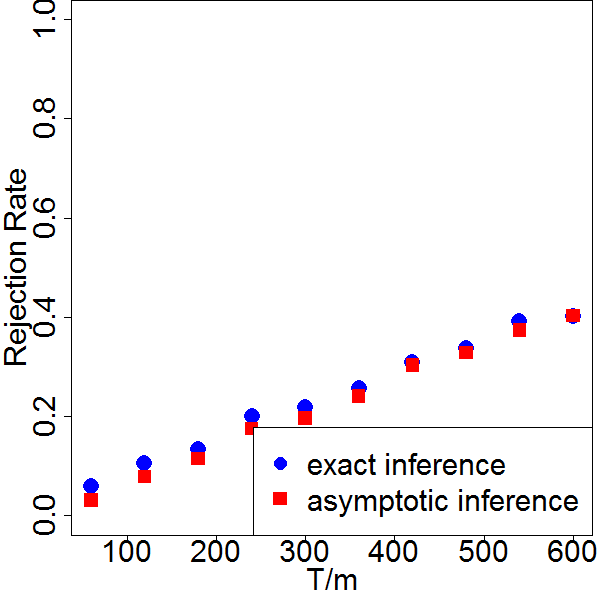}
\end{subfigure}\hfill
\begin{subfigure}{.33\textwidth}
\centering
\includegraphics[width=\textwidth]{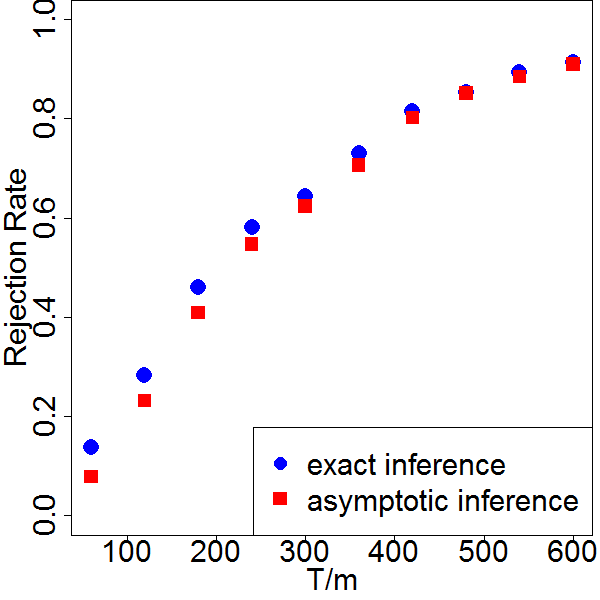}
\end{subfigure}\hfill
\begin{subfigure}{.33\textwidth}
\centering
\includegraphics[width=\textwidth]{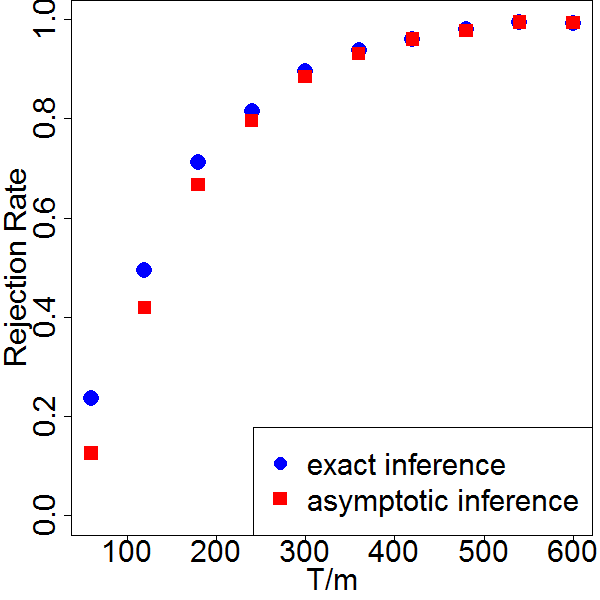}
\end{subfigure}
\caption{Rejection rates and their dependence on $T/m$, when the rejection threshold is $0.01$. Left: $\delta=1$; Middle: $\delta=2$; Right: $\delta=3$}
\label{fig:additional:RejectionRates001}
\end{figure}

In Section~\ref{sec:simu:RejectionRates} we have provided simulation results for the rejection rates when the rejection threshold is $0.1$.
In this section we provide additional simulation results for the rejection rates when the rejection threshold is replaced by $0.05$ and $0.01$. See Figures~\ref{fig:additional:RejectionRates005} and~\ref{fig:additional:RejectionRates001}.

The blue dots are rejection rates under exact inference; the red dots are under asymptotic inference.
Similar to the simulation results in Section~\ref{sec:simu:RejectionRates}, we would ideally wish to reject both the Fisher's null hypothesis \eqref{Fisher null} and the Neyman's null hypothesis \eqref{Neyman null}. Both figures illustrate such rejection rates.

Besides the three observations we make in Section~\ref{sec:simu:RejectionRates} (namely, dependence on $T/m$, between two inference methods, and dependence on the signal-to-noise ratio), we make an extra observation here.
When we decrease the rejection threshold, we expect to reject the Neyman's null hypothesis under smaller $p$-values.
As a result, as we decrease the rejection threshold, the rejection rates should be smaller, which is supported by our simulation results in Figures~\ref{fig:additional:RejectionRates005} and~\ref{fig:additional:RejectionRates001}.

\end{document}